\documentclass[10pt]{article}

\usepackage{amsmath}
\usepackage{mathtools}
\usepackage{stackrel}
\usepackage{leftidx}
\usepackage[amsmath,amsthm,thmmarks]{ntheorem}
\usepackage{scalerel}

\usepackage{authblk}

\usepackage{euscript}
\usepackage[charter]{mathdesign}
\let\mathbb\undefined
\usepackage{bbold}

\usepackage[shortlabels]{enumitem}

\usepackage{hyperref}
\usepackage{cleveref}
\hypersetup{colorlinks=true, linkcolor=red!50!black, citecolor=green!50!black, urlcolor=blue!80!black}

\usepackage{xcolor}
\usepackage{tcolorbox}
\usepackage{graphicx}
\usepackage{subfigure}

\usepackage[all,2cell,arrow,matrix]{xy} \UseAllTwocells \SilentMatrices
\usepackage{tikz-cd}
\usepackage{tikz}
\usetikzlibrary{arrows,arrows.meta,decorations.markings,decorations.pathreplacing,calc}

\usepackage{verbatim}
\usepackage{comment}

\usepackage{epstopdf} 

\usepackage[nottoc]{tocbibind} 
\usepackage{appendix}

\usepackage{geometry}
\tikzset{->-/.style={decoration={markings,mark=at position #1 with {\arrow{Stealth}}},postaction={decorate}},->-/.default=0.55}

\theoremstyle{definition}

\newtheorem{thm}{Theorem}[section]

{
\theoremsymbol{\mbox{${\blacksquare}$}}
\newtheorem{defn}[thm]{Definition}
}
{
\theoremsymbol{\mbox{$\heartsuit$}}

}
{
\theoremsymbol{\mbox{$\diamondsuit$}}
\newtheorem{rem}[thm]{Remark}
}
\qedsymbol{\mbox{$\square$}}

\numberwithin{equation}{section}
\numberwithin{thm}{section}

\newcommand\be            {\begin{equation}}
\newcommand\ee            {\end{equation}}
\newcommand\bea           {\begin{eqnarray}}
\newcommand\eea         {\end{eqnarray}}
\newcommand\bnu          {\begin{enumerate}}
\newcommand\enu          {\end{enumerate}}
\newcommand\bit          {\begin{itemize}}
\newcommand\eit          {\end{itemize}}

\newcommand{\pf}{\begin{proof}}
\newcommand{\epf}{\qed\end{proof}}

\makeatletter
\providecommand{\leftsquigarrow}{%
  \mathrel{\mathpalette\reflect@squig\relax}%
}
\newcommand{\reflect@squig}[2]{%
  \reflectbox{$\m@th#1\rightsquigarrow$}%
}
\makeatother

\DeclareSymbolFont{usualmathcal}{OMS}{cmsy}{m}{n}
\DeclareSymbolFontAlphabet{\mathcal}{usualmathcal}

\newcommand\Zb			{\mathbb{Z}}

\newcommand\CC			{\EuScript{C}}

\newcommand\CM			{\EuScript{M}}

\newcommand{\FZ}			{\text{\usefont{U}{euf}{m}{n}Z}}

\newcommand\SM			{\mathsf{M}}

\newcommand{\one}			{\mathbb{1}}

\newcommand\vect			{\mathrm{Vec}}

\newcommand\rep			{\mathrm{Rep}}

\newcommand\sfc        {$\mathrm{SFC}$}

\newcommand{\bscale}	{0.7}
\makeatletter
\newcommand{\ec}[2][]	{{\@ec{#1 |}{#2}}}
\newcommand{\bc}[2][]	{{\@ec{#1}{#2}}}
\newcommand{\@ec}[2]	{\mathchoice
  {\displaystyle \raise.9ex\hbox{$\scaleobj{\bscale}{#1}$} {#2}}%
  {\textstyle \raise.9ex\hbox{$\scaleobj{\bscale}{#1}$} {#2}}%
  {\scriptstyle \raise.55ex\hbox{$\scriptstyle \scaleobj{\bscale}{#1}$} {#2}}%
  {\scriptscriptstyle \raise.38ex\hbox{$\scriptscriptstyle \scaleobj{\bscale}{#1}$} {#2}}%
}
\makeatother

\newcommand{\TC}{\mathsf{TC}}

\newcommand{\sfa}{\mathsf{a}}
\newcommand{\sfb}{\mathsf{b}}
\renewcommand{\sfc}{\mathsf{c}}
\newcommand{\sm}{\mathsf{m}}
\newcommand{\loc}{\mathrm{loc}}

\begin{document}

\title{String Condensations in 3+1D and Lagrangian Algebras}
\author[1,2]{Jiaheng Zhao}
\author[3,4]{Jia-Qi Lou}
\author[5,6]{Zhi-Hao Zhang}
\author[3,4,7]{Ling-Yan Hung \thanks{Email: \href{mailto:lyhung@fudan.edu.cn}{\tt lyhung@fudan.edu.cn}}}
\author[6,8,9]{Liang Kong \thanks{Email: \href{mailto:kongl@sustech.edu.cn}{\tt kongl@sustech.edu.cn}}}
\author[10]{Yin Tian \thanks{Email: \href{mailto:yintian@bnu.edu.cn}{\tt yintian@bnu.edu.cn}}}
\affil[1]{Academy of Mathematics and Systems Science, Chinese Academy of Sciences, Beijing 100190, China}
\affil[2]{University of Chinese Academy of Sciences, Beijing 100049, China}
\affil[3]{State Key Laboratory of Surface Physics, Fudan University, 200433 Shanghai, China}
\affil[4]{Department of Physics and Center for Field Theory and Particle Physics, \authorcr
Fudan University, Shanghai 200433, China}
\affil[5]{Wu Wen-Tsun Key Laboratory of Mathematics of Chinese Academy of Sciences, \authorcr
School of Mathematical Sciences,
University of Science and Technology of China, Hefei, 230026, China}
\affil[6]{Shenzhen Institute for Quantum Science and Engineering, \authorcr
Southern University of Science and Technology, Shenzhen, 518055, China}
\affil[7]{Yau Mathematical Sciences Center (YMSC), Tsinghua University, Beijing, 100084, China}
\affil[8]{International Quantum Academy, Shenzhen 518048, China}
\affil[9]{Guangdong Provincial Key Laboratory of Quantum Science and Engineering, \authorcr
Southern University of Science and Technology, Shenzhen, 518055, China}
\affil[10]{Laboratory of Mathematics and Complex Systems, School of Mathematical Sciences, \authorcr Beijing Normal University, Beijing 100875, China}
\date{\vspace{-7ex}}

\maketitle

\begin{abstract}
We present three Lagrangian algebras in the modular 2-category associated to the 3+1D $\Zb_2$ topological order and discuss their physical interpretations, connecting algebras with gapped boundary conditions, and interestingly, maps 
(braided autoequivalences) exchanging algebras with bulk domain walls.
A Lagrangian algebra, together with its modules and local modules, encapsulates detailed physical data of strings condensing at a gapped boundary. In particular, the condensed strings can terminate at boundaries in non-trivial ways. This phenomenon has no lower dimensional analogue and corresponds to novel mathematical structures associated to higher algebras.
We provide a layered construction and also explicit lattice realizations of these boundaries and illustrate the correspondence between physics and mathematics of these boundary conditions. 
This is a first detailed study of the mathematics of Lagrangian algebras in modular 2-categories and their corresponding physics, that brings together rich phenomena of string condensations, gapped boundaries and domain walls in 3+1D topological orders.
\end{abstract}

\tableofcontents


\section{Introduction}

The idea of anyon condensations \cite{Bais:2002pb, Bais_2009, Bais_2009prb} plays an important role in the study of 2+1D (spacetime dimension) or 2d (space dimension) topological orders, especially in the study of gapped boundaries and defects \cite{Bais:2002pb, Bais_2009, Bais_2009prb, Kitaev_2012, Lev13, Barkeshli:2012pr, Barkeshli:2013yta}. It provided precise descriptions of bulk-boundary duality, and led to novel applications of defects in constructing quantum gates based on (projective) braiding properties between defects \cite{Barkeshli:2012pr, Barkeshli:2013yta, Jian_2014, iriscong}, opening up new possible realizations of robust quantum computing.
The mathematical theory of anyon condensations is based on condensable algebras and their representations in modular 1-categories \cite{Kong:2013aya}. How to generalize it to higher dimensions, particularly 3d, is a fundamental theoretical question that would promise myriad possibilities of applications. Earlier discussions of string condensations or membrane condensations \cite{Kong:2014qka, Lan:2018vjb} lacked mathematical precision and were limited mostly to physical intuitions. 
While there were explicit 3d lattice model realizations of the Dijkgraaf-Witten theories \cite{Dijkgraaf:1989pz, Hamma:2004ud, Wan:2014woa, BD21} and their gapped boundaries \cite{Wan:2014woa,Wang:2018qvd,Kong_2020, BD21, Del22}, boundary and bulk excitations constructions, and other physical interpretations of these gapped boundaries in terms of string condensations and condensable algebras are yet to be understood. 
Only very recently, the modular 2-categories 
associated to 3+1D Dijkgraaf-Witten theories were explicitly computed \cite{Kong:2019brm}. 
This progress makes an explicit study of the precise relation between Lagrangian algebras, string condensations and gapped boundaries in 3+1D possible.

Every 3d topological order has infinitely many gapped boundaries (or Lagrangian algebras) because we can always stack a gapped boundary with a 2d topological order and introduce some coupling (or condensation) between two layers to get a new one. In this work, we focus on three gapped boundaries of the 3d $\Zb_2$ topological order. More precisely, we explicitly construct three Lagrangian algebras $A_e,A_1,A_2$ in the modular 2-category $\TC$ associated to the 3d $\Zb_2$ topological order and show that these algebras, together with their modules and local modules, embody the physics of three string condensations which produce three gapped boundaries of the 3d $\Zb_2$ topological order, respectively\footnote{The relation between two of the algebras and associated boundaries were sketched in \cite{Kong_2020}. 
}. By \cite{KLWZZ20,KLWZZ20a,KZ22}, the 3d $\Zb_2$ topological order with the gapped boundaries associated to $A_e,A_1,A_2$ are holographically dual to the trivial symmetry-breaking order, the trivial SPT order and the non-trivial SPT order with $\Zb_2$ onsite symmetry in 2+1D, respectively. We discuss more general gapped boundaries and Lagrangian algebras in Remark \ref{rem:boundary_Lagrangian_TC} and \ref{rem:more_Lagrangian}.



\section{A physicist's sketch of the modular 2-category associated to the 3d toric code model}

The 3+1D toric code model realizes physically the $\mathbb{Z}_2$ topological order. 
Its macroscopic observables forms a modular 2-category $\TC$ (see \cite{Kong_2020} for details), which consists of three levels of data. 
\begin{enumerate}
\item At ground level -- we have a set of objects $\sfa,\sfb, \ldots$. $\TC$ has 4 different simple objects $\{\one, \one_c, \sm, \sm_c\}$. 
Physically, they are four different kinds of string-like excitations (or 1-codimensional defects) --  $\one$ is the trivial string, $\one_c$ the electric charges condensed along a string (i.e., a $\one_c$-string), $\sm$ the magnetic string (or an $\sm$-string), and $\sm_c$ a `dyonic string' (or an $\sm_c$-string). Direct sum of simple objects (e.g. $\sfa\oplus \sfb$) are called composite strings.  
\item  At level one -- we have a set $\hom_\TC(\sfa,\sfb)$ of 1-morphisms (denoted by arrows $f,g \colon \sfa \to \sfb$) between any two objects $\sfa$ and $\sfb$. Physically, they are 0d domain walls (i.e., defects of codimension 2) connecting an $\sfa$-string with a $\sfb$-string. In particular, 1-morphisms in $\hom_\TC(\one,\one)$ are usual particles. We illustrate all simple 1-morphisms in the following diagram: 
\begin{equation}\label{eq:TC-1}
\xymatrix @C=2em{
\one \ar@(ul,ur)[]^{\{1_\one,e\}} \ar@/^/[rr]^{\{ x\}} & & \one_c \ar@(ul,ur)[]^{\{ 1_{\one_c},z\}} \ar@/^/[ll]^{\{ y\}}
} 
\quad
\xymatrix @C=2em{
\sm \ar@(ul,ur)[]^{\{1_\sm,e\}} \ar@/^/[rr]^{\{ x \}} & & \sm_c \ar@(ul,ur)[]^{\{ 1_{\sm_c},z\}} \ar@/^/[ll]^{\{ y\}}
}
\end{equation}
where $1_\sfa$ is the trivial particle on the $\sfa$-string, the $z$-particle on the $\one_c$-string can be obtained by winding an $\sm$-string around $\one_c$ \cite{Kong_2020}. There is no domain wall connecting $\one,\one_c$ with $m,m_c$. Namely, $\TC$ splits into two connecting component. These 1-morphisms can be fused (composed) along strings. The rules of composition in the first connecting component are $e\circ e = 1_\one$, $z \circ z = 1_{\one_c}$, $x \circ y = 1_{\one_c} \oplus z$ and $y\circ x = 1_\one \oplus e$. 

\item At level two -- we have a vector space $\hom_\TC(f,g)$ of 2-morphisms between any two 1-morphisms $f$ and $g$. We denote a 2-morphism by a 2-arrow $\beta \colon f\Rightarrow g$. Physically, 2-morphisms are codimensional-3 defects (also called instantons). We have $\hom_\TC(f,g)=\mathbb{C} \cdot \delta_{f,g}$ if both $f$ and $g$ are simple. 
\end{enumerate}

String-like excitations can be fused as follows: 
\begin{align}
&\one_c \otimes \one_c = \one_c \oplus \one_c, \quad \sm \otimes \sm = \one, \nonumber \\
&\sm_c \otimes \sm_c = \one_c \oplus \one_c, \quad \one_c \otimes \sm = \sm_c. \label{eq:fusion-rules}
\end{align}
Fusions are associative up to some linear maps characterized by the 15j-symbols, which are trivial for $\TC$. 

Two 2-morphisms can fuse ``vertically'' -- when two defects on a string are slid close together, and ``horizontally'' when two parallel strings are fused (defects on them also merge to become defects on the fused string). 

Particles or strings can braid around strings. The braiding structure in $\TC$ was summarized in \cite[Example 3.8]{Kong:2019brm}. Its most important ingredient is the Aharanov-Bohm phase `-1' whenever $e$ winds around $\sm$. 

Mathematically, the braided fusion 2-category $\TC$ is given by the Drinfeld center $\FZ_1(2\rep(\Zb_2))$ or $\FZ_1(2\vect_{\Zb_2})$. These categorical data can be visualized explicitly in the 3d toric code model \cite{Kong_2020}, where excitations are effected by string and membrane operators. 
In Appendix \ref{sec:3dlattice}, we review the 3d toric code model as a $\mathbb{Z}_2$ lattice gauge theory with spin-$1/2$ degrees of freedom populating edges of a triangulation of space. 

\section{Lagrangian algebras in \texorpdfstring{$\TC$}{TC} and their lattice realizations}

We are now ready to discuss the topological boundaries of the 3d toric code model based on the condensation of Lagrangian algebras in $\TC$. In 2d, a Lagrangian algebra in the modular 1-category associated to a 2d topological order control an anyon condensation that produces a gapped boundary \cite{Kong:2013aya}. In 3d, a Lagrangian algebra in modular 2-category also control or define a string condensation that produces a gapped boundary.

To define an algebra in a modular 2-category $\mathcal{C}$, we need a composite object $A \in \mathcal{C}$, the direct summands of which play the roles of condensed strings. To define the algebraic structure on $A$, one needs a 1-morphism $\mu \colon A \otimes A \to A$ describing the multiplication of the algebra and a unit 1-morphism $u \colon \one \to A$.  In 3d, the multiplication of $A$ is associative up to a 2-morphism $\alpha$ called the 2-associator. Similarly, the multiplication is unital up to 2-morphisms $\lambda,\rho$ called the left and right 2-unitors, which are often omitted in the description of an algebra. For a commutative algebra, there is also a 2-commutator $\beta$. Altogether, the collection of data is given by $(A, u, \mu, \alpha, \beta)$. Again, they satisfy a set of compatibility conditions analogous to those satisfied in anyon condensation in 2d 
(see Definition \ref{def:algebra} \& \ref{def:commutative-algebra} in Appendix \ref{appendix:commutative_algebra} for their defining axioms). 
Boundary excitations are described by $A$-modules. An $A$-module is also a composite object in $\mathcal{C}$ equipped with an $A$-action 
(see Definition \ref{def:module} in Appendix \ref{sec:modules} for the definition). 
These $A$-modules form a 2-category denoted by $\mathcal{C}_A$. If composing the $A$-action with a double braiding leaves the $A$-action invariant, then the $A$-module is deemed ``local'' 
(see Definition \ref{def_local module} in Appendix \ref{sec:modules} for the definition). 
All local $A$-modules also form a 2-category $\mathcal{C}_A^{\loc}$, which describes the topological defects in the condensed phase. A Lagrangian algebra is a commutative algebra $A$ whose only local module is $A$ itself (and its direct sums thereof) -- signalling that the condensed phase is indeed a vacuum state with no other deconfined excitations. 


Although there are infinitely many gapped boundaries of the 3d toric code model, in this work we only focus on three simplest gapped boundary conditions \cite{Wang:2018qvd}. Following \cite{Kong_2020}, we call them the rough boundary, smooth boundary and twisted smooth boundary in this work. 
By \cite{KLWZZ20,KLWZZ20a,KZ22}, the gapped boundaries of the 3d toric code model are holographically dual to anomaly-free 2d gapped quantum liquids with $\Zb_2$-symmetry, including $\Zb_2$ SPT/SET orders and symmetry-breaking orders. The smooth boundary and twisted smooth boundary are corresponding to the trivial and nontrivial 2d $\Zb_2$ SPT order, respectively, and the rough boundary is corresponding to the $\Zb_2$ symmetry-breaking order. In this section, we demonstrate that above three boundaries are incarnations of the three string condensations, or equivalently, three Lagrangian algebras in $\TC$, which are defined below with the full set of data $(A,u,\mu, \alpha, \beta)$. 
We give their (local) modules but leave the detailed calculation of the modules in Appendix \ref{sec:modules}.  

\medskip
\subsection{The algebra \texorpdfstring{$A_e$}{Ae} and the rough boundary}

The algebra $A_e = \one_c$ or $\one_c$-string is itself a Lagrangian algebra in $\TC$. 
Its multiplication, trivially associative, is defined by
\[
\one_c \otimes \one_c = \one_c  \oplus \one_c \xrightarrow{1_{\one_c} \oplus 0} \one_c.
\]
The unit 1-morphism is defined by $x \colon \one \to \one_c$. The 2-associator, 2-unitors and 2-commutator are identity 2-morphisms. 
There are two simple $A_e$-modules: $\one_c$ and $\sm_c$. The former corresponds to the trivial string excitation on the boundary and the latter corresponds to the fact that the magnetic strings and its bound state with the electric strings have non-trivial braiding with the condensate, and together form a non-trivial stringy excitation on the boundary. 
 Since $e$ particles could  disappear into the boundary, the non-trivial morphisms existing on $\sm$ and $\sm_c$ become trivial at the boundary, confirming the mathematical result that $\mathrm{hom}_{\TC_{A_e}}(\sm_c,\sm_c) = \mathrm{Vec}$. 

This rough boundary is the direct analogue of the well-known rough boundary in the 2d toric code model \cite{Kitaev_2012}. 

\subsection{The algebra \texorpdfstring{$A_1$}{A1} and the smooth boundary}

The second Lagrangian  algebra is $A_1 =  (\one \oplus \sm)_1$, where we use the subscript $1$ because there are two distinct algebra structures on $\one \oplus \sm$. The multiplication 1-morphism of $\mu_1 \colon A_1\otimes A_1 \to A_1$ is defined component-wise as follows: 
\[
\begin{array}{c}
\begin{tikzcd}
\one \otimes \one \ar[d,"1_\one"] \\
\one
\end{tikzcd}
\end{array}
\begin{array}{c}
\begin{tikzcd}
\one \otimes \sm \ar[d,"1_{\sm}"] \\
\sm
\end{tikzcd}
\end{array}
\begin{array}{c}
\begin{tikzcd}
\sm \otimes \one \ar[d,"1_{\sm}"] \\
\sm
\end{tikzcd}
\end{array}
\begin{array}{c}
\begin{tikzcd}
\sm \otimes \sm \ar[d,"1_{\one}"] \\
\one
\end{tikzcd}
\end{array}
\]
The unit morphism $u_1$ of $A_1$ is $\one \xrightarrow{ 1\oplus 0} \one \oplus \sm$. The 2-associator and 2-unitors are identity 2-morphisms. The 2-commutator is trivial on all components except $\beta_{\sm,\sm}=\pm 1$, 
which define two commutative algebra structures that are isomorphic to each other. More explicitly, the commutative algebra isomorphism 
(see Definition \ref{def:alg-hom} in Appendix \ref{appendix:commutative_algebra} for the definition) 
is defined by the 1-isomorphism
\begin{equation} \label{eq:alg-hom-A1}
\one \oplus \sm \xrightarrow{1_\one \oplus e} \one \oplus \sm
\end{equation}
and $\eta,\xi$ are identity 2-morphisms. Note that an $\sm$-string in the bulk can end at the boundary, and the end point can carry either a bosonic $m$ charge or a fermionic $f$ charge. The commutative algebra homomorphism defined in \eqref{eq:alg-hom-A1} means that  moving an $e$-particle along an $\sm$-string to its end point at the boundary can change the bosonic $m$ charge to the fermionic $f=e\otimes m$ charge and vise versa. This algebra $A_1$ defines the condensation of magnetic strings analogue to the smooth boundary in 2d. 

There is again only one non-trivial simple $A_1$-module $X_e = \one_c \oplus \sm_c = (\one\oplus \sm) \otimes \one_c$, which corresponds to moving a $\one_c$-string to the boundary. 
Electric particle $e$ in the bulk can move to the boundary and become  $\tilde e = (\one \oplus \sm) \otimes e$. They play the role of defects between the condensate $A_1$ and $A_1$. The defects $\{x,y\}$ are mapped to $\{\tilde x, \tilde y\}$ that are defects between $A_1$ and $X_e$. The $z$ excitations living on $\one_c$ become excitations $\tilde z$ on $X_e$. 
The lattice realization are illustrated in Appendix \ref{sec:lattice_construction}. 
The edges along the boundary surface can fluctuate freely and this is the analogue of the ``smooth'' boundary in 2d toric code model.   One can see that $\sm$-strings can end on the boundary without energy cost, and a string parallel to the boundary can be annihilated entirely there.

\subsection{The algebra \texorpdfstring{$A_2$}{A2} and the twisted smooth boundary}

The algebra $A_2 = (\one \oplus \sm)_2$ has the same multiplication 1-morphism and unit 1-morphism as those of $A_1$.  
The 2-associator $\alpha$ has only one non-trivial component: 
\begin{equation} \label{eq:alpha}
\xymatrix{\begin{array}{c}
\begin{tikzpicture}[scale = 0.4]
\draw[black,thick](0,0)--(1,1)--(1,2);
\draw[black,thick](0,0)--(-1,1)--(-1,1.5)--(-1.5,2);
\draw[black,thick](-1,1.5)--(-0.5,2);
\draw[black,thick](0,0)--(0,-1);
\node[black] at (-1.5,2.3) {\scriptsize $\sm$};
\node[black] at (-0.5,2.3) {\scriptsize $\sm$};
\node[black] at (1,2.3) {\scriptsize $\sm$};
\end{tikzpicture}
\end{array}
\ar@{=>}[rr]^{\alpha_{\sm,\sm,\sm}=-1} 
&&
\begin{array}{c}
\begin{tikzpicture}[scale = 0.4]
\draw[black,thick](0,0)--(1,1)--(1,1.5);
\draw[black,thick](0,0)--(-1,1)--(-1,2);
\draw[black,thick](1,1.5)--(1.5,2);
\draw[black,thick](1,1.5)--(0.5,2);
\draw[black,thick](0,0)--(0,-1);
\node[black] at (1.5,2.3) {\scriptsize $\sm$};
\node[black] at (0.5,2.3) {\scriptsize $\sm$};
\node[black] at (-1,2.3) {\scriptsize $\sm$};
\end{tikzpicture}
\end{array}}.
\end{equation}
The 2-commutator $\beta$ is trivial on all components except $\beta_{\sm,\sm}=\pm i$. These two choices of $\beta_{\sm,\sm}$ define two commutative algebra structures that are isomorphic. Again, the commutative algebra isomorphism is defined by \eqref{eq:alg-hom-A1} and $\eta,\xi$ are identity 2-morphisms.

The algebra $A_2$ does not have an analogue in 2+1D. It provides a new way to condense the magnetic strings. 
Now we give a physical interpretation of the non-trivial 2-associator $\alpha_{\sm,\sm,\sm}=-1$ and $\beta_{\sm,\sm}=\pm i$. Note that an $\sm$-string in the bulk can end at the twist smooth boundary. The end point is again non-trivial and now carries either a {\it semion} $s$ or {\it anti-semion}  $\bar s$ charge with self-statistics $i$ and $-i$, respectively. This is precisely the physical meanings of $\beta_{\sm,\sm}=\pm i$. Moving an $e$-particle to the boundary can change the end point from the semion to the anti-semion and vise versa. This is again the physical realization of the commutative algebra homomorphism defined in \eqref{eq:alg-hom-A1}, analogous to the case of the algebra $A_1$. It further suggests that the $e$ particle becomes a bound state of the $s \bar s$ bound state at the boundary.  

Consider three $\sm$-strings ending at the twisted smooth boundary with three endings attached to three semions. The end points have trivial mutual statistics. The semions at the boundary are known to have fusion satisfying the associativity up to a non-trivial 3-cocycles in $H^3(\mathbb{Z}_2, U(1))$. This 3-cocycle is inherited by the $\sm$-string that ends with a boundary semion, physically realizing the 2-associator $\alpha_{\sm,\sm,\sm}=-1$ of $A_2$.

The above physical interpretation can be visualised explicitly in lattice realizations. It was known that in the 3d toric code model, apart from the rough and smooth boundaries, there is also a twisted smooth boundary in which the vertex term $A_v$ in the boundary Hamiltonian is modified by a 3-cocycle $\alpha \in H^3(\mathbb{Z}_2, U(1))$ \cite{Wang:2018qvd}. 
We make use of membrane operators constructed in the bulk and string operators constructed at the boundary and demonstrate the above physical picture in Appendix \ref{sec:lattice_construction}. 

\subsection{An invertible domain wall exchanging \texorpdfstring{$A_1$}{A1} and \texorpdfstring{$A_2$}{A2}}

There is a braided autoequivalence $\phi$ of $\TC$ fixing 
the connecting component consisting of $\{\one,\one_c\}$ 
and exchanging the algebras $A_1$ and $A_2$.  It is involutive (i.e., $\phi^2 \simeq \mathrm{Id}_\TC$) and is defined by the identity functor together with a braided monoidal structure such that $\phi(\sfa)\otimes \phi(\sfb) \to \phi(\sfa\otimes\sfb)$ is the identity 1-morphism and the 2-isomorphism $\phi_{\sfa,\sfb,\sfc}$ between two composed 1-morphisms from $(\phi(\sfa)\otimes\phi(\sfb)) \otimes \phi(\sfc)$ to $\phi(\sfa\otimes (\sfb\otimes \sfc))$ is non-trivial only for the component $\phi_{\sm,\sm,\sm}=-1$. Physically, $\phi$ should correspond to a nontrivial gapped invertible 2d domain wall $M_\phi$ in the 3d toric code model\footnote{The lattice model realization of $M_\phi$ is not yet available.}. Fusing $M_\phi$ with the smooth boundary gives the twisted smooth boundary (see \cite{KLWZZ20} for more discussions). According to \cite{KLWZZ20,JFR23,KZ21a}, this $\phi$ precisely corresponds to the non-trivial 2+1D $\Zb_2$ SPT order, or equivalently, the non-trivial minimal modular extension of $\rep(\Zb_2)$. 




\section{Boundary conditions from layered construction}

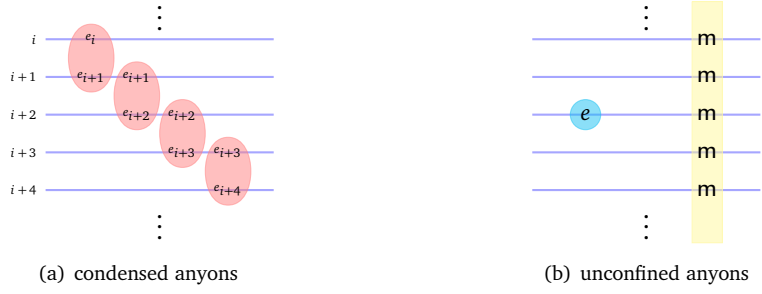
\begin{figure}[htbp]  
    \centering
    \subfigure[condensed anyons]
    {    
        \begin{minipage}[t]{0.4\linewidth}
            \centering
            \begin{tikzpicture}
                \draw[thick, blue!50!white, opacity=0.7] (0,0) -- (3,0);
                \draw[thick, blue!50!white, opacity=0.7] (0,0.5) -- (3,0.5);
                \draw[thick, blue!50!white, opacity=0.7] (0,1) -- (3,1);
                \draw[thick, blue!50!white, opacity=0.7] (0,1.5) -- (3,1.5);
                \draw[thick, blue!50!white, opacity=0.7] (0,2) -- (3,2);
                \node[above] at (1.5,2) {$\vdots$};
                \node[below] at (1.5,0) {$\vdots$};
                \node[left] at (0,0) {\tiny$i+4$};
                \node[left] at (0,0.5) {\tiny$i+3$};
                \node[left] at (0,1) {\tiny$i+2$};
                \node[left] at (0,1.5) {\tiny$i+1$};
                \node[left] at (0,2) {\tiny$i$};
                \draw[red!50!white, fill=red!50!white, opacity=0.5] (0.3,1.75) to [out=90, in=180] (0.6,2.2) to [out=0, in=90] (0.9, 1.75) to [out=270, in=0] (0.6, 1.3) to [out=180, in=270] (0.3, 1.75);
                \draw[red!50!white, fill=red!50!white, opacity=0.5] (0.9,1.25) to [out=90, in=180] (1.2,1.7) to [out=0, in=90] (1.5, 1.25) to [out=270, in=0] (1.2, 0.8) to [out=180, in=270] (0.9, 1.25);
                \draw[red!50!white, fill=red!50!white, opacity=0.5] (1.5,0.75) to [out=90, in=180] (1.8,1.2) to [out=0, in=90] (2.1, 0.75) to [out=270, in=0] (1.8, 0.3) to [out=180, in=270] (1.5, 0.75);
                \draw[red!50!white, fill=red!50!white, opacity=0.5] (2.1,0.25) to [out=90, in=180] (2.4,0.7) to [out=0, in=90] (2.7, 0.25) to [out=270, in=0] (2.4, -0.2) to [out=180, in=270] (2.1, 0.25);
                \node at (0.6,2) {\tiny$e_{i}$};
                \node at (0.6,1.5) {\tiny$e_{i+1}$};
                \node at (1.2,1.5) {\tiny$e_{i+1}$};
                \node at (1.2,1) {\tiny$e_{i+2}$};
                \node at (1.8,1) {\tiny$e_{i+2}$};
                \node at (1.8,0.5) {\tiny$e_{i+3}$};
                \node at (2.4,0.5) {\tiny$e_{i+3}$};
                \node at (2.4,0) {\tiny$e_{i+4}$};
            \end{tikzpicture}
        \end{minipage}
    }
    \subfigure[unconfined anyons]
    {
        \begin{minipage}[t]{0.4\linewidth}
            \centering
            \begin{tikzpicture}
                \draw[thick, blue!50!white, opacity=0.7] (0,0) -- (3,0);
                \draw[thick, blue!50!white, opacity=0.7] (0,0.5) -- (3,0.5);
                \draw[thick, blue!50!white, opacity=0.7] (0,1) -- (3,1);
                \draw[thick, blue!50!white, opacity=0.7] (0,1.5) -- (3,1.5);
                \draw[thick, blue!50!white, opacity=0.7] (0,2) -- (3,2);
                \node[above] at (1.5,2) {$\vdots$};
                \node[below] at (1.5,0) {$\vdots$};
                \draw[yellow!50!white, fill=yellow!50!white, opacity=0.5] (2.1, 2.5) -- (2.1, -0.7) -- (2.5, -0.7) -- (2.5, 2.5) -- (2.1, 2.5);
                \draw[cyan!70!white, fill=cyan!70!white, opacity=0.5] (0.7,1) circle [radius=0.2cm];
                \node at (0.7,1) {\small$e$};
                \node at (2.3,0) {\small$\sm$};
                \node at (2.3,0.5) {\small$\sm$};
                \node at (2.3,1) {\small$\sm$};
                \node at (2.3,1.5) {\small$\sm$};
                \node at (2.3,2) {\small$\sm$};
            \end{tikzpicture}
        \end{minipage}
    }
    \caption{
        {Layered construction of the 3d $\Zb_2$ topological order: 
        (a) Each blue line represents a layer of the $\Zb_2$ topological order and the numbers on the left label these layers. 
        Each red circle represents a tensor product state between the electric charges of consecutive layers.
        The condensation of these tensor product states will introduce coupling between consecutive layers.
        (b) Besides the trivial string, the unconfined anyons are the $e$-particle and the $\sm$-string.}
    } \label{fig:layer1}
\end{figure}

\begin{figure}[htbp]  
    \centering
    \subfigure[The rough boundary]
    {
        \begin{minipage}[t]{0.3\linewidth}
            \centering
            \begin{tikzpicture}
                \draw[thick, blue!50!white, opacity=0.7] (0,0) -- (3,0);
                \draw[thick, blue!50!white, opacity=0.7] (0,0.5) -- (3,0.5);
                \draw[thick, blue!50!white, opacity=0.7] (0,1) -- (3,1);
                \draw[thick, blue!50!white, opacity=0.7] (0,1.5) -- (3,1.5);
                \draw[thick, blue!50!white, opacity=0.7] (0,2) -- (3,2);
                \node[below] at (1.5,0) {$\vdots$};
                \node[left] at (0,0) {\tiny$5$};
                \node[left] at (0,0.5) {\tiny$4$};
                \node[left] at (0,1) {\tiny$3$};
                \node[left] at (0,1.5) {\tiny$2$};
                \node[left] at (0,2) {\tiny$b$};
                \draw[red!50!white, fill=red!50!white, opacity=0.5] (0.3,1.75) to [out=90, in=180] (0.6,2.2) to [out=0, in=90] (0.9, 1.75) to [out=270, in=0] (0.6, 1.3) to [out=180, in=270] (0.3, 1.75);
                \draw[red!50!white, fill=red!50!white, opacity=0.5] (0.9,1.25) to [out=90, in=180] (1.2,1.7) to [out=0, in=90] (1.5, 1.25) to [out=270, in=0] (1.2, 0.8) to [out=180, in=270] (0.9, 1.25);
                \draw[red!50!white, fill=red!50!white, opacity=0.5] (1.5,0.75) to [out=90, in=180] (1.8,1.2) to [out=0, in=90] (2.1, 0.75) to [out=270, in=0] (1.8, 0.3) to [out=180, in=270] (1.5, 0.75);
                \draw[red!50!white, fill=red!50!white, opacity=0.5] (2.1,0.25) to [out=90, in=180] (2.4,0.7) to [out=0, in=90] (2.7, 0.25) to [out=270, in=0] (2.4, -0.2) to [out=180, in=270] (2.1, 0.25);
                \draw[red!50!white, fill=red!50!white, opacity=0.5] (2.4,2) circle [radius=0.2cm];
                \node at (0.6,2) {\tiny$e_{b}$};
                \node at (0.6,1.5) {\tiny$e_{2}$};
                \node at (1.2,1.5) {\tiny$e_{2}$};
                \node at (1.2,1) {\tiny$e_{3}$};
                \node at (1.8,1) {\tiny$e_{3}$};
                \node at (1.8,0.5) {\tiny$e_{4}$};
                \node at (2.4,0.5) {\tiny$e_{4}$};
                \node at (2.4,0) {\tiny$e_{5}$};
                \node at (2.4,2) {\tiny$e_{b}$};
            \end{tikzpicture}
        \end{minipage}
    }
    \subfigure[The smooth boundary]
    {
        \begin{minipage}[t]{0.3\linewidth}
            \centering
            \begin{tikzpicture}
                \draw[thick, blue!50!white, opacity=0.7] (0,0) -- (3,0);
                \draw[thick, blue!50!white, opacity=0.7] (0,0.5) -- (3,0.5);
                \draw[thick, blue!50!white, opacity=0.7] (0,1) -- (3,1);
                \draw[thick, blue!50!white, opacity=0.7] (0,1.5) -- (3,1.5);
                \draw[thick, blue!50!white, opacity=0.7] (0,2) -- (3,2);
                \node[below] at (1.5,0) {$\vdots$};
                \node[left] at (0,0) {\tiny$5$};
                \node[left] at (0,0.5) {\tiny$4$};
                \node[left] at (0,1) {\tiny$3$};
                \node[left] at (0,1.5) {\tiny$2$};
                \node[left] at (0,2) {\tiny$b$};
                \draw[red!50!white, fill=red!50!white, opacity=0.5] (0.3,1.75) to [out=90, in=180] (0.6,2.2) to [out=0, in=90] (0.9, 1.75) to [out=270, in=0] (0.6, 1.3) to [out=180, in=270] (0.3, 1.75);
                \draw[red!50!white, fill=red!50!white, opacity=0.5] (0.9,1.25) to [out=90, in=180] (1.2,1.7) to [out=0, in=90] (1.5, 1.25) to [out=270, in=0] (1.2, 0.8) to [out=180, in=270] (0.9, 1.25);
                \draw[red!50!white, fill=red!50!white, opacity=0.5] (1.5,0.75) to [out=90, in=180] (1.8,1.2) to [out=0, in=90] (2.1, 0.75) to [out=270, in=0] (1.8, 0.3) to [out=180, in=270] (1.5, 0.75);
                \draw[red!50!white, fill=red!50!white, opacity=0.5] (2.1,0.25) to [out=90, in=180] (2.4,0.7) to [out=0, in=90] (2.7, 0.25) to [out=270, in=0] (2.4, -0.2) to [out=180, in=270] (2.1, 0.25);
                \node at (0.6,2) {\tiny$e_{b}$};
                \node at (0.6,1.5) {\tiny$e_{2}$};
                \node at (1.2,1.5) {\tiny$e_{2}$};
                \node at (1.2,1) {\tiny$e_{3}$};
                \node at (1.8,1) {\tiny$e_{3}$};
                \node at (1.8,0.5) {\tiny$e_{4}$};
                \node at (2.4,0.5) {\tiny$e_{4}$};
                \node at (2.4,0) {\tiny$e_{5}$};
            \end{tikzpicture}
        \end{minipage}
    }
    \subfigure[The twisted smooth boundary]
    {
        \begin{minipage}[t]{0.3\linewidth}
            \centering
            \begin{tikzpicture}
                \draw[thick, blue!50!white, opacity=0.7] (0,0) -- (3,0);
                \draw[thick, blue!50!white, opacity=0.7] (0,0.5) -- (3,0.5);
                \draw[thick, blue!50!white, opacity=0.7] (0,1) -- (3,1);
                \draw[thick, blue!50!white, opacity=0.7] (0,1.5) -- (3,1.5);
                \draw[thick, orange!50!white, opacity=0.7] (0,2) -- (3,2);
                \node[below] at (1.5,0) {$\vdots$};
                \node[left] at (0,0) {\tiny$5$};
                \node[left] at (0,0.5) {\tiny$4$};
                \node[left] at (0,1) {\tiny$3$};
                \node[left] at (0,1.5) {\tiny$2$};
                \node[left] at (0,2) {\tiny$b$};
                \draw[red!50!white, fill=red!50!white, opacity=0.5] (0.3,1.75) to [out=90, in=180] (0.6,2.2) to [out=0, in=90] (0.9, 1.75) to [out=270, in=0] (0.6, 1.3) to [out=180, in=270] (0.3, 1.75);
                \draw[red!50!white, fill=red!50!white, opacity=0.5] (0.9,1.25) to [out=90, in=180] (1.2,1.7) to [out=0, in=90] (1.5, 1.25) to [out=270, in=0] (1.2, 0.8) to [out=180, in=270] (0.9, 1.25);
                \draw[red!50!white, fill=red!50!white, opacity=0.5] (1.5,0.75) to [out=90, in=180] (1.8,1.2) to [out=0, in=90] (2.1, 0.75) to [out=270, in=0] (1.8, 0.3) to [out=180, in=270] (1.5, 0.75);
                \draw[red!50!white, fill=red!50!white, opacity=0.5] (2.1,0.25) to [out=90, in=180] (2.4,0.7) to [out=0, in=90] (2.7, 0.25) to [out=270, in=0] (2.4, -0.2) to [out=180, in=270] (2.1, 0.25);
                \node at (0.6,2) {\tiny$(s\bar s)_{b}$};
                \node at (0.6,1.5) {\tiny$e_{2}$};
                \node at (1.2,1.5) {\tiny$e_{2}$};
                \node at (1.2,1) {\tiny$e_{3}$};
                \node at (1.8,1) {\tiny$e_{3}$};
                \node at (1.8,0.5) {\tiny$e_{4}$};
                \node at (2.4,0.5) {\tiny$e_{4}$};
                \node at (2.4,0) {\tiny$e_{5}$};
            \end{tikzpicture}
        \end{minipage}
    }
    \caption{
       {Boundary conditions from layered construction:
        The layer labeled by b is the boundary layer.
        (a) Besides the tensor product states of electric charges, 
        $e_b$ itself is also condensed to make sure that the electric charges can be absorbed into the boundary.
        (b) Only the tensor product states of electric charges are condensed.
        Then one can readily see that the $\sm$-string can end on the boundary.
        (c) Here the boundary layer is a double semion order instead of a $\Zb_2$ order.
        The related condensed tensor product state is $s\bar s\times e$. 
        An $\sm$-string can still end on this boundary as long as its end point on the boundary layer is either $s$ or $\bar s$.}
    }  \label{fig:layered}
\end{figure}
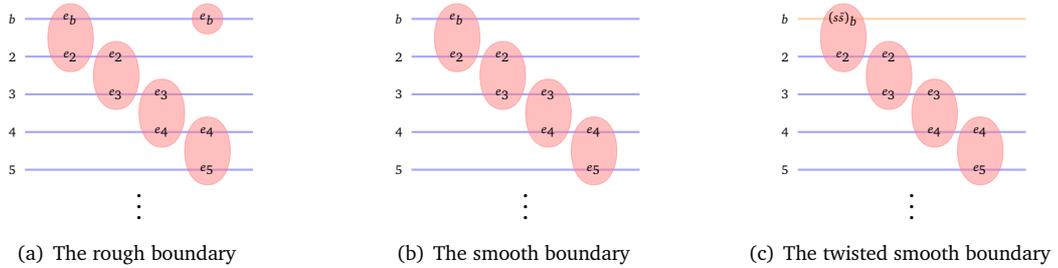

These boundary conditions can also be understood in more intuitive terms via the layered construction of topological orders from lower dimensional ones \cite{Jian_2014}. As illustrated in Figure \ref{fig:layer1}, the 3d $\Zb_2$ topological order can be constructed from stacking 2d $\Zb_2$ topological orders. Consecutive layers are glued together by condensing anyons -- in this case the tensor product state of $e_1 \times e_2$, where the subscript indicates the layer the electric charge belongs to. The pattern is repeated in an indefinite number of vertically stacked 2d $\Zb_2$ topological orders. The $e$ particles and the $\sm$-strings are unconfined anyons under the condensation of paired electric charges in the stack. To describe different boundary conditions, it comes down to special treatment at the last layer as illustrated in Figure \ref{fig:layered}.

\bit
\item \textbf{The rough boundary}: It is obtained where the last layer labeled $b$ of the stack is still a 2d $\Zb_2$ topological order. In addition to condensing the interlayer $e_b \times e_2$, we condense also $e_b$ separately.
In this case, the $\sm$-string is confined at the boundary.
An $e$-particle moved to the boundary would become part of the condensate and disappear.
\item \textbf{The smooth boundary}: The setup is similar to the rough boundary, except that $e_b$ remains uncondensed.
Collecting the unconfined excitations at the boundary, one can see that $\sm$ strings can end on the boundary, even though isolated $\sm$ at the boundary is confined.
Electric charges remain independent particles at the boundary, furnishing the non-trivial $\one_c \oplus \sm_c$ excitations at the boundary, and also the automorphism in $A_1$.
\item \textbf{The twisted smooth boundary}: This is achieved with the last layer replaced by a doubled semion order. To glue this layer with the next layer of a $\Zb_2$ topological order we condense the tensor product excitation $s \bar s \times e$. One can check that an $\sm$-string can end on the boundary layer with its end point attached to a semion or an anti-semion, thus remaining unconfined. 
Since $s\bar s \times e$ is condensed, a single $s\bar s$-particle can hop into the bulk and become an $e$-particle (see Figure \ref{fig:layered}). 
The boundary free $s\bar s$ particle realizes the automorphism structure of $A_2$. 
\eit

More general boundary conditions can be obtained by stacking an anomaly-free 2d topological order on the last layer and introducing some coupling (or condensation). We do not give any explicit construction of such generic gapped boundaries (see a few works \cite{JTX22,Luo22} appeared after this work). However, we provide more examples of the fusion 2-categories and Lagrangian algebras associated to the gapped boundaries of the 3d $\mathbb{Z}_2$ topological order in the following remark.

\begin{rem} \label{rem:boundary_Lagrangian_TC}
By \cite{KLWZZ20,KLWZZ20a,KZ22}, a 3d finite gauge theory (with a gauge group $G$) with a gapped boundary is holographically dual to a 2d SPT/SET order or spontaneous symmetry-breaking (SSB) order with a global symmetry $G$. As a consequence, 2d $\mathbb{Z}_2$ SPT/SET/SSB orders provide more examples of the gapped boundaries of the 3d $\mathbb{Z}_2$ topological order. 

A 2d SPT/SET order with a finite group $G$ symmetry can be mathematically described by a braided fusion category $\CC$ with the M\"{u}ger center given by $\rep(G)$ and a minimal modular extension of $\CC$ \cite{LKW16a}. The data of a minimal modular extension is equivalent to that of a braided equivalence $\FZ_1(\Sigma \CC) \simeq \FZ_1(2\rep(G))$ \cite{KLWZZ20,JFR23}, where $\Sigma\EuScript{C} \coloneqq \mathrm{RMod}_{\EuScript{C}}(2\mathrm{Vec})$ is the delooping of $\CC$ \cite{DR18}. According to \cite{KLWZZ20,JFR23} and the fact that $H^4(\Zb_2,U(1))$ is trivial, every braided fusion category $\CC$ with the M\"{u}ger center given by $\rep(\Zb_2)$ admits a minimal modular extesion, and thus $\Sigma \CC$ can be realized as the 2-category of topological defects on a gapped boundary of the 3d $\mathbb{Z}_2$ topological order by the boundary-bulk relation \cite{KWZ15,KWZ17}. In particular, when $\CC = \rep(\Zb_2)$, there are two minimal modular extensions given by $\FZ_1(\rep(\Zb_2)) \simeq \FZ_1(\vect_{\Zb_2})$ and $\FZ_1(\vect_{\Zb_2}^\omega)$, where $\omega$ represents the nontrivial cohomology class in $H^3(\Zb_2,U(1)) \simeq \Zb_2$. These two minimal modular extensions describe the trivial and nontrivial 2d $\Zb_2$ SPT order respectively, and the corresponding 2d gapped boundaries of the 3d toric code model are the smooth and twisted smooth boundaries respectively.

Given a gapped boundary condition $X$, one can physically determine the associated Lagrangian algebra by cutting a cylinder-shaped hole in the bulk and imposing the boundary condition $X$ on the hole. This hole viewed from far away is precisely the (usually composite) string or the underlying object of the associated Lagrangian algebra. By viewing $\TC \simeq \FZ_1(2\rep(\Zb_2))$ as the 2-category of braided modules over $\rep(\Zb_2)$ \cite{DN21}, the Lagrangian algebras in $\TC$ are given by nondegenerate braided fusion categories $\CM$ equipped with braided functors $\rep(\Zb_2) \to \CM$ \cite{JFR23}. Then the Lagrangian algebras associated to the rough boundary, smooth boundary and twisted smooth boundary are given by the forgetful functor $\rep(\Zb_2) \xrightarrow{w} \vect$, the trivial minimal modular extension $\rep(\Zb_2) \hookrightarrow \FZ_1(\rep(\Zb_2))$ and the nontrivial minimal modular extension $\rep(\Zb_2) \hookrightarrow \FZ_1(\vect_{\Zb_2}^\omega)$, respectively. The gapped boundaries associated to a minimal modular extension $(\CC \hookrightarrow \CM)$ of the braided fusion category $\CC$ with the M\"{u}ger center given by $\rep(\Zb_2)$ can be understood as the Lagrangian algebra associated to the composed braided functor $(\rep(\Zb_2) \hookrightarrow \CC \hookrightarrow \CM)$. If the braided functor factors as $(\rep(\Zb_2) \simeq \rep(\Zb_2)\boxtimes \vect \xrightarrow{w\boxtimes \one_\CM} \vect \boxtimes \CM \simeq \CM)$, where $\one_\CM$ denotes the unique braided functor from $\vect$ to $\CM$, the associated gapped boundary of the 2d toric code model is obtained by stacking the 2d topological order associated to the modular 1-category $\CM$ on the rough boundary. 
\end{rem}


\section{Conclusions}

In this paper, we present three Lagrangian algebras in $\TC$, which is the modular 2-category associated to the 3d $\Zb_2$ topological order. Similar to the 2d case, these Lagrangian algebras characterize distinct topological boundaries of the bulk topological order. We provide the physical meanings of these algebras, illustrating how the mathematical structure of these generalized Lagrangian algebras corresponds to properties of string condensations characterizing the gapped boundaries.  Particularly, there are new structures in these higher algebras unknown to lower dimensions, namely the non-trivial associator and commutators. They signify non-trivial topological charges that can be attached to end points of strings belonging to the condensate that can terminate at the boundary. When there is more than one way of terminating these condensed strings at a given topological boundary, one finds correspondingly multiple algebra characterizing the same boundary that are related to each other by a {\it commutative algebra homomorphism}. Such a homomorphism between commutative algebras is physically realized as bulk charges that can be pushed to the boundary and be attached to the end points of strings, changing their braiding and fusion properties there. 
We also show how the module categories correspond to bulk excitations that are confined at the boundary. These mathematical structures can either be visualised in a layered construction of the 3d $\Zb_2$ topological order, or an explicit lattice realization.
As in 2d, the notion of a Lagrangian algebra in 3d supplies the correct mathematical structure that describes the condensation of strings characterizing each gapped boundary. We believe the string condensation picture we presented here applies to all 3d topological orders. 

\medskip

\noindent {\bf Acknowledgements} 
We would like to thank Hao Xu for helpful discussions about Lagrangian algebras in a braided fusion 2-category. LYH acknowledges the support of NSFC (Grant No.~11922502, 11875111) and the Shanghai Municipal Science and Technology Major Project (Shanghai Grant No.~2019SHZDZX01). YT is partially supported by the NSFC grant No.~11971256. LK and ZHZ are supported by NSFC (Grant No.~11971219) and by Guangdong Provincial Key Laboratory (Grant No.~2019B121203002) and by Guangdong Basic and Applied Basic Research Foundation (Grant No.~2020B1515120100). ZHZ is also supported by Wu Wen-Tsun Key Laboratory of Mathematics at USTC of Chinese Academy of Sciences.

\appendix
\appendixpage


\section{Modular 2-category \texorpdfstring{$\TC$}{TC}} \label{sec:review2cat}

It was shown in \cite{Kong_2020} that all topological defects of codimension 2 and higher form a braided fusion 2-category $\TC=\mathcal{Z}(2\vect_{\Zb_2})$, which was explicitly computed in \cite{Kong:2019brm}. We briefly summarize the braided 2-category structures of $\TC$ below. 
\begin{itemize}

\item There are four simple objects in $\TC$: $\one,\one_c, \sm,\sm_c$. All other objects are direct sums of simple objects. 

\item For each pair of objects $\sfa$ and $\sfb$, there is a 1-category $\hom_\TC(\sfa,\sfb)$. The objects in the $\hom_\TC(\sfa,\sfb)$ are 1-morphisms in $\TC$, denoted by $f,g \colon \sfa \to \sfb$. The 1-morphisms in $\hom_\TC(\sfa,\sfb)$ are called 2-morphisms in $\TC$, denoted by $\phi, \psi \colon f \Rightarrow g$. We illustrate these categories $\hom_\TC(\sfa,\sfb)$ for simple objects $\sfa, \sfb$ in the following graph,
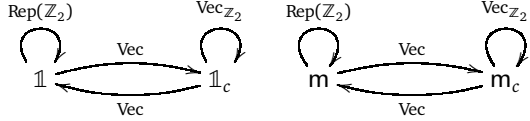
\begin{equation*} 
\xymatrix{
\one \ar@(ul,ur)[]^{\mathrm{Rep}(\mathbb{Z}_2)} \ar@/^/[rr]^{\mathrm{Vec}} & & \one_c \ar@(ul,ur)[]^{\mathrm{Vec}_{\mathbb{Z}_2}} \ar@/^/[ll]^{\mathrm{Vec}}
} 
\quad
\xymatrix{
\sm \ar@(ul,ur)[]^{\mathrm{Rep}(\mathbb{Z}_2)} \ar@/^/[rr]^{\mathrm{Vec}} & & \sm_c \ar@(ul,ur)[]^{\mathrm{Vec}_{\mathbb{Z}_2}} \ar@/^/[ll]^{\mathrm{Vec}}
}
\end{equation*}
where $\rep(\Zb_2)$ is the 1-category of finite dimensional representations of the group $\Zb_2$, $\vect$ is the 1-category of finite dimensional vector spaces and $\vect_{\Zb_2}$ is the 1-category of finite dimensional $\Zb_2$-graded vector spaces. 
The simple 1-morphisms in $\TC$ are illustrated in \eqref{eq:TC-1}. 

Note that $\TC$ is disconnected, in the sense that the only 1-morphisms between $\one,\one_c$ and $\sm,\sm_c$ are zero morphisms. In other words, $\TC$ splits into a direct sum: 
\begin{equation} \label{eq:decomposation}
\TC = 2\rep(\Zb_2)_0 \boxplus 2\rep(\Zb_2)_1.
\end{equation}
where two connecting components are entirely same and denoted by $2\rep(\Zb_2)$, and two subscripts $0$ and $1$ endow $\TC$ with a $\Zb_2$-grading. 

\item Composition of 1-morphisms: The rules of composition $2\rep(\Zb_2)_0$ are $e\circ e = 1_\one$, $z \circ z = 1_{\one_c}$, $x \circ y = 1_{\one_c} \oplus z$ and $y\circ x = 1_\one \oplus e$. Those in $2\rep(\Zb_2)_1$ are similar. 
\end{itemize}
The monoidal structure of $\TC$ is strict and is defined by the fusion rules in \eqref{eq:fusion-rules}. 
The braiding structure can be found in \cite[Example 3.8]{Kong:2019brm}. Moreover, it was shown in \cite{Kong:2019brm} that the braidings of $\TC$ are non-degenerate. The precise definition of a modular 2-category is not yet known. However, a reasonable definition of a modular 2-category should include $\TC$ as an example. Therefore, we use the term `modular 2-category' freely.

\section{Commutative algebras in a braided monoidal 2-category} \label{appendix:commutative_algebra}

The definition of an algebra in a semistrict monoidal 2-category (i.e., a pseudo-monoid in a Gray monoid) can be found in \cite{DS97}, and that of an algebra in a weak monoidal 2-category can be found in \cite{Dec21}. For simplicity, we hide some necessary coherence data (e.g. 1-associators) in the following definitions. 

\begin{defn} \label{def:algebra}
Let $\SM$ be a braided monoidal 2-category with the tensor product $\otimes$ and the tensor unit $\one$. An algebra in $\SM$ is a sextuple $(A, u, \mu,\alpha, \lambda, \rho)$, where $A \in \mathrm{ob}(\SM)$, $u \colon \one \to A$ and $\mu \colon A \otimes A \to A$ are 1-morphisms, and $\alpha, \lambda, \rho$ are invertible 2-morphisms (called the 2-associator, the left 2-unitor and the right 2-unitor, respectively) as depicted in the following diagrams
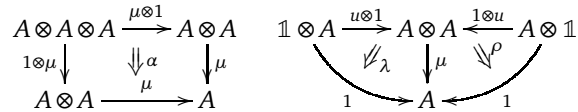
\begin{align*}
    {    \xymatrix @R=0.2in @C=0.25in{
            A \otimes A \otimes A \ar[r]^-{\mu \otimes 1} \ar[d]_-{1 \otimes \mu} & A \otimes A \ar[d]^\mu \\
             \rtwocell<\omit>{<-3>\alpha}  A \otimes A \ar[r]^-{\mu} & A 
        }\quad
        \xymatrix @R=0.2in @C=0.25in{
            \one \otimes A \ar[r]^-{u \otimes 1} \ar@/^{-2ex}/[rd]_-{1} \drtwocell<\omit>{<-0.7>\lambda} & A \otimes A \ar[d]^-{\mu} & A \otimes \one \ar[l]_-{1 \otimes u} \ar@/^{2ex}/[ld]^-{1} \\
            & A \urtwocell<\omit>{<-0.7>\rho}&
        }} 
    \end{align*}
or equivalently, by the following graphs: 
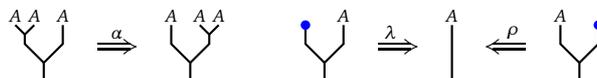
\[
\xymatrix @R=0.2in @C=0.2in{
\begin{array}{c}
\begin{tikzpicture}[scale = 0.25]
\draw[black,thick](0,0)--(1,1)--(1,2);
\draw[black,thick](0,0)--(-1,1)--(-1,1.5)--(-1.5,2);
\draw[black,thick](-1,1.5)--(-0.5,2);
\draw[black,thick](0,0)--(0,-1);
\node[black] at (-1.5,2.5) {\scriptsize $A$};
\node[black] at (-0.5,2.5) {\scriptsize $A$};
\node[black] at (1,2.5) {\scriptsize $A$};
\end{tikzpicture}
\end{array}
\ar@{=>}[r]^\alpha
&
\begin{array}{c}
\begin{tikzpicture}[scale = 0.25]
\draw[black,thick](0,0)--(1,1)--(1,1.5);
\draw[black,thick](0,0)--(-1,1)--(-1,2);
\draw[black,thick](1,1.5)--(1.5,2);
\draw[black,thick](1,1.5)--(0.5,2);
\draw[black,thick](0,0)--(0,-1);
\node[black] at (1.5,2.5) {\scriptsize $A$};
\node[black] at (0.5,2.5) {\scriptsize $A$};
\node[black] at (-1,2.5) {\scriptsize $A$};
\end{tikzpicture}
\end{array}
}
\quad
\xymatrix@R=0.2in @C=0.2in{
\begin{array}{c}
\begin{tikzpicture}[scale = 0.25]
\draw[black,thick](0,0)--(1,1)--(1,2);
\draw[black,thick](0,0)--(-1,1)--(-1,2);
\draw[black,thick](0,0)--(0,-1);
\fill[blue] (-1,2) circle (0.25);
\node[black] at (1,2.5) {\scriptsize $A$};
\end{tikzpicture}
\end{array}
\ar@{=>}[r]^\lambda
&
\begin{array}{c}
\begin{tikzpicture}[scale = 0.25]
\draw[black,thick](0,2)--(0,-1);
\node[black] at (0,2.5) {\scriptsize $A$};
\end{tikzpicture}
\end{array}
&
\begin{array}{c}
\begin{tikzpicture}[scale = 0.25]
\draw[black,thick](0,0)--(1,1)--(1,2);
\draw[black,thick](0,0)--(-1,1)--(-1,2);
\draw[black,thick](0,0)--(0,-1);
\fill[blue] (1,2) circle (0.25);
\node[black] at (-1,2.5) {\scriptsize $A$};
\end{tikzpicture}
\end{array}
\ar@{=>}[l]_\rho
}
\]
such that the following diagrams are commutative.
\begin{equation}
\raisebox{2em}{\xymatrix @R=0.17in @C=0.2in{
\begin{array}{c}
\begin{tikzpicture}[scale = 0.2]
\draw[black,thick](0,0)--(1.5,1.5)--(1.5,3);
\draw[black,thick](0,0)--(-1,1)--(-2,2)--(-2,3)--(-2.5,3.5);
\draw[black,thick](-2,3)--(-1.5,3.5);
\draw[black,thick](-1,1)--(0,2)--(0,3);
\draw[black,thick](0,0)--(0,-2);
\end{tikzpicture}
\end{array}
\ar@{=>}[rr]^\alpha \ar@{=>}[d]_\alpha
&&
\begin{array}{c}
\begin{tikzpicture}[scale = 0.2]
\draw[black,thick](0,0)--(1.5,1.5)--(1.5,3);
\draw[black,thick](0,0)--(-1,1)--(-1,1.5)--(-2,2.5)--(-2,3);
\draw[black,thick](-1,1.5)--(0,2.5)--(0,3)--(0.5,3.5);
\draw[black,thick](0,3)--(-0.5,3.5);
\draw[black,thick](0,0)--(0,-2);
\end{tikzpicture}
\end{array}
\ar@{=>}[d]^\alpha
\\
\begin{array}{c}
\begin{tikzpicture}[scale = 0.2]
\draw[black,thick](0,0)--(1.5,1.5)--(1.5,3)--(2,3.5);
\draw[black,thick](1.5,3)--(1,3.5);
\draw[black,thick](0,0)--(-1.5,1.5)--(-1.5,3)--(-2,3.5);
\draw[black,thick](-1.5,3)--(-1,3.5);
\draw[black,thick](0,0)--(0,-2);
\end{tikzpicture}
\end{array}
\ar@{=>}[r]^\alpha 
& 
\begin{array}{c}
\begin{tikzpicture}[scale = 0.2]
\draw[black,thick](0,0)--(-1.5,1.5)--(-1.5,3);
\draw[black,thick](0,0)--(1,1)--(2,2)--(2,3)--(2.5,3.5);
\draw[black,thick](2,3)--(1.5,3.5);
\draw[black,thick](1,1)--(0,2)--(0,3);
\draw[black,thick](0,0)--(0,-2);
\end{tikzpicture}
\end{array}
&
\begin{array}{c}
\begin{tikzpicture}[scale = 0.2]
\draw[black,thick](0,0)--(-1.5,1.5)--(-1.5,3);
\draw[black,thick](0,0)--(1,1)--(1,1.5)--(2,2.5)--(2,3);
\draw[black,thick](1,1.5)--(0,2.5)--(0,3)--(-0.5,3.5);
\draw[black,thick](0,3)--(0.5,3.5);
\draw[black,thick](0,0)--(0,-2);
\end{tikzpicture}
\end{array}
\ar@{=>}[l]_\alpha 
}}
\end{equation}

\begin{equation}
\raisebox{2em}{\xymatrix @R=0.15in @C=0.2in{
\begin{array}{c}
\begin{tikzpicture}[scale = 0.2]
\draw[black,thick](0,0)--(1,1)--(1,2);
\draw[black,thick](0,0)--(-1,1)--(-1,1.5)--(-1.5,2);
\draw[black,thick](-1,1.5)--(-0.5,2);
\draw[black,thick](0,0)--(0,-1);
\fill[black] (-0.5,2) circle (0.2);
\end{tikzpicture}
\end{array}
\ar@{=>}[rr]^\alpha \ar@{=>}[rd]_\rho
&&
\begin{array}{c}
\begin{tikzpicture}[scale = 0.2]
\draw[black,thick](0,0)--(1,1)--(1,1.5);
\draw[black,thick](0,0)--(-1,1)--(-1,2);
\draw[black,thick](1,1.5)--(1.5,2);
\draw[black,thick](1,1.5)--(0.5,2);
\fill[black] (0.5,2) circle (0.2);
\draw[black,thick](0,0)--(0,-1);
\end{tikzpicture}
\end{array}
\ar@{=>}[ld]^\lambda
\\
&
\begin{array}{c}
\begin{tikzpicture}[scale = 0.2]
\draw[black,thick](0,0)--(1.5,1.5);
\draw[black,thick](0,0)--(-1.5,1.5);
\draw[black,thick](0,0)--(0,-2);
\end{tikzpicture}
\end{array}
}}
\end{equation}
We often abbreviate an algebra to $(A, u, m)$ or $A$.
\end{defn}

\begin{defn} \label{def:commutative-algebra}
A commutative algebra (also called a braided pseudo-mononid in \cite{DS97}) in $\SM$ is an algebra $A$ equipped with a 2-isomorphism (called the 2-commutator) $\beta$
\begin{equation} \label{eq:commute}
\xymatrix @R=1em @C=2em{
\begin{array}{c}
\begin{tikzpicture}[scale = 0.3]
\draw[black,thick](0,0)--(1,1);
\draw[black,thick](0,0)--(-1,1);
\draw[black,thick](0,0)--(0,-1.5);
\node[black] at (-1,1.5) {\scriptsize $A$};
\node[black] at (1,1.5) {\scriptsize $A$};
\end{tikzpicture}
\end{array}
\ar@{=>}[r]^\beta
&
\begin{array}{c}
\begin{tikzpicture}[scale = 0.3]
\draw[black,thick](0,0)--(-0.5,0.5)--(0.5,1.5);
\draw[white,line width =3](0,0)--(0.5,0.5)--(-0.5,1.5);
\draw[black,thick](0,0)--(0.5,0.5)--(-0.5,1.5);
\draw[black,thick](0,0)--(0,-1);
\node[black] at (-0.5,2) {\scriptsize $A$};
\node[black] at (0.5,2) {\scriptsize $A$};
\end{tikzpicture}
\end{array},
}
\end{equation}
such that the following two diagrams are commutative, 
\begin{equation}\label{diagram:commutativity}
\raisebox{2em}{\xymatrix @R=1em @C=2em{
\begin{array}{c}
\begin{tikzpicture}[scale = 0.2]
\draw[black,thick](0,0)--(1,1)--(1,1.5);
\draw[black,thick](0,0)--(-1,1)--(-1,2);
\draw[black,thick](1,1.5)--(1.5,2);
\draw[black,thick](1,1.5)--(0.5,2);
\draw[black,thick](0,0)--(0,-1);
\end{tikzpicture}
\end{array}
\ar@{=>}[d]_\beta 
&
\begin{array}{c}
\begin{tikzpicture}[scale = 0.2]
\draw[black,thick](0,0)--(1,1)--(1,2);
\draw[black,thick](0,0)--(-1,1)--(-1,1.5)--(-1.5,2);
\draw[black,thick](-1,1.5)--(-0.5,2);
\draw[black,thick](0,0)--(0,-1);
\end{tikzpicture}
\end{array}
\ar@{=>}[r]^\beta
\ar@{=>}[l]_\alpha 
&
\begin{array}{c}
\begin{tikzpicture}[scale = 0.15]
\draw[black,thick](-1,1.5)--(-1.5,2)--(-0.5,3);
\draw[white,line width = 3](-0.5,2)--(-1.5,3);
\draw[black,thick](0,0)--(1,1)--(1,2.5);
\draw[black,thick](0,0)--(-1,1)--(-1,1.5)--(-0.5,2)--(-1.5,3);
\draw[black,thick](0,0)--(0,-2);
\end{tikzpicture}
\end{array}
\ar@{=>}[rd]^\alpha
&
\\
\begin{array}{c}
\begin{tikzpicture}[scale = 0.15]
\draw[black,thick](0,0)--(-1,1)--(1,3)--(1,3.5)--(1.5,4);
\draw[white,line width = 3](1,1)--(-1,3);
\draw[black,thick](0,0)--(1,1)--(-1,3)--(-1,4);
\draw[black,thick](1,3.5)--(0.5,4);
\draw[black,thick](0,0)--(0,-2);
\end{tikzpicture}
\end{array}
\ar@{=>}[r]
&
\begin{array}{c}
\begin{tikzpicture}[scale = 0.15]
\draw[black,thick](0,0)--(-1,1)--(-1,1.5)--(1,3.5);
\draw[black,thick](-1,1.5)--(-1.5,2)--(0.5,4);
\draw[white,line width = 3](1,2)--(-1,4);
\draw[black,thick](0,0)--(1,1)--(1,2)--(-1,4);
\draw[black,thick](0,0)--(0,-2);
\end{tikzpicture}
\end{array}
\ar@{=>}[r]^\alpha
&
\begin{array}{c}
\begin{tikzpicture}[scale = 0.15]
\draw[black,thick](1,1.5)--(0.5,2)--(2,3.5);
\draw[black,thick](0,0)--(-1,1)--(-1,2)--(1,4);
\draw[white,line width = 3](1.5,2)--(-0.5,4);
\draw[black,thick](0,0)--(1,1)--(1,1.5)--(1.5,2)--(-0.5,4);
\draw[black,thick](0,0)--(0,-2);
\end{tikzpicture}
\end{array}
&
\begin{array}{c}
\begin{tikzpicture}[scale = 0.15]
\draw[black,thick](0,0)--(-1,1)--(-1,2)--(0.5,3.5);
\draw[white,line width = 3](1,2)--(-0.5,3.5);
\draw[black,thick](0,0)--(1,1)--(1,2)--(-0.5,3.5);
\draw[black,thick](1,2)--(2,3);
\draw[black,thick](0,0)--(0,-2);
\end{tikzpicture}
\end{array}
\ar@{=>}[l]_\beta 
}}
\end{equation}

\begin{equation}\label{diagram:commutativity_2}
\raisebox{2em}{\xymatrix @R=1em @C=2em{
\begin{array}{c}
\begin{tikzpicture}[scale = 0.2]
\draw[black,thick](0,0)--(-1,1)--(-1,1.5);
\draw[black,thick](0,0)--(1,1)--(1,2);
\draw[black,thick](-1,1.5)--(-1.5,2);
\draw[black,thick](-1,1.5)--(-0.5,2);
\draw[black,thick](0,0)--(0,-1);
\end{tikzpicture}
\end{array}
\ar@{=>}[d]_\beta 
&
\begin{array}{c}
\begin{tikzpicture}[scale = 0.2]
\draw[black,thick](0,0)--(-1,1)--(-1,2);
\draw[black,thick](0,0)--(1,1)--(1,1.5)--(1.5,2);
\draw[black,thick](1,1.5)--(0.5,2);
\draw[black,thick](0,0)--(0,-1);
\end{tikzpicture}
\end{array}
\ar@{=>}[r]^\beta
\ar@{=>}[l]_\alpha 
&
\begin{array}{c}
\begin{tikzpicture}[scale = 0.15]
\draw[black,thick](0,0)--(1,1)--(1,1.5)--(0.5,2)--(1.5,3);
\draw[white,line width = 3](1.5,2)--(0.5,3);
\draw[black,thick](0,0)--(-1,1)--(-1,2.5);
\draw[black,thick](1,1.5)--(1.5,2)--(0.5,3);
\draw[black,thick](0,0)--(0,-2);
\end{tikzpicture}
\end{array}
\ar@{=>}[rd]^\alpha
&
\\
\begin{array}{c}
\begin{tikzpicture}[scale = 0.15]
\draw[black,thick](0,0)--(-1,1)--(1,3)--(1,4);
\draw[white,line width = 3](1,1)--(-1,3);
\draw[black,thick](0,0)--(1,1)--(-1,3)--(-1,3.5)--(-1.5,4);
\draw[black,thick](-1,3.5)--(-0.5,4);
\draw[black,thick](0,0)--(0,-2);
\end{tikzpicture}
\end{array}
\ar@{=>}[r]
&
\begin{array}{c}
\begin{tikzpicture}[scale = 0.15]
\draw[black,thick](0,0)--(-1,1)--(-1,2)--(1,4);
\draw[white,line width = 3](1,1.5)--(-1,3.5);
\draw[white,line width = 3](1.5,2)--(-0.5,4);
\draw[black,thick](0,0)--(1,1)--(1,1.5)--(-1,3.5);
\draw[black,thick](1,1.5)--(1.5,2)--(-0.5,4);
\draw[black,thick](0,0)--(0,-2);
\end{tikzpicture}
\end{array}
\ar@{=>}[r]^\alpha
&
\begin{array}{c}
\begin{tikzpicture}[scale = 0.15]
\draw[black,thick](0,0)--(-1,1)--(-1,1.5)--(-1.5,2)--(0.5,4);
\draw[white,line width = 3](-0.5,2)--(-2,3.5);
\draw[white,line width = 3](1,2)--(-1,4);
\draw[black,thick](-1,1.5)--(-0.5,2)--(-2,3.5);
\draw[black,thick](0,0)--(1,1)--(1,2)--(-1,4);
\draw[black,thick](0,0)--(0,-2);
\end{tikzpicture}
\end{array}
&
\begin{array}{c}
\begin{tikzpicture}[scale = 0.15]
\draw[black,thick](0,0)--(-1,1)--(-1,2)--(0.5,3.5);
\draw[white,line width = 3](1,2)--(-0.5,3.5);
\draw[black,thick](0,0)--(1,1)--(1,2)--(-0.5,3.5);
\draw[black,thick](-1,2)--(-2,3);
\draw[black,thick](0,0)--(0,-2);
\end{tikzpicture}
\end{array}
\ar@{=>}[l]_\beta
}}
\end{equation}
\end{defn}

\begin{defn} \label{def:alg-hom}
Let $A$ and $B$ be two commutative algebras in $\SM$. An algebra homomorphism $A \to B$ is a triple $(f,\eta,\xi)$, where $f \colon A\to B$ is a 1-morphism, $\eta \colon f\circ u_A \Rightarrow u_B$ and $\xi \colon \mu_B \circ (f\otimes f) \Rightarrow f \circ \mu_A$ are 2-isomorphisms as illustrated below,
\[
\begin{tikzcd}
\begin{array}{c}
\begin{tikzpicture}[scale = 0.2]
\draw[->-,black,thick](0,2.5)--(0,-2.5);
\fill[black](0,2.5) circle (0.3);
\node[black] at (-1,1) {\scriptsize $A$};
\node[black] at (-1,-2.3) {\scriptsize$B$};
\node[black] at (1,0) {\scriptsize $f$};
\end{tikzpicture}
\end{array}
\ar[r,Rightarrow,"\eta"]
&
\begin{array}{c}
\begin{tikzpicture}[scale = 0.2]
\draw[black,thick](0,1.5)--(0,-1.5);
\fill[blue](0,1.5) circle (0.3);
\node[black] at (1,-1.8) {\scriptsize $B$};
\end{tikzpicture}
\end{array}
\end{tikzcd}
,\quad\quad
\begin{tikzcd}
\begin{array}{c}
\begin{tikzpicture}[scale = 0.15]
\draw[->-,black,thick](-2,2)--(0,0);
\draw[->-,black,thick](2,2)--(0,0);
\draw[black,thick](0,0)--(0,-3);
\node[black] at (-2.3,2.3) {\scriptsize$A$};
\node[black] at (2.3,2.3) {\scriptsize $A$};
\node[black] at (0.7,-3.5) {\scriptsize $B$};
\end{tikzpicture}
\end{array}
\ar[r,Rightarrow,"\xi"]
&
\begin{array}{c}
\begin{tikzpicture}[scale = 0.15]
\draw[black,thick](-2,2)--(0,0);
\draw[black,thick](2,2)--(0,0);
\draw[->-,black,thick](0,0)--(0,-3);
\node[black] at (-2.3,2.3) {\scriptsize $A$};
\node[black] at (2.3,2.3) {\scriptsize $A$};
\node[black] at (0.7,-3.5) {\scriptsize $B$};
\end{tikzpicture}
\end{array}
\end{tikzcd}
\]
such that the following diagrams are commutative:  
\begin{equation} 
\raisebox{2em}{\xymatrix @R=1em @C=2em{
\begin{array}{c}
\begin{tikzpicture}[scale = 0.15]
\draw[->-,black,thick](-2,2)--(0,0);
\draw[->-,black,thick](2,2)--(0,0);
\draw[black,thick](0,0)--(0,-3);
\fill[blue](-2,2) circle (0.3);
\end{tikzpicture}
\end{array}
\ar@{=>}[r]^\xi \ar@{=>}[d]_\eta
& 
\begin{array}{c}
\begin{tikzpicture}[scale = 0.15]
\draw[black,thick](-2,2)--(0,0);
\draw[black,thick](2,2)--(0,0);
\draw[->-,black,thick](0,0)--(0,-3);
\fill[blue](-2,2) circle (0.3);
\end{tikzpicture}
\end{array}
\ar@{=>}[d]^{\lambda_A}
\\
\begin{array}{c}
\begin{tikzpicture}[scale = 0.15]
\draw[black,thick](-2,2)--(0,0);
\draw[->-,black,thick](2,2)--(0,0);
\draw[black,thick](0,0)--(0,-3);
\fill[blue](-2,2) circle (0.3);
\end{tikzpicture}
\end{array}
\ar@{=>}[r]^{\lambda_B}
&
\begin{array}{c}
\begin{tikzpicture}[scale = 0.15]
\draw[->-,black,thick](0,2)--(0,-2);
\end{tikzpicture}
\end{array}
}}
\quad
\raisebox{2em}{\xymatrix @R=1em @C=2em{
\begin{array}{c}
\begin{tikzpicture}[scale = 0.15]
\draw[->-,black,thick](-2,2)--(0,0);
\draw[->-,black,thick](2,2)--(0,0);
\draw[black,thick](0,0)--(0,-3);
\fill[blue](2,2) circle (0.3);
\end{tikzpicture}
\end{array}
\ar@{=>}[r]^\xi \ar@{=>}[d]_\eta
& 
\begin{array}{c}
\begin{tikzpicture}[scale = 0.15]
\draw[black,thick](-2,2)--(0,0);
\draw[black,thick](2,2)--(0,0);
\draw[->-,black,thick](0,0)--(0,-3);
\fill[blue](2,2) circle (0.3);
\end{tikzpicture}
\end{array}
\ar@{=>}[d]^{\rho_A}
\\
\begin{array}{c}
\begin{tikzpicture}[scale = 0.15]
\draw[->-,black,thick](-2,2)--(0,0);
\draw[black,thick](2,2)--(0,0);
\draw[black,thick](0,0)--(0,-3);
\fill[blue](2,2) circle (0.3);
\end{tikzpicture}
\end{array}
\ar@{=>}[r]^{\rho_B}
&
\begin{array}{c}
\begin{tikzpicture}[scale = 0.15]
\draw[->-,black,thick](0,2)--(0,-2);
\end{tikzpicture}
\end{array}
}}
\end{equation}
\begin{equation}
\raisebox{2em}{\xymatrix @R=1em @C=2em{
\begin{array}{c}
\begin{tikzpicture}[scale = 0.2]
\draw[->-,black,thick](1.5,3.5)--(1.5,1.5);
\draw[->-,black,thick](-2.5,3.5)--(-1,2);
\draw[->-,black,thick](0.5,3.5)--(-1,2);
\draw[black,thick](-1,2)--(-1,1);
\draw[black,thick](-1,1)--(0,0);
\draw[black,thick](1.5,1.5)--(0,0);
\draw[black,thick](0,0)--(0,-2);
\end{tikzpicture}
\end{array}
\ar@{=>}[r]^\xi
\ar@{=>}[d]_{\alpha_B}
&
\begin{array}{c}
\begin{tikzpicture}[scale = 0.2]
\draw[->-,black,thick](1.5,3.5)--(1.5,1.5);
\draw[black,thick](-2.5,3.5)--(-1,2);
\draw[black,thick](0.5,3.5)--(-1,2);
\draw[black,thick](-1,2)--(-1,1);
\draw[->-,black,thick](-1,1)--(0,0);
\draw[black,thick](1.5,1.5)--(0,0);
\draw[black,thick](0,0)--(0,-2);
\end{tikzpicture}
\end{array}
\ar@{=>}[r]^\xi
& 
\begin{array}{c}
\begin{tikzpicture}[scale = 0.2]
\draw[black,thick](1.5,3.5)--(1.5,1.5);
\draw[black,thick](-2.5,3.5)--(-1,2);
\draw[black,thick](0.5,3.5)--(-1,2);
\draw[black,thick](-1,2)--(-1,1);
\draw[black,thick](-1,1)--(0,0);
\draw[black,thick](1.5,1.5)--(0,0);
\draw[->-,black,thick](0,0)--(0,-2);
\end{tikzpicture}
\end{array}
\ar@{=>}[d]^{\alpha_A}
\\
\begin{array}{c}
\begin{tikzpicture}[scale = 0.2]
\draw[->-,black,thick](-1.5,3.5)--(-1.5,1.5);
\draw[->-,black,thick](2.5,3.5)--(1,2);
\draw[->-,black,thick](-0.5,3.5)--(1,2);
\draw[black,thick](1,2)--(1,1);
\draw[black,thick](1,1)--(0,0);
\draw[black,thick](-1.5,1.5)--(0,0);
\draw[black,thick](0,0)--(0,-2);
\end{tikzpicture}
\end{array}
\ar@{=>}[r]^\xi
&
\begin{array}{c}
\begin{tikzpicture}[scale = 0.2]
\draw[->-,black,thick](-1.5,3.5)--(-1.5,1.5);
\draw[black,thick](2.5,3.5)--(1,2);
\draw[black,thick](-0.5,3.5)--(1,2);
\draw[black,thick](1,2)--(1,1);
\draw[->-,black,thick](1,1)--(0,0);
\draw[black,thick](-1.5,1.5)--(0,0);
\draw[black,thick](0,0)--(0,-2);
\end{tikzpicture}
\end{array}
\ar@{=>}[r]^\xi
&
\begin{array}{c}
\begin{tikzpicture}[scale = 0.2]
\draw[black,thick](-1.5,3.5)--(-1.5,1.5);
\draw[black,thick](2.5,3.5)--(1,2);
\draw[black,thick](-0.5,3.5)--(1,2);
\draw[black,thick](1,2)--(1,1);
\draw[black,thick](1,1)--(0,0);
\draw[black,thick](-1.5,1.5)--(0,0);
\draw[->-,black,thick](0,0)--(0,-2);
\end{tikzpicture}
\end{array}
}}
\end{equation}
\begin{equation}
\raisebox{2em}{\xymatrix @R=1em @C=2em{
\begin{array}{c}
\begin{tikzpicture}[scale = 0.15]
\draw[->-,black,thick](-2,2)--(0,0);
\draw[->-,black,thick](2,2)--(0,0);
\draw[black,thick](0,0)--(0,-3);
\end{tikzpicture}
\end{array}
\ar@{=>}[rr]^\xi \ar@{=>}[d]_-\beta
& &
\begin{array}{c}
\begin{tikzpicture}[scale = 0.15]
\draw[black,thick](-2,2)--(0,0);
\draw[black,thick](2,2)--(0,0);
\draw[->-,black,thick](0,0)--(0,-3);
\end{tikzpicture}
\end{array}
\ar@{=>}[d]^-\beta
\\
\begin{array}{c}
\begin{tikzpicture}[scale = 0.15]
\draw[->-,black,thick](2,5)--(0,3);
\draw[black,thick](0,3)--(-1.5,1.5);
\draw[white,line width = 3](-0.5,2.5)--(0.5,3.5);
\draw[black,thick](1.5,1.5)--(0,0);
\draw[black,thick](-1.5,1.5)--(0,0);
\draw[black,thick](0,0)--(0,-3);
\draw[->-,black,thick](-2,5)--(0,3);
\draw[black,thick](0,3)--(1.5,1.5);
\end{tikzpicture}
\end{array}
\ar@{=>}[r]
& 
\begin{array}{c}
\begin{tikzpicture}[scale = 0.15]
\draw[black,thick](1,4)--(-1.5,1.5);
\draw[white,line width = 3](-1,4)--(1.5,1.5);
\draw[->-,black,thick](-1.5,1.5)--(0,0);
\draw[->-,black,thick](1.5,1.5)--(0,0);
\draw[black,thick](0,0)--(0,-3);
\draw[black,thick](-1,4)--(1.5,1.5);
\end{tikzpicture}
\end{array}
\ar@{=>}[r]^\xi
&
\begin{array}{c}
\begin{tikzpicture}[scale = 0.15]
\draw[black,thick](1,4)--(-1.5,1.5);
\draw[white,line width = 3](-1,4)--(1.5,1.5);
\draw[black,thick](-1.5,1.5)--(0,0);
\draw[black,thick](1.5,1.5)--(0,0);
\draw[->-,black,thick](0,0)--(0,-3);
\draw[black,thick](-1,4)--(1.5,1.5);
\end{tikzpicture}
\end{array}
}}
\end{equation}
\end{defn}

\section{Module categories of the Lagrangian algebras} \label{sec:modules}
The notions in this subsection are standard (see for example \cite{DN21,JFR23,Dec21}).
\begin{defn} \label{def:module}
Let $A$ be an algebra in a monoidal 2-category. A right $A$-module is a quadruple $(M,\mu_M,\alpha_M,r_M)$, where $M$ is a an object, the 1-morphism $\mu_M \colon M \otimes A \to M$ defines the right $A$-action on $M$ and is depicted as follows,
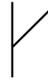
\[
\begin{tikzcd}
\begin{array}{c}
\begin{tikzpicture}[scale = 0.2]
\draw[black,thick](0,3)--(0,-2);
\draw[black,thick](2.5,2.5)--(0,0);
\end{tikzpicture}
\end{array}
\end{tikzcd}
\]
and the 2-associator $\alpha_M$ and the right 2-unitor $r_M$ are invertible 2-morphisms depicted as follows,
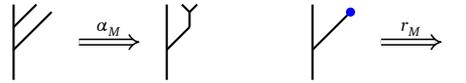
\[
\begin{tikzcd}
\begin{array}{c}
\begin{tikzpicture}[scale = 0.2]
\draw[black,thick](0,3)--(0,-2);
\draw[black,thick](1.5,3)--(0,1.5);
\draw[black,thick](2.5,2.5)--(0,0);
\end{tikzpicture}
\end{array}
\ar[r,Rightarrow,"\alpha_M"]
&
\begin{array}{c}
\begin{tikzpicture}[scale = 0.2]
\draw[black,thick](0,3)--(0,-2);
\draw[black,thick](1.5,1.5)--(0,0);
\draw[black,thick](1.5,2.5)--(1.5,1.5);
\draw[black,thick](2,3)--(1.5,2.5);
\draw[black,thick](1,3)--(1.5,2.5);
\end{tikzpicture}
\end{array}
&
\begin{array}{c}
\begin{tikzpicture}[scale = 0.2]
\draw[black,thick](0,3)--(0,-2);
\draw[black,thick](2.5,2.5)--(0,0);
\fill[blue](2.5,2.5) circle (0.3);
\end{tikzpicture}
\end{array}
\ar[r,Rightarrow,"r_M"]
&
\begin{array}{c}
\begin{tikzpicture}[scale = 0.2]
\draw[black,thick](0,3)--(0,-2);
\end{tikzpicture}
\end{array}
\end{tikzcd}
\]
such that the following diagrams are commutative.
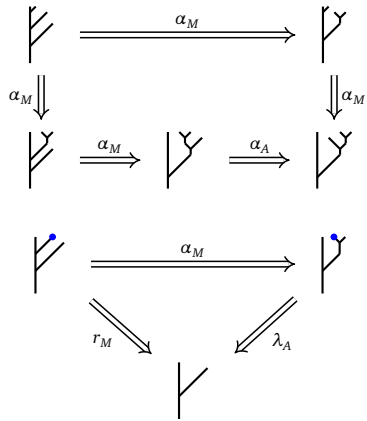
\[
\begin{tikzcd}
\begin{array}{c}
\begin{tikzpicture}[scale = 0.15]
\draw[black,thick](0,3)--(0,-2);
\draw[black,thick](0.5,3)--(0,2.5);
\draw[black,thick](1.5,2.5)--(0,1);
\draw[black,thick](2,1.5)--(0,-0.5);
\end{tikzpicture}
\end{array}
\ar[rr,Rightarrow,"\alpha_M"]
\ar[d,Rightarrow,"\alpha_M"']
&&
\begin{array}{c}
\begin{tikzpicture}[scale = 0.15]
\draw[black,thick](0,3)--(0,-2);
\draw[black,thick](0.5,3)--(0,2.5);
\draw[black,thick](1.5,1.5)--(0,0);
\draw[black,thick](1.5,2)--(1.5,1.5);
\draw[black,thick](2,2.5)--(1.5,2);
\draw[black,thick](1,2.5)--(1.5,2);
\end{tikzpicture}
\end{array}
\ar[d,Rightarrow,"\alpha_M"]
\\
\begin{array}{c}
\begin{tikzpicture}[scale = 0.15]
\draw[black,thick](0,3)--(0,-2);
\draw[black,thick](2,1.5)--(0,-0.5);
\draw[black,thick](1.5,2)--(0,0.5);
\draw[black,thick](1.5,2.5)--(1.5,2);
\draw[black,thick](1,3)--(1.5,2.5);
\draw[black,thick](2,3)--(1.5,2.5);
\end{tikzpicture}
\end{array}
\ar[r,Rightarrow,"\alpha_M"]
&
\begin{array}{c}
\begin{tikzpicture}[scale = 0.15]
\draw[black,thick](0,3)--(0,-2);
\draw[black,thick](2,1)--(0,-1);
\draw[black,thick](2,1.5)--(2,1);
\draw[black,thick](1.5,2)--(2,1.5);
\draw[black,thick](1.5,2.5)--(1.5,2);
\draw[black,thick](1,3)--(1.5,2.5);
\draw[black,thick](2,3)--(1.5,2.5);
\draw[black,thick](3,2.5)--(2,1.5);
\end{tikzpicture}
\end{array}
\ar[r,Rightarrow,"\alpha_A"]
&
\begin{array}{c}
\begin{tikzpicture}[scale = 0.15]
\draw[black,thick](0,3)--(0,-2);
\draw[black,thick](2,1)--(0,-1);
\draw[black,thick](2,1.5)--(2,1);
\draw[black,thick](2.5,2)--(2,1.5);
\draw[black,thick](2.5,2.5)--(2.5,2);
\draw[black,thick](3,3)--(2.5,2.5);
\draw[black,thick](2,3)--(2.5,2.5);
\draw[black,thick](1,2.5)--(2,1.5);
\end{tikzpicture}
\end{array}
\end{tikzcd}
\]
\[
\begin{tikzcd}
\begin{array}{c}
\begin{tikzpicture}[scale = 0.15]
\draw[black,thick](0,3)--(0,-2);
\draw[black,thick](1.5,3)--(0,1.5);
\draw[black,thick](2.5,2.5)--(0,0);
\fill[blue](1.5,3) circle (0.3);
\end{tikzpicture}
\end{array}
\ar[rr,Rightarrow,"\alpha_M"]
\ar[rd,Rightarrow,"r_M"']
&&
\begin{array}{c}
\begin{tikzpicture}[scale = 0.15]
\draw[black,thick](0,3)--(0,-2);
\draw[black,thick](1.5,1.5)--(0,0);
\draw[black,thick](1.5,2.5)--(1.5,1.5);
\draw[black,thick](2,3)--(1.5,2.5);
\draw[black,thick](1,3)--(1.5,2.5);
\fill[blue](1,3) circle (0.3);
\end{tikzpicture}
\end{array}
\ar[ld,Rightarrow,"\lambda_A"]
\\
&
\begin{array}{c}
\begin{tikzpicture}[scale = 0.15]
\draw[black,thick](0,3)--(0,-2);
\draw[black,thick](2.5,2.5)--(0,0);
\end{tikzpicture}
\end{array}
\end{tikzcd}
\]
Here $\alpha_A$ is the associator of $A$, and $\lambda_A$ is the left unitor of $A$. For convenience we simply write $M$ for the quadruple. The definition of a left $A$-module is similar. 
\end{defn}

\begin{defn}
Let $M,N$ be right $A$-modules. An $A$-module 1-map from $M$ to $N$ is a pair $(f,\phi)$ where $f \colon M \to N$ is a 1-morphism and $\phi$ is an invertible 2-morphism depicted as:
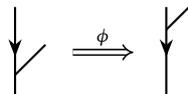
\[
\begin{tikzcd}
\begin{array}{c}
\begin{tikzpicture}[scale = 0.2]
\draw[->-,black,thick](0,3)--(0,-3);
\draw[black,thick](2,0.5)--(0,-1.5);
\end{tikzpicture}
\end{array}
\ar[r,Rightarrow,"\phi"]
&
\begin{array}{c}
\begin{tikzpicture}[scale = 0.2]
\draw[->-,black,thick](0,3)--(0,-3);
\draw[black,thick](1.5,3)--(0,1.5);
\end{tikzpicture}
\end{array}
\end{tikzcd}
\]
such that the following diagrams are commutative:
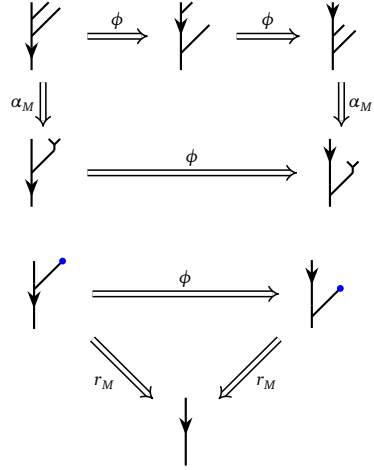
\[
\begin{tikzcd}
\begin{array}{c}
\begin{tikzpicture}[scale = 0.15]
\draw[black,thick](0,3)--(0,-1);
\draw[black,thick](1.5,3)--(0,1.5);
\draw[black,thick](2.5,2.5)--(0,0);
\draw[->-,black,thick](0,-1)--(0,-3);
\end{tikzpicture}
\end{array}
\ar[r,Rightarrow,"\phi"]
\ar[d,Rightarrow,"\alpha_M"']
&
\begin{array}{c}
\begin{tikzpicture}[scale = 0.15]
\draw[black,thick](0,3)--(0,2);
\draw[black,thick](1,3)--(0,2);
\draw[->-,black,thick](0,2)--(0,-1.5);
\draw[black,thick](2.5,1)--(0,-1.5);
\draw[black,thick](0,-1.5)--(0,-3);
\end{tikzpicture}
\end{array}
\ar[r,Rightarrow,"\phi"]
& 
\begin{array}{c}
\begin{tikzpicture}[scale = 0.15]
\draw[->-,black,thick](0,3)--(0,0);
\draw[black,thick](0,0)--(0,-3);
\draw[black,thick](1,1)--(0,0);
\draw[black,thick](2,0.5)--(0,-1.5);
\end{tikzpicture}
\end{array}
\ar[d,Rightarrow,"\alpha_M"]
\\
\begin{array}{c}
\begin{tikzpicture}[scale = 0.15]
\draw[black,thick](0,3)--(0,-1);
\draw[black,thick](2,2)--(0,0);
\draw[black,thick](2,2.5)--(2,2);
\draw[black,thick](2.5,3)--(2,2.5);
\draw[black,thick](1.5,3)--(2,2.5);
\draw[->-,black,thick](0,-1)--(0,-3);
\end{tikzpicture}
\end{array}
\ar[rr,Rightarrow,"\phi"]
& &
\begin{array}{c}
\begin{tikzpicture}[scale = 0.15]
\draw[->-,black,thick](0,3)--(0,-1);
\draw[black,thick](2,0)--(0,-2);
\draw[black,thick](2,0.5)--(2,0);
\draw[black,thick](2.5,1)--(2,0.5);
\draw[black,thick](1.5,1)--(2,0.5);
\draw[black,thick](0,-1)--(0,-3);
\end{tikzpicture}
\end{array}
\end{tikzcd}
\]
\[
\begin{tikzcd}
\begin{array}{c}
\begin{tikzpicture}[scale = 0.15]
\draw[black,thick](0,3)--(0,1);
\draw[black,thick](2.5,3)--(0,0.5);
\fill[blue](2.5,3) circle (0.3);
\draw[->-,black,thick](0,1)--(0,-3);
\end{tikzpicture}
\end{array}
\ar[rr,Rightarrow,"\phi"]
\ar[rd,Rightarrow,"r_M"']
&&
\begin{array}{c}
\begin{tikzpicture}[scale = 0.15]
\draw[->-,black,thick](0,3)--(0,-1);
\draw[black,thick](2.5,0.5)--(0,-2);
\fill[blue](2.5,0.5) circle (0.3);
\draw[black,thick](0,-1)--(0,-3);
\end{tikzpicture}
\end{array}
\ar[ld,Rightarrow,"r_M"]
\\
&
\begin{array}{c}
\begin{tikzpicture}[scale = 0.15]
\draw[->-,black,thick](0,3)--(0,-3);
\end{tikzpicture}
\end{array}
\end{tikzcd}
\]
\end{defn}

\begin{defn}
Let $(f,\phi)$ and $(g,\phi^\prime)$ be two $A$-module 1-maps from $M$ to $N$. An $A$-module 2-map from $(f,\phi)$ to $(g,\phi^\prime)$ is a 2-morphism $\theta \colon f \Rightarrow g$ such that the following diagram is commutative:
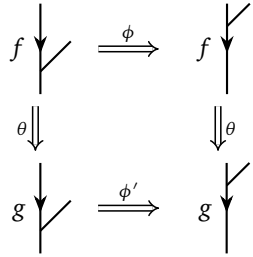
\[
\begin{tikzcd}
\begin{array}{c}
\begin{tikzpicture}[scale = 0.2]
\draw[->-,black,thick](0,3)--(0,-3);
\draw[black,thick](2,0.5)--(0,-1.5);
\node[left] at (0,0) {$f$};
\end{tikzpicture}
\end{array}
\ar[r,Rightarrow,"\phi"]
\ar[d,Rightarrow,"\theta"']
&
\begin{array}{c}
\begin{tikzpicture}[scale = 0.2]
\draw[->-,black,thick](0,3)--(0,-3);
\draw[black,thick](1.5,3)--(0,1.5);
\node[left] at (0,0) {$f$};
\end{tikzpicture}
\end{array}
\ar[d,Rightarrow,"\theta"]
\\
\begin{array}{c}
\begin{tikzpicture}[scale = 0.2]
\draw[->-,black,thick](0,3)--(0,-3);
\draw[black,thick](2,0.5)--(0,-1.5);
\node[left] at (0,0) {$g$};
\end{tikzpicture}
\end{array}
\ar[r,Rightarrow,"\phi^\prime"]
&
\begin{array}{c}
\begin{tikzpicture}[scale = 0.2]
\draw[->-,black,thick](0,3)--(0,-3);
\draw[black,thick](1.5,3)--(0,1.5);
\node[left] at (0,0) {$g$};
\end{tikzpicture}
\end{array}
\end{tikzcd}
\]
\end{defn}

\begin{defn}
The 2-category of right $A$-modules is defined as follows: the objects are right $A$-modules, 1-morphisms are $A$-module 1-maps, 2-morphisms are $A$-module 2-maps. This 2-category is denoted by $\SM_A$.
\end{defn}

\begin{defn}
An $A$-$A$-bimodule $M$ is a left $A$-module and a right $A$-module, equipped with a 2-isomorphism 
\[
\delta \colon \mu_M^L \circ (1 \otimes \mu_M^R) \Rightarrow  \mu_M^R \circ (\mu_M^L \otimes 1)
\] 
such that $(\mu_M^R, \delta)$ defines a left $A$-module 1-map and $(\mu_M^L,\delta^{-1})$ defines a right $A$-module 1-map. Graphically $\delta$ can be depicted as follows:
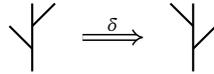
\[
\begin{tikzcd}
\begin{array}{c}
\begin{tikzpicture}[scale = 0.15]
\draw[black,thick](0,3)--(0,-3);
\draw[black,thick](2,3)--(0,1);
\draw[black,thick](-2,1)--(0,-1);
\end{tikzpicture}
\end{array}
\ar[r,Rightarrow,"\delta"]
&
\begin{array}{c}
\begin{tikzpicture}[scale = 0.15]
\draw[black,thick](0,3)--(0,-3);
\draw[black,thick](-2,3)--(0,1);
\draw[black,thick](2,1)--(0,-1);
\end{tikzpicture}
\end{array}
\end{tikzcd}
\]
\end{defn}

\begin{defn}
Let $M,N$ be $A$-$A$-bimodules. An $A$-$A$-bimodule 1-map is a triple $(f,\phi^l,\phi^r)$ such that $(f,\phi^l)/(f,\phi^r)$ is a left/right $A$-module 1-map, and the following diagram is commutative:
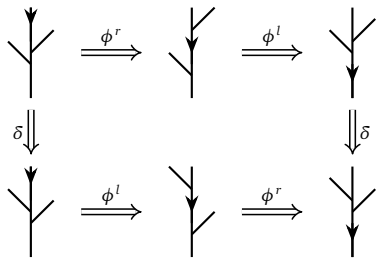
\[
\begin{tikzcd}
\begin{array}{c}
\begin{tikzpicture}[scale = 0.15]
\draw[->-,black,thick](0,4)--(0,1);
\draw[black,thick](0,4)--(0,-4);
\draw[black,thick](2,2)--(0,0);
\draw[black,thick](-2,1)--(0,-1);
\end{tikzpicture}
\end{array}
\ar[r,Rightarrow,"\phi^r"]
\ar[d,Rightarrow,"\delta"']
&
\begin{array}{c}
\begin{tikzpicture}[scale = 0.15]
\draw[->-,black,thick](0,1)--(0,-1);
\draw[black,thick](0,4)--(0,-4);
\draw[black,thick](2,4)--(0,2);
\draw[black,thick](-2,0)--(0,-2);
\end{tikzpicture}
\end{array}
\ar[r,Rightarrow,"\phi^l"]
&
\begin{array}{c}
\begin{tikzpicture}[scale = 0.15]
\draw[->-,black,thick](0,-1)--(0,-4);
\draw[black,thick](0,4)--(0,-4);
\draw[black,thick](2,3)--(0,1);
\draw[black,thick](-2,2)--(0,0);
\end{tikzpicture}
\end{array}
\ar[d,Rightarrow,"\delta"]
\\
\begin{array}{c}
\begin{tikzpicture}[scale = 0.15]
\draw[->-,black,thick](0,4)--(0,1);
\draw[black,thick](0,4)--(0,-4);
\draw[black,thick](-2,2)--(0,0);
\draw[black,thick](2,1)--(0,-1);
\end{tikzpicture}
\end{array}
\ar[r,Rightarrow,"\phi^l"]
&
\begin{array}{c}
\begin{tikzpicture}[scale = 0.15]
\draw[->-,black,thick](0,1)--(0,-1);
\draw[black,thick](0,4)--(0,-4);
\draw[black,thick](-2,4)--(0,2);
\draw[black,thick](2,0)--(0,-2);
\end{tikzpicture}
\end{array}
\ar[r,Rightarrow,"\phi^r"]
&
\begin{array}{c}
\begin{tikzpicture}[scale = 0.15]
\draw[->-,black,thick](0,-1)--(0,-4);
\draw[black,thick](0,4)--(0,-4);
\draw[black,thick](-2,3)--(0,1);
\draw[black,thick](2,2)--(0,0);
\end{tikzpicture}
\end{array}
\end{tikzcd}
\]
\end{defn}

If $A$ is a commutative algebra in $\SM$, then a right $A$-module $M$ is also equipped with 2 different structures of left $A$-modules, defined in the following two diagrams respectively:
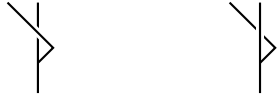
\[
\begin{tikzcd}
\begin{array}{c}
\begin{tikzpicture}[scale = 0.2]
\draw[black,thick](0,3)--(0,-3);
\draw[white,line width = 3](-0.5,1.5)--(0.5,0.5);
\draw[black,thick](-2,3)--(1,0)--(0,-1);
\end{tikzpicture}
\end{array}
&&
\begin{array}{c}
\begin{tikzpicture}[scale = 0.2]
\draw[black,thick](-2,3)--(1,0)--(0,-1);
\draw[white,line width = 3](0,0)--(0,2);
\draw[black,thick](0,3)--(0,-3);
\end{tikzpicture}
\end{array}
\end{tikzcd}
\]
and we denote the them by $M_+$ and $M_-$ respectively. The 2-associator of $M_+$ can be written as the composition of the following 2-morphisms:
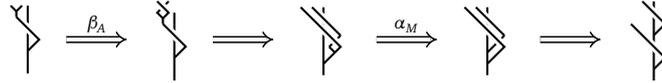
\[
\begin{tikzcd}
\begin{array}{c}
\begin{tikzpicture}[scale = 0.15]
\draw[black,thick](0,3)--(0,-3);
\draw[white,line width = 3](-0.5,1.5)--(0.5,0.5);
\draw[black,thick](-1,2)--(1,0)--(0,-1);
\draw[black,thick](-1,2.5)--(-1,2);
\draw[black,thick](-1.5,3)--(-1,2.5);
\draw[black,thick](-0.5,3)--(-1,2.5);
\end{tikzpicture}
\end{array}
\ar[r,Rightarrow,"\beta_A"]
&
\begin{array}{c}
\begin{tikzpicture}[scale = 0.15]
\draw[black,thick](0,3)--(0,-3);
\draw[white,line width = 3](-0.5,1.5)--(0.5,0.5);
\draw[black,thick](-1,2)--(1,0)--(0,-1);
\draw[black,thick](-1,2.5)--(-1,2);
\draw[black,thick](-1,2.5)--(-1.5,3)--(-0.5,4);
\draw[white,line width = 3](-0.5,3)--(-1.5,4);
\draw[black,thick](-1,2.5)--(-0.5,3)--(-1.5,4);
\end{tikzpicture}
\end{array}
\ar[r,Rightarrow]
&
\begin{array}{c}
\begin{tikzpicture}[scale = 0.15]
\draw[black,thick](0,3)--(0,-3);
\draw[black,thick](0.5,-0.5)--(1.5,0.5);
\draw[white,line width = 3](-1,2)--(2,-1);
\draw[white,line width = 3](-1,3)--(1,1);
\draw[black,thick](-1,3)--(1.5,0.5);
\draw[black,thick](-2,3)--(1.5,-0.5)--(0,-2);
\draw[black,thick](0.5,-0.5)--(1,-1);
\end{tikzpicture}
\end{array}
\ar[r,Rightarrow,"\alpha_M"]
&
\begin{array}{c}
\begin{tikzpicture}[scale = 0.15]
\draw[black,thick](0,3)--(0,-3);
\draw[black,thick](0.5,-0.5)--(1.5,0.5);
\draw[white,line width = 3](-1,2)--(2,-1);
\draw[white,line width = 3](-1,3)--(1,1);
\draw[black,thick](-1,3)--(1.5,0.5);
\draw[black,thick](-2,3)--(1.5,-0.5)--(0,-2);
\draw[black,thick](0.5,-0.5)--(0,-1);
\end{tikzpicture}
\end{array}
\ar[r,Rightarrow]
&
\begin{array}{c}
\begin{tikzpicture}[scale = 0.15]
\draw[black,thick](0,4)--(0,-3);
\draw[white,line width = 3](-2,2)--(1,-1);
\draw[white,line width = 3](-1,4)--(1,2);
\draw[black,thick](-2,2)--(1,-1)--(0,-2);
\draw[black,thick](-1,4)--(1,2)--(0,1);
\end{tikzpicture}
\end{array}
\end{tikzcd}
\]
The 2-associator of $M_-$ can be written down in a similar way. We denote these two associators by $\alpha_M^+$ and $\alpha_M^-$ respectively. Both left $A$-module structures are compatible with the right $A$-module structure in the sense that they can be upgraded to two $A$-bimodule structures. This compatibility for $M_+$ is defined by the composition of the following 2-isomorphisms.  
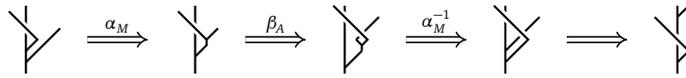
\[
\begin{tikzcd}
\begin{array}{c}
\begin{tikzpicture}[scale = 0.15]
\draw[black,thick](0,3)--(0,-3);
\draw[white,line width = 3](-1.5,2.5)--(1,0);
\draw[black,thick](-1.5,2.5)--(1,0)--(0,-1);
\draw[black,thick](3,1)--(0,-2);
\end{tikzpicture}
\end{array}
\ar[r,Rightarrow,"\alpha_M"]
&
\begin{array}{c}
\begin{tikzpicture}[scale = 0.15]
\draw[black,thick](0,3)--(0,-3);
\draw[white,line width = 3](-1.5,2.5)--(1,0);
\draw[black,thick](-1.5,2.5)--(1,0)--(1,-0.5)--(0,-1.5);
\draw[black,thick](2,1)--(1,0);
\end{tikzpicture}
\end{array}
\ar[r,Rightarrow,"\beta_A"]
&
\begin{array}{c}
\begin{tikzpicture}[scale = 0.15]
\draw[black,thick](0,3)--(0,-3);
\draw[black,thick](1,0)--(3,2);
\draw[white,line width = 3](-0.5,2.5)--(2,0);
\draw[black,thick](-1,3)--(2,0)--(1.5,-0.5);
\draw[black,thick](1,0)--(1.5,-0.5);
\draw[black,thick](1.5,-0.5)--(1.5,-1);
\draw[black,thick](1.5,-1)--(0,-2.5);
\end{tikzpicture}
\end{array}
\ar[r,Rightarrow,"\alpha_M^{-1}"]
&
\begin{array}{c}
\begin{tikzpicture}[scale = 0.15]
\draw[black,thick](0,3)--(0,-3);
\draw[black,thick](0,-1)--(3,2);
\draw[white,line width = 3](-0.5,2.5)--(2,0);
\draw[black,thick](-1,3)--(2,0)--(0,-2);
\end{tikzpicture}
\end{array}
\ar[r,Rightarrow]
&
\begin{array}{c}
\begin{tikzpicture}[scale = 0.15]
\draw[black,thick](0,3)--(0,-3);
\draw[white,line width = 3](-0.5,0.5)--(0.5,-0.5);
\draw[black,thick](-2,2)--(1,-1)--(0,-2);
\draw[black,thick](1,3)--(0,2);
\end{tikzpicture}
\end{array}
\end{tikzcd}
\]
The bimodule constraint for $M_-$ is can be written down similarly. We denote the two bimodule constraints by $\theta_M^+$ and $\theta_M^-$ respectively.

\begin{defn} \label{def_local module}
Let $A$ be a commutative algebra in a braided monoidal 2-category $\SM$. A local $A$-module is a pair $(M,\gamma)$, where $M$ (actually a quadruple) is a right module over $A$, and $\gamma$ is an invertible 2-morphism depicted as follows,
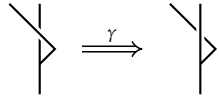
\[
\begin{tikzcd}
\begin{array}{c}
\begin{tikzpicture}[scale = 0.2]
\draw[black,thick](0,3)--(0,-3);
\draw[white,line width = 3](-0.5,1.5)--(0.5,0.5);
\draw[black,thick](-2,3)--(1,0)--(0,-1);
\end{tikzpicture}
\end{array}
\ar[r,Rightarrow,"\gamma"]
&
\begin{array}{c}
\begin{tikzpicture}[scale = 0.2]
\draw[black,thick](-2,3)--(1,0)--(0,-1);
\draw[white,line width = 3](0,0)--(0,2);
\draw[black,thick](0,3)--(0,-3);
\end{tikzpicture}
\end{array}
\end{tikzcd}
\] 
such that the following diagrams are commutative:
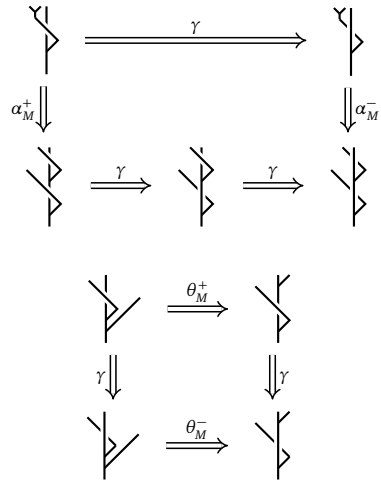
\[
\begin{tikzcd}
\begin{array}{c}
\begin{tikzpicture}[scale = 0.15]
\draw[black,thick](0,3)--(0,-3);
\draw[white,line width = 3](-0.5,1.5)--(0.5,0.5);
\draw[black,thick](-1,2)--(1,0)--(0,-1);
\draw[black,thick](-1,2.5)--(-1,2);
\draw[black,thick](-1.5,3)--(-1,2.5);
\draw[black,thick](-0.5,3)--(-1,2.5);
\end{tikzpicture}
\end{array}
\ar[d,Rightarrow,"\alpha_M^+"']
\ar[rr,Rightarrow,"\gamma"]
&&
\begin{array}{c}
\begin{tikzpicture}[scale = 0.15]
\draw[black,thick](-1,2)--(1,0)--(0,-1);
\draw[black,thick](-1,2.5)--(-1,2);
\draw[black,thick](-1.5,3)--(-1,2.5);
\draw[black,thick](-0.5,3)--(-1,2.5);
\draw[white,line width = 3](0,0)--(0,3);
\draw[black,thick](0,3)--(0,-3);
\end{tikzpicture}
\end{array}
\ar[d,Rightarrow,"\alpha_M^-"]
\\
\begin{array}{c}
\begin{tikzpicture}[scale = 0.15]
\draw[black,thick](0,4)--(0,-3);
\draw[white,line width = 3](-2,2)--(1,-1);
\draw[white,line width = 3](-1,4)--(1,2);
\draw[black,thick](-2,2)--(1,-1)--(0,-2);
\draw[black,thick](-1,4)--(1,2)--(0,1);
\end{tikzpicture}
\end{array}
\ar[r,Rightarrow,"\gamma"]
&
\begin{array}{c}
\begin{tikzpicture}[scale = 0.15]
\draw[black,thick](-2,2)--(1,-1)--(0,-2);
\draw[white,line width = 3](0,-0.5)--(0,0.5);
\draw[black,thick](0,4)--(0,-3);
\draw[white,line width = 3](-1,4)--(1,2);
\draw[black,thick](-1,4)--(1,2)--(0,1);
\end{tikzpicture}
\end{array}
\ar[r,Rightarrow,"\gamma"]
&
\begin{array}{c}
\begin{tikzpicture}[scale = 0.15]
\draw[black,thick](-2,2)--(1,-1)--(0,-2);
\draw[white,line width = 3](0,-0.5)--(0,0.5);
\draw[black,thick](-1,4)--(1,2)--(0,1);
\draw[white,line width = 3](0,2)--(0,4);
\draw[black,thick](0,4)--(0,-3);
\end{tikzpicture}
\end{array}
\end{tikzcd}
\]
\[
\begin{tikzcd}
\begin{array}{c}
\begin{tikzpicture}[scale = 0.15]
\draw[black,thick](0,3)--(0,-3);
\draw[white,line width = 3](-1.5,2.5)--(1,0);
\draw[black,thick](-1.5,2.5)--(1,0)--(0,-1);
\draw[black,thick](3,1)--(0,-2);
\end{tikzpicture}
\end{array}
\ar[r,Rightarrow,"\theta_M^+"]
\ar[d,Rightarrow,"\gamma"']
&
\begin{array}{c}
\begin{tikzpicture}[scale = 0.15]
\draw[black,thick](0,3)--(0,-3);
\draw[white,line width = 3](-0.5,0.5)--(0.5,-0.5);
\draw[black,thick](-2,2)--(1,-1)--(0,-2);
\draw[black,thick](1,3)--(0,2);
\end{tikzpicture}
\end{array}
\ar[d,Rightarrow,"\gamma"]
\\
\begin{array}{c}
\begin{tikzpicture}[scale = 0.15]
\draw[black,thick](-1.5,2.5)--(1,0);
\draw[white,line width = 3](0,0.5)--(0,1.5);
\draw[black,thick](0,3)--(0,-3);
\draw[black,thick](1,0)--(0,-1);
\draw[black,thick](3,1)--(0,-2);
\end{tikzpicture}
\end{array}
\ar[r,Rightarrow,"\theta_M^-"]
&
\begin{array}{c}
\begin{tikzpicture}[scale = 0.15]
\draw[black,thick](-2,2)--(1,-1);
\draw[white,line width = 3](0,-0.5)--(0,0.5); 
\draw[black,thick](1,-1)--(0,-2);
\draw[black,thick](1,3)--(0,2);
\draw[black,thick](0,3)--(0,-3);
\end{tikzpicture}
\end{array}
\end{tikzcd}
\]
\end{defn}

\begin{rem}
The notion of a local $A$-module can be equivalently defined as follows.

If $M$ be a right $A$-module, then $(1_M,1_{\mu_M})$ is automatically a right $A$-module 1-map. A local module over $A$ is a pair $(M,\gamma)$ where $\gamma$ is a 2-morphism depicted as
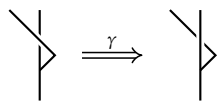
\[
\begin{tikzcd}
\begin{array}{c}
\begin{tikzpicture}[scale = 0.2]
\draw[black,thick](0,3)--(0,-3);
\draw[white,line width = 3](-0.5,1.5)--(0.5,0.5);
\draw[black,thick](-2,3)--(1,0)--(0,-1);
\end{tikzpicture}
\end{array}
\ar[r,Rightarrow,"\gamma"]
&
\begin{array}{c}
\begin{tikzpicture}[scale = 0.2]
\draw[black,thick](-2,3)--(1,0)--(0,-1);
\draw[white,line width = 3](0,0)--(0,2);
\draw[black,thick](0,3)--(0,-3);
\end{tikzpicture}
\end{array}
\end{tikzcd}
\]
such that the triple $(1_M,\gamma,1_{\mu_M})$ defines a bimodule equivalence
\[
(1_M,\gamma,1_{\mu_M}) \colon M_+ \to M_-.
\]
If we write down the compatibility relations in detail, we obtain Definition \ref{def_local module}.
\end{rem}

\begin{defn}
Let $(M,\gamma)$ and $(M^\prime,\gamma^\prime)$ be local $A$-modules. A local $A$-module 1-map is a right $A$-module 1-map $(f,\phi) \colon M \to M^\prime$ such that the following diagram is commutative.
\[
\begin{tikzcd}
\begin{array}{c}
\begin{tikzpicture}[scale = 0.12]
\draw[->-,black,thick](0,3)--(0,2);
\draw[black,thick](0,5)--(0,-5);
\draw[white,line width = 3](-4,4)--(2,-2);
\draw[black,thick](-4,4)--(2,-2);
\draw[black,thick](2,-2)--(0,-4);
\end{tikzpicture}
\end{array}
\ar[r,Rightarrow]
\ar[d,Rightarrow,"\gamma"']
&
\begin{array}{c}
\begin{tikzpicture}[scale = 0.12]
\draw[black,thick](0,5)--(0,-1);
\draw[->-,black,thick](0,1)--(0,-3);
\draw[black,thick](0,-3)--(0,-5);
\draw[white,line width = 3](-1,4)--(1.5,1.5);
\draw[black,thick](-2,5)--(3,0);
\draw[black,thick](3,0)--(0,-3);
\end{tikzpicture}
\end{array}
\ar[r,Rightarrow,"\phi"]
&
\begin{array}{c}
\begin{tikzpicture}[scale = 0.12]
\draw[black,thick](0,5)--(0,-1);
\draw[->-,black,thick](0,-1)--(0,-5);
\draw[white,line width = 3](-1,4)--(1.5,1.5);
\draw[black,thick](-2,5)--(1.5,1.5);
\draw[black,thick](1.5,1.5)--(0,0);
\end{tikzpicture}
\end{array}
\ar[d,Rightarrow,"\gamma"]
\\
\begin{array}{c}
\begin{tikzpicture}[scale = 0.12]
\draw[black,thick](-4,4)--(2,-2);
\draw[white,line width = 3](0,5)--(0,-5);
\draw[->-,black,thick](0,3)--(0,2);
\draw[black,thick](0,5)--(0,-5);
\draw[black,thick](2,-2)--(0,-4);
\end{tikzpicture}
\end{array}
\ar[r,Rightarrow]
&
\begin{array}{c}
\begin{tikzpicture}[scale = 0.12]
\draw[black,thick](-2,5)--(3,0);
\draw[white,line width = 3](0,5)--(0,-5);
\draw[black,thick](0,5)--(0,-1);
\draw[->-,black,thick](0,1)--(0,-3);
\draw[black,thick](0,-3)--(0,-5);
\draw[black,thick](3,0)--(0,-3);
\end{tikzpicture}
\end{array}
\ar[r,Rightarrow,"\phi"]
&
\begin{array}{c}
\begin{tikzpicture}[scale = 0.12]
\draw[black,thick](-2,5)--(1.5,1.5);
\draw[white,line width = 3](0,5)--(0,-5);
\draw[black,thick](0,5)--(0,-1);
\draw[->-,black,thick](0,-1)--(0,-5);
\draw[black,thick](1.5,1.5)--(0,0);
\end{tikzpicture}
\end{array}
\end{tikzcd}
\]
\end{defn}

\begin{defn}
Let $A$ be a commutative algebra in a braided monoidal 2-category $\SM$. The 2-category of local $A$-modules is defined as follows: the objects are local $A$-modules, 1-morphisms are local $A$-module maps and 2-morphisms are $A$-module 2-maps. This category is denoted by $\SM_{A}^{\loc}$
\end{defn}

Now we compute the 2-categories of the right modules of the three Lagrangian algebras $A_e, A_1$ and $A_2$. The following theorem was proved in \cite[Lemma 3.2.13]{Dec21}.

\begin{thm}
Let $\SM$ be a monoidal 2-category and $A \in \SM$ be an algebra. Recall that $\SM_A$ is the category of right $A$-modules. Then there is a free generation functor $F_A = - \otimes A \colon \SM \to \SM_A$ and a forgetful functor $U \colon \SM_A \to \SM$. The pair $(F_A,U)$ forms a 2-adjunction
\[
F_A: \SM  \leftrightarrows \SM_A: U
\] 
More explicitly, given $X \in \SM$ and $M \in \SM_A$, there is an equivalence of categories:
\[
\mathrm{hom}_{\SM_A}(X \otimes A,M) = \mathrm{hom}_{\SM}(X,M).
\]
\end{thm}

Recall that the 2-category $\TC=2\mathrm{Rep}(\mathbb{Z}_2) \boxplus 2\mathrm{Rep}(\mathbb{Z}_2)$ can be illustrated in the following graph:
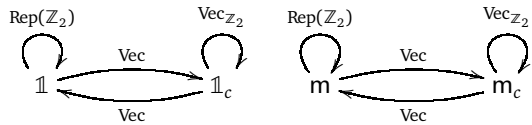
\begin{equation} 
\xymatrix{
\one \ar@(ul,ur)[]^{\mathrm{Rep}(\mathbb{Z}_2)} \ar@/^/[rr]^{\mathrm{Vec}} & & \one_c \ar@(ul,ur)[]^{\mathrm{Vec}_{\mathbb{Z}_2}} \ar@/^/[ll]^{\mathrm{Vec}}
} 
\quad
\xymatrix{
\sm \ar@(ul,ur)[]^{\mathrm{Rep}(\mathbb{Z}_2)} \ar@/^/[rr]^{\mathrm{Vec}} & & \sm_c \ar@(ul,ur)[]^{\mathrm{Vec}_{\mathbb{Z}_2}} \ar@/^/[ll]^{\mathrm{Vec}}
}   \nonumber
\end{equation}

\begin{enumerate}[(1)]

\item The 2-category $\TC_{A_e} = \TC_{\one_c}$:
\begin{itemize}
\item There are two simple objects (up to equivalence) in $\TC_{A_e}$, that is $\one_c$ itself and $\sm_c =  \sm \otimes \one_c$. The action of $\one_c$ on them is obvious.
\item $\mathrm{hom}_{\TC_{\one_c}} (\one_c,\one_c) = \mathrm{hom}_{\TC}(\one,\one_c) = \mathrm{Vec}$. \item $\mathrm{hom}_{\TC_{\one_c}} (\one_c,\sm_c) = \mathrm{hom}_{\TC}(\one,\sm_c) = 0$.
\item $\mathrm{hom}_{\TC_{\one_c}} (\sm_c,\one_c) = \mathrm{hom}_{\TC}(\sm,\one_c) = 0$.
\item $\mathrm{hom}_{\TC_{\one_c}} (\sm_c,\sm_c) = \mathrm{hom}_{\TC}(\sm,\sm_c) = \mathrm{Vec}$.
\item We conclude that $\TC_{A_e} \simeq 2 \mathrm{Vec}_{\mathbb{Z}_2}$.
\end{itemize}

\item The 2-category $\TC_{A_1}$:
\begin{itemize}
\item There are two simple objects (up to equivalence) in $\TC_{A_1}$, that is $\one \oplus \sm$ itself and $\one_c \oplus \sm_c =   \one_c \otimes (\one \oplus \sm)$.

\item $\mathrm{hom}_{\TC_{A_1}}(A_1,A_1) = \mathrm{hom}_{\TC}(\one , \one \oplus m) = \mathrm{Rep}(\mathbb{Z}_2) $.
\item $\mathrm{hom}_{\TC_{A_1}}(A_1,\one_c \otimes A_1) = \mathrm{hom}_{\TC}(\one, \one_c \oplus \sm_c) = \mathrm{Vec}$.
\item $\mathrm{hom}_{\TC_{A_1}}(\one_c \otimes  A_1,A_1) = \mathrm{hom}_{\TC}(\one_c, \one \oplus \sm) = \mathrm{Vec}$.
\item $\mathrm{hom}_{\TC_{A_1}}(\one_c \otimes A_1,\one_c \otimes A_1) = \mathrm{hom}_{\TC} (\one_c, \one_c \oplus \sm_c) = \mathrm{Rep}(\mathbb{Z}_2)$.
\item We conclude that $\TC_{A_1} \simeq 2 \mathrm{Rep}(\mathbb{Z}_2)$.
\end{itemize}

\item The 2-category $\TC_{A_2}$:
\begin{itemize}
\item There are two simple objects (up to equivalence) in $\TC_{A_2}$, that is $\one \oplus \sm$ itself and $\one_c \oplus \sm_c = (\one \oplus \sm) \otimes \one_c$.

\item These two simple modules are still free. Similar to the case of $A_1$, we conclude that $\TC_{A_2} \simeq 2 \mathrm{Rep}(\mathbb{Z}_2)$.

\end{itemize}

\end{enumerate}

It remains to show that the 2-categories of local modules of these three algebras are trivial. We have the following theorem, the proof of which is tautological. 

\begin{thm}
Let $A$ be a commutative algebra in a braided monoidal 2-category $\SM$ and $X \in \SM$ be an object. If the double braiding $c_{A,X} \circ c_{X,A}$ is not isomorphic to the identity 1-morphism, then the free right $A$-module $X \otimes A$ is not a local $A$-module.
\end{thm}

Since the double braiding of $\one_c$ and $\sm$ is not isomorphic to the identity 1-morphism, the free module $A_e \otimes \sm$ is not a local $A_e$-module. For the same reason, the free module $\one_c \otimes A_1$ is not a local $A_1$-module, and the $\one_c \otimes A_2$ is not a local $A_2$-module. We obtain the following results. 
\begin{itemize}
\item $\TC_{A_e}^{\loc} \simeq 2 \mathrm{Vec}$. This is because $\TC_{A_e}^{\loc}$ is generated by a single object $A_e = \one_c$ and the hom-category of the generator is $\mathrm{Vec}$. 
\item $\TC_{A_1}^{\loc} \simeq 2 \mathrm{Vec}$. $\TC_{A_1}^{\loc}$ is generated by a single element $A_1 = \one \oplus \sm$. The non-trivial 1-morphism $(e,e_\sm) \colon \one \oplus \sm \to \one \oplus \sm$ is not a morphism of local $A_1$-modules.
\item $\TC_{A_2}^{\loc} \simeq 2 \mathrm{Vec}$. $\TC_{A_2}^{\loc}$ is generated by a single element $A_2 = \one \oplus \sm$. The non-trivial 1-morphism $(e,e_\sm) \colon \one \oplus \sm \to \one \oplus \sm$ is not a morphism of local $A_2$-modules.
\end{itemize}

\begin{rem} \label{rem:more_Lagrangian}
The Lagrangian algebras in $2\mathrm{Vec}$ are non-degenerate braided fusion 1-categories \cite{DN21,JFR23}. Physically, the boundary of the trivial 3+1D topological order induced by the condensation of a non-degenerate braided fusion 1-category $B \in 2\vect$ is precisely the anomaly-free 2+1D topological order associated to $B$.

Given a Lagrangian algebra $A \in \TC$, we obtain a new Lagrangian algebra $A \boxtimes B \in \TC \boxtimes 2\mathrm{Vec} \simeq \TC$ for a non-degenerate braided fusion 1-category $B$. Physically, the boundary corresponding to $A \boxtimes B$ is the stacking of the boundary corresponding to $A$ and the anomaly-free 2+1D topological order corresponding to $B$.
\end{rem}


\section{The 3d toric code -- Lattice model} \label{sec:3dlattice}

The 3+1D Dijkgraaf-Witten model is defined on a triangulation $\Gamma$ of an orientable 3-manifold with vertices assigned by ordered labels and it is independent of this assignment as long as the relative order is kept during the calculation. Each edge $<a,b>$ with adjacent vertices $a$ and $b$ is assigned by a group element of $\mathbb{Z}_2$, which can also be viewed as putting a space of spin-1/2, $\mathcal{H}_e=\mathbb{C}^2$, on each edge. The spin pointing up and down corresponds to $\mathbb{Z}_2$ elements $1$ and $\sm$ respectively. Hence the total Hilbert space consists of all possible configurations of group elements on the edges, i.e., $\mathcal{H}_{tot}=\otimes_e\mathcal{H}_e$.

\begin{figure}[htbp!]
\centering
\begin{tikzpicture}
\draw (0,0) -- (0,3) -- (3,3) -- (3,0) -- (0,0);
\draw (0,3) -- (1,4) -- (4,4) -- (3,3);
\draw (4,4) -- (4,1) -- (3,0);
\draw[dashed] (1,4) -- (1,1) -- (0,0);
\draw[dashed] (1,1) -- (4,1);
\draw[red] (0,3) -- (3,0) -- (4,4) -- (0,3);
\draw[dashed, red] (0,0) -- (1,4) -- (4,1) -- (0,0);
\draw[dashed, red] (1,4) -- (3,0);
\end{tikzpicture}
\caption{Trangulation of the cube}\label{tcube}
\end{figure}
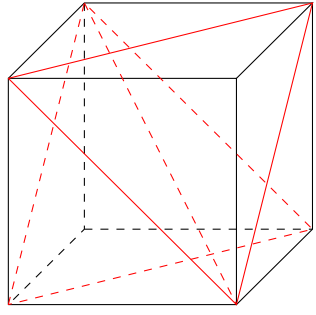

For simplicity we first take the 3d square lattice and then trangulate each cube as Figure \ref{tcube}.

\begin{figure}[htbp!]
\centering
\begin{tikzpicture}[scale=0.7]
\draw (0,0) -- (0,3) -- (3,3) -- (3,0) -- (0,0);
\draw (0,3) -- (1,4) -- (4,4) -- (3,3);
\draw (4,4) -- (4,1) -- (3,0);
\draw[dashed] (1,4) -- (1,1) -- (0,0);
\draw[dashed] (1,1) -- (4,1);
\draw[dashed] (0,0) -- (1,4) -- (4,1) -- (0,0);
\draw[dashed] (1,4) -- (3,0);
\draw (4,4) -- (4,7);
\draw (4,4) -- (7,4);
\draw (4,4) -- (5,5);
\draw (4,4) -- (1,7);
\draw (4,4) -- (7,1);
\draw (4,4) -- (6,0);
\draw (4,4) -- (5,8);
\draw (4,4) -- (2,8);
\draw (4,4) -- (8,5);
\draw [dashed, blue] (0,3) -- (0,6) -- (3,6) -- (3,3);
\draw [dashed, blue] (0,6) -- (1,7) -- (1,4);
\draw [dashed, blue] (1,7) -- (4,7) -- (3,6);
\draw [dashed, blue] (1,7) -- (2,8) -- (5,8) -- (4,7);
\draw [dashed, blue] (5,8) -- (5,5) -- (8,5) -- (8,8) -- (5,8);
\draw [dashed, blue] (8,8) -- (7,7) -- (4,7);
\draw [dashed, blue] (7,7) -- (7,4) -- (8,5);
\draw [dashed, blue] (2,8) -- (2,2) -- (8,2) -- (8,5);
\draw [dashed, blue] (5,5) -- (2,5) -- (1,4);
\draw [dashed, blue] (5,5) -- (5,2) -- (4,1);
\draw [dashed, blue] (2,2) -- (1,1);
\draw [dashed, blue] (3,6) -- (6,6) -- (7,7);
\draw [dashed, blue] (6,6) -- (6,0) -- (3,0);
\draw [dashed, blue] (3,3) -- (6,3) -- (7,4);
\draw [dashed, blue] (7,4) -- (7,1) -- (4,1);
\draw [dashed, blue] (6,0) -- (8,2);
\draw[fill=red, opacity=0.5] (0,3) -- (3,0) -- (4,4) -- (0,3);
\node[below] at (3,0) {2};
\node[below] at (6,0) {3};
\node[below] at (4,1) {5};
\node[below] at (7,1) {6};
\node[above] at (0,3) {10};
\node[above] at (3,3) {11};
\node[above] at (1,4) {13};
\node[above] at (4,4) {14};
\node[above] at (7,4) {15};
\node[above] at (5,5) {17};
\node[above] at (8,5) {18};
\node[above] at (1,7) {22};
\node[above] at (4,7) {23};
\node[above] at (2,8) {25};
\node[above] at (5,8) {26};
\draw [blue, fill=blue] (4,4) circle [radius=0.08cm];
\end{tikzpicture}
\caption{Local operators, $A_{14}$ and $B'_{[2,10,14]}$, in the bulk of 3+1D toric code}\label{lotc}
\end{figure}
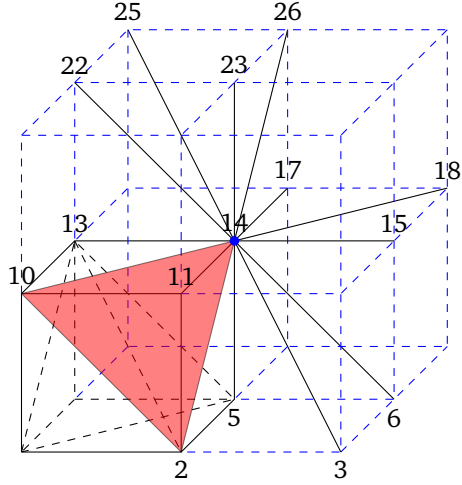

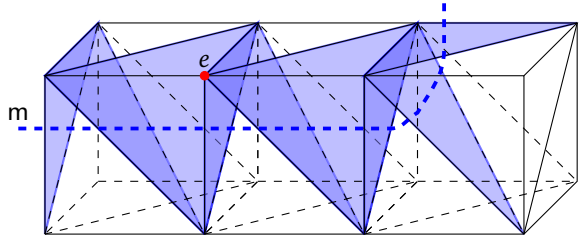
\begin{figure}[htbp!]
\centering
\begin{tikzpicture}[scale=0.7]
\draw[thick, blue, fill=blue!50!white, opacity=0.5] (0,0) -- (0,3) -- (1,4) -- (0,0);
\draw[thick, blue, fill=blue!50!white, opacity=0.5] (3,0) -- (0,3) -- (1,4) -- (3,0);
\draw[thick, blue, fill=blue!50!white, opacity=0.5] (3,0) -- (0,3) -- (4,4) -- (3,0);
\draw (0,0) -- (0,3) -- (3,3) -- (3,0) -- (0,0);
\draw (0,3) -- (1,4) -- (4,4) -- (3,3);
\draw[dashed] (4,4) -- (4,1) -- (3,0);
\draw[dashed] (1,4) -- (1,1) -- (0,0);
\draw[dashed] (1,1) -- (4,1);
\draw (4,4) -- (0,3) -- (3,0);
\draw[dashed] (4,4) -- (3,0);
\draw[dashed] (0,0) -- (1,4) -- (4,1) -- (0,0);
\draw[dashed] (1,4) -- (3,0);

\begin{scope}[xshift=3cm,yshift=0cm]{
\draw[thick, blue, fill=blue!50!white, opacity=0.5] (0,0) -- (0,3) -- (1,4) -- (0,0);
\draw[thick, blue, fill=blue!50!white, opacity=0.5] (3,0) -- (0,3) -- (1,4) -- (3,0);
\draw[thick, blue, fill=blue!50!white, opacity=0.5] (3,0) -- (0,3) -- (4,4) -- (3,0);
\draw (0,0) -- (0,3) -- (3,3) -- (3,0) -- (0,0);
\draw (0,3) -- (1,4) -- (4,4) -- (3,3);
\draw[dashed] (4,4) -- (4,1) -- (3,0);
\draw[dashed] (1,4) -- (1,1) -- (0,0);
\draw[dashed] (1,1) -- (4,1);
\draw (4,4) -- (0,3) -- (3,0);
\draw[dashed] (4,4) -- (3,0);
\draw[dashed] (0,0) -- (1,4) -- (4,1) -- (0,0);
\draw[dashed] (1,4) -- (3,0);
}
\end{scope}

\begin{scope}[xshift=6cm,yshift=0cm]{
\draw[thick, blue, fill=blue!50!white, opacity=0.5] (0,0) -- (0,3) -- (1,4) -- (0,0);
\draw[thick, blue, fill=blue!50!white, opacity=0.5] (4,4) -- (0,3) -- (1,4) -- (4,4);
\draw[thick, blue, fill=blue!50!white, opacity=0.5] (3,0) -- (0,3) -- (1,4) -- (3,0);
\draw (0,0) -- (0,3) -- (3,3) -- (3,0) -- (0,0);
\draw (0,3) -- (1,4) -- (4,4) -- (3,3);
\draw (4,4) -- (4,1) -- (3,0);
\draw[dashed] (1,4) -- (1,1) -- (0,0);
\draw[dashed] (1,1) -- (4,1);
\draw (4,4) -- (0,3) -- (3,0);
\draw (4,4) -- (3,0);
\draw[dashed] (0,0) -- (1,4) -- (4,1) -- (0,0);
\draw[dashed] (1,4) -- (3,0);
}
\end{scope}
\draw[line width=0.05cm, dashed, blue] (-0.5,2) -- (6.5,2) to [out=0, in =210] (7,2.3) to [out=60, in=270] (7.5,3.5) -- (7.5, 4.4);
\draw[red, fill=red] (3,3) circle [radius=0.08cm];
\node[above] at (3,3) {$e$};
\node[above] at (-0.5,2) {$\sm$};
\end{tikzpicture}
\caption{The $e$-particle (red dot) and $\sm$-string (blue plaquettes) in 3+1D toric code}\label{eandm}
\end{figure}

In the bulk, for every vertex $i$ and plaquette ${[i,j,k]}$ we define a vertex operator $A_i^g$ and a plaquette operator $B_{[i,j,k]}$ acting on every adjacent edges respectively. They are defined as
\begin{equation}
\begin{aligned}
&\begin{tikzpicture}[scale=0.7]
\node at (-1,0) {\scriptsize{$A_i^g$}};
\draw (-0.25,0.866) -- (-0.25,-0.866);
\draw (2.25,0.866) -- (2.5,0) -- (2.25,-0.866);
\begin{scope}[xshift=0cm, yshift=0.2cm]{
\draw (2,-0.866) -- (0.5,-1.5) -- (0,-0.866);
\draw (1,0.866) -- (0,-0.866);
\draw (1,0.866) -- (2,-0.866);
\draw (1,0.866) -- (0.5,-1.5);
\draw[dashed] (0,-0.866) -- (2,-0.866);
\draw[dashed] (0.9,-0.5) -- (2,-0.866);
\draw[dashed] (0.9,-0.5) -- (0.5,-1.5);
\draw[dashed] (0.9,-0.5) -- (0,-0.866);
\draw[dashed] (0.9,-0.5) -- (1,0.866);
\node[right] at (1,0.866) {$l$};
\node[below] at (-0.1,-0.866) {$h$};
\node[below] at (2,-0.866) {$k$};
\node[right] at (0.65,-1.6) {$j$};
\node[right] at (0.9,-0.3) {$i$};
\draw[blue, fill=blue, opacity=0.5] (0.9,-0.5) circle [radius=0.05cm];
}
\end{scope}
\begin{scope}[xshift=2cm, yshift=-3.5cm]{
\node at (0,1.5) {\scriptsize{$=\frac{{\omega(<h,i'>,<i',i>,<i,j>,<j,l>)\omega(<i',i>,<i,j>,<i,k>,<k,l>)}}{{\omega(<h,i'>,<i',i>,<i,j>,<j,k>)\omega(<h,i'>,<i',i>,<i,k>,<k,l>)}}$}};
\node at (-1.5,0) {\scriptsize{$\times\delta_{<i',i>,g}$}};
\draw (-0.25,0.866) -- (-0.25,-0.866);
\draw (2.25,0.866) -- (2.5,0) -- (2.25,-0.866);
\begin{scope}[xshift=0cm, yshift=0.2cm]{
\draw (2,-0.866) -- (0.5,-1.5) -- (0,-0.866);
\draw (1,0.866) -- (0,-0.866);
\draw (1,0.866) -- (2,-0.866);
\draw (1,0.866) -- (0.5,-1.5);
\draw[dashed] (0,-0.866) -- (2,-0.866);
\draw[dashed] (0,-0.866) -- (2,-0.866);
\draw[dashed] (0.9,-0.5) -- (2,-0.866);
\draw[dashed] (0.9,-0.5) -- (0.5,-1.5);
\draw[dashed] (0.9,-0.5) -- (0,-0.866);
\draw[dashed] (0.9,-0.5) -- (1,0.866);
\node[right] at (1,0.866) {$l$};
\node[below] at (-0.1,-0.866) {$h$};
\node[below] at (2,-0.866) {$k$};
\node[right] at (0.65,-1.6) {$j$};
\node[right] at (0.9,-0.3) {$i'$};
\draw[blue, fill=blue, opacity=0.5] (0.9,-0.5) circle [radius=0.05cm];
}
\end{scope}
}
\end{scope}
\end{tikzpicture}\\
&\begin{tikzpicture}[scale=0.7]
\node at (-1,0) {\scriptsize{$B_{[i,j,k]}$}};
\draw (-0.25,0.866) -- (-0.25,-0.866);
\draw (0,-0.866) -- (2,-0.866) -- (1,0.866) -- (0,-0.866);
\draw (2.25,0.866) -- (2.5,0) -- (2.25,-0.866);
\node[right] at (1,0.866) {$i$};
\node[below] at (0,-0.866) {$j$};
\node[below] at (2,-0.866) {$k$};
\begin{scope}[xshift=7cm,yshift=0cm]{
\node at (-2.4,0) {\scriptsize{$=\delta_{1,<i,j>\cdot<j,k>\cdot<k,i>}$}};
\draw (-0.25,0.866) -- (-0.25,-0.866);
\draw (0,-0.866) -- (2,-0.866) -- (1,0.866) -- (0,-0.866);
\draw (2.25,0.866) -- (2.5,0) -- (2.25,-0.866);
\node[right] at (1,0.866) {$i$};
\node[below] at (0,-0.866) {$j$};
\node[below] at (2,-0.866) {$k$};
}
\end{scope}
\end{tikzpicture}
\end{aligned}
\end{equation}
where $h<i<j<k<l$ and $<i',j>=<i',i>\cdot<i,j>$. When the 4-cocycle $\omega$ is trivial, one can readily see that $A_i^1$ is trivial and $A_i^m$ is exactly the vertex operator $A_i=\prod_{<i,j>}\sigma_x^{<i,j>}$ of 3+1D toric code meanwhile $B_{[i,j,k]}$ is also related with the plaquette operator $B'_{[i,j,k]}=\sigma_z^{<i,j>}\sigma_z^{<i,k>}\sigma_z^{<j,k>}$ by $B_{[i,j,k]}=(B'_{[i,j,k]}+1)/2$. The Hamiltonian of the 3+1D Dijkgraaf-Witten model is given by
\begin{equation}
H_{DW}=-\frac{{1}}{{\left|G\right|}}\sum_i\sum_{g\in G}A_i^g-\sum_{[i,j,k]}B_{[i,j,k]}
\end{equation}
when $G=\mathbb{Z}_2$ with trivial 4-cocycle, it reduces to
\begin{equation}
H_{DW}=-\frac{{1}}{{2}}(\sum_i(1+A_i)+\sum_{[i,j,k]}(1+B'_{[i,j,k]}))
\end{equation}
which is exactly the familiar 3+1D toric code. The set of operators $\{A_i,\ B'_{[i,j,k]}\}$ is much convenient to visualize on the lattice, for example in Figure \ref{lotc}
\begin{equation}
\begin{aligned}
&A_{14}=\sigma_x^{<14,2>}\sigma_x^{<14,3>}\sigma_x^{<14,5>}\sigma_x^{<14,6>}\sigma_x^{<14,10>}\sigma_x^{<14,11>}\\
&\times\sigma_x^{<14,13>}\sigma_x^{<14,15>}\sigma_x^{<14,17>}\sigma_x^{<14,18>}\sigma_x^{<14,22>}\sigma_x^{<14,23>}\\
&\times\sigma_x^{<14,23>}\sigma_x^{<14,25>}\sigma_x^{<14,26>}\\
&B'_{[2,10,14]}=\sigma_z^{<2,10>}\sigma_z^{<2,14>}\sigma_z^{<10,14>}
\end{aligned}
\end{equation}\\
All the vertex operators $A_i$ and plaquette operators $B'_{[i,j,k]}$ commute with each other and have eigenvalues $\pm1$. Hence the total Hilbert can be decomposed into the common eigenspaces of these operators. Then the subspace of ground state corresponds to the common eigenspace which all the $A_i$ and $B'_{[i,j,k]}$ have +1 eigenvalue.\\
The topological excitations in this model have already been worked out in \cite{Kong_2020}. The well known $e$-particle and $\sm$-string are shown in Figure \ref{eandm}. Similar to the 2+1D toric code the state with an $e$-particle at vertex $v$ is the state which all the $A_i$ and $B'_{[i,j,k]}$ have eigenvalue +1 except $A_v$ which has eigenvalue -1. A pair of $e$-particles can be created on the end points of a string operator $\prod_{<i,j>\in P}\sigma_z^{<i,j>}$, where P is a continuous path on the edges. However the $\sm$-string is somewhat not similar to the $\sm$-particle in the 2+1D case which is defined by $B'_{[i,j,k]}=-1$ at plaquette $[i,j,k]$. Due to the constraint $\prod_{[i,j,k]\in t}B'_{[i,j,k]}=1$, where $t$ is any tetrahedron, there must be an even number of plaquettes where $B'_{[i,j,k]}=-1$ in a tetrahedron. Hence it forms a string which can not be broken in the bulk. The $\sm$-string can be created on the boundary of an membrane operator, $\prod_{<i,j>\in M}\sigma_x^{<i,j>}$, where M is a membrane and $<i,j>$ are the edges which the membrane cut through. The $e$-particle and $\sm$-string are both self-dual. Namely
\begin{equation}
\sm\otimes \sm=\one\quad e\circ e=1_{\one}
\end{equation}
where $\one$ and $1_{\one}$ are the trivial string and the trivial 0d excitation on the trivial string respectively.

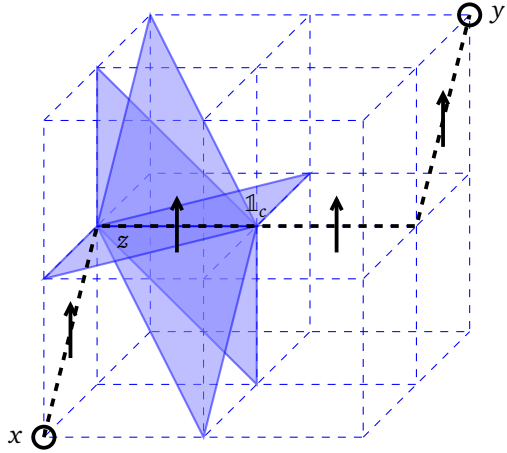
\begin{figure}[htbp!]
\centering
\begin{tikzpicture}[scale=0.7]
\draw [dashed, blue] (0,0) -- (0,3) -- (3,3) -- (3,0) -- (0,0);
\draw [dashed, blue] (0,3) -- (1,4) -- (4,4) -- (3,3);
\draw [dashed, blue] (4,4) -- (4,1) -- (3,0);
\draw [dashed, blue] (1,4) -- (1,1) -- (0,0);
\draw [dashed, blue] (1,1) -- (4,1);
\draw [dashed, blue] (4,4) -- (4,7);
\draw [dashed, blue] (5,5) -- (4,4) -- (7,4);
\draw [dashed, blue] (0,3) -- (0,6) -- (3,6) -- (3,3);
\draw [dashed, blue] (0,6) -- (1,7) -- (1,4);
\draw [dashed, blue] (1,7) -- (4,7) -- (3,6);
\draw [dashed, blue] (1,7) -- (2,8) -- (5,8) -- (4,7);
\draw [dashed, blue] (5,8) -- (5,5) -- (8,5) -- (8,8) -- (5,8);
\draw [dashed, blue] (8,8) -- (7,7) -- (4,7);
\draw [dashed, blue] (7,7) -- (7,4) -- (8,5);
\draw [dashed, blue] (2,8) -- (2,2) -- (8,2) -- (8,5);
\draw [dashed, blue] (5,5) -- (2,5) -- (1,4);
\draw [dashed, blue] (5,5) -- (5,2) -- (4,1);
\draw [dashed, blue] (2,2) -- (1,1);
\draw [dashed, blue] (3,6) -- (6,6) -- (7,7);
\draw [dashed, blue] (6,6) -- (6,0) -- (3,0);
\draw [dashed, blue] (3,3) -- (6,3) -- (7,4);
\draw [dashed, blue] (7,4) -- (7,1) -- (4,1);
\draw [dashed, blue] (6,0) -- (8,2);
\draw [thick, blue, fill=blue!50!white, opacity=0.5] (1,4) -- (4,4) -- (4,1) -- (1,4);
\draw [thick, blue, fill=blue!50!white, opacity=0.5] (1,4) -- (4,4) -- (3,0) -- (1,4);
\draw [thick, blue, fill=blue!50!white, opacity=0.5] (1,4) -- (4,4) -- (0,3) -- (1,4);
\draw [thick, blue, fill=blue!50!white, opacity=0.5] (1,4) -- (4,4) -- (1,7) -- (1,4);
\draw [thick, blue, fill=blue!50!white, opacity=0.5] (1,4) -- (4,4) -- (2,8) -- (1,4);
\draw [thick, blue, fill=blue!50!white, opacity=0.5] (1,4) -- (4,4) -- (5,5) -- (1,4);
\draw [dashed, line width=0.05cm] (0,0) -- (1,4) -- (7,4) -- (8,8);
\draw [line width=0.05cm] (0,0) circle [radius=0.2cm];
\draw [line width=0.05cm] (8,8) circle [radius=0.2cm];
\node [above] at (4,4) {$\one_c$};
\node [left] at (-0.2,0) {$x$};
\node [right] at (8.2,8) {$y$};
\node [below] at (1.5,4) {$z$};
\begin{scope}[xshift=1cm, yshift=1cm]{
\draw[line width=0.05cm] (4.5,2.5) -- (4.5,3.5);
\draw[line width=0.05cm] (4.4,3.3) -- (4.5,3.5) -- (4.6,3.3);
}
\end{scope}
\begin{scope}[xshift=-2cm, yshift=1cm]{
\draw[line width=0.05cm] (4.5,2.5) -- (4.5,3.5);
\draw[line width=0.05cm] (4.4,3.3) -- (4.5,3.5) -- (4.6,3.3);
}
\end{scope}
\begin{scope}[xshift=3cm, yshift=3cm]{
\draw[line width=0.05cm] (4.5,2.5) -- (4.5,3.5);
\draw[line width=0.05cm] (4.4,3.3) -- (4.5,3.5) -- (4.6,3.3);
}
\end{scope}
\begin{scope}[xshift=-4cm, yshift=-1cm]{
\draw[line width=0.05cm] (4.5,2.5) -- (4.5,3.5);
\draw[line width=0.05cm] (4.4,3.3) -- (4.5,3.5) -- (4.6,3.3);
}
\end{scope}
\end{tikzpicture}
\caption{The condensation descendant $\one_c$-string which is denoted in black dashed line. The 0d excitations between $\one_c$ and $\one$ strings: $x$, $y$ are denoted as circles and the 0d excitation between $\one_c$ and $\one_c$ strings: $z$ is an $\sm$-string winding around the $\one_c$-string}\label{1_c}
\end{figure}

Besides these elementary excitations, there are condensation descendants, $\one_c$ and $\sm_c$ strings. The local space on the edges along $\one_c$ is $\mathbb{C}$ or equivalently the spins on these edges are all fixed to be pointing up. The $\one_c$-string along a path $P$ is created by adding to the Hamiltonian the $A_i/2$ and projection operators, $(\sigma_z+1)/2$, on the vertices and edges respectively along $P$. The $e$-particle condenses on the $\one_c$-string. One can check that if an $e$-particle moves onto the $\one_c$-string, it will disappear. Hence it can be viewed as a condensation of $1_{\one}\oplus e$ on the trivial string $\one$, a condensation descendant of $\one$. $\sm_c$-string is the fusion of $\sm$ and $\one_c$ strings. It can also be viewed as the condensation descendant, $1_{\one}\oplus e$, of the $\sm$-string. Unlike the $\sm$-string, the $\one_c$-string can have end points, which are the 0d defects between $\one_c$ and $\one$. The 0d defect from $\one$ to $\one_c$ is denoted as $x \colon \one \rightarrow \one_c$ and the one from $\one_c$ to $\one$ is denoted as $y \colon \one_c\rightarrow \one$. There is also one non-trivial 0d defect on the $\one_c$-string, $z$, which is a $\sm$-string winding around the $\one_c$-string as depicted as in Figure \ref{1_c}.

The fusion rules are given by \cite{Kong_2020}
\begin{equation}
\begin{aligned}
&\one_c\otimes\one_c=\one_c\oplus\one_c\quad\one_c\otimes \sm = \sm_c\quad z\circ z=1_{\one_c}\\
&x\circ y=1_{\one}\oplus e\quad y\circ x=1_{\one_c}\oplus z
\end{aligned}
\end{equation}
where $1_{\one_c}$ is the trivial 0d defect on $\one_c$. The other fusion rules can be deduced from these.

\section{Lattice realization of the twisted smooth boundary}  \label{sec:lattice_construction}

In this section, we would like to give the complete details of the lattice realization of the three boundary conditions. 

\subsection{The (twisted) smooth boundary condition}

The boundary Hamiltonian was given in \cite{Hu_2013} and in addition to reviewing that, we will bring together results of string operators of the Levin-Wen model constructed before \cite{Levin_2005} to obtain a membrane operator in the 3d model that could end at the boundary.

\begin{figure}[htbp!]
\centering
\begin{tikzpicture}
\draw[red, fill=red, opacity=0.5] (2,4) -- (3,4) -- (2,3) -- (2,4);
\foreach \x in {1,2,3,4,5}
\draw (\x,0.5) -- (\x,5.5);
\foreach \y in {1,2,3,4,5}
\draw (0.5,\y) -- (5.5,\y);
\foreach \w in {1,2,3,4}
\foreach \v in {1,2,3,4}
\draw (\w+1,\v+1) -- (\w,\v);
\foreach \z in {1,2,3,4,5}
\node[left] at (\z,5) {$l_{\z}$};
\foreach \z in {6,7,8,9,10}
\node[left] at (\z-5,4) {$l_{\z}$};
\foreach \z in {11,12,13,14,15}
\node[left] at (\z-10,3) {$l_{\z}$};
\foreach \z in {16,17,18,19,20}
\node[left] at (\z-15,2) {$l_{\z}$};
\foreach \z in {21,22,23,24,25}
\node[left] at (\z-20,1) {$l_{\z}$};
\draw[blue, fill=blue, opacity=0.5] (3,3) circle [radius=0.08cm];
\end{tikzpicture}
\caption{The twisted smooth boundary (hiding the bulk edges) and the local operators, $\tilde A_{l_{13}}$ and $\tilde B_{[l_{7},l_{8},l_{12}]}$, on the boundary}
\label{tsbdy}
\end{figure}
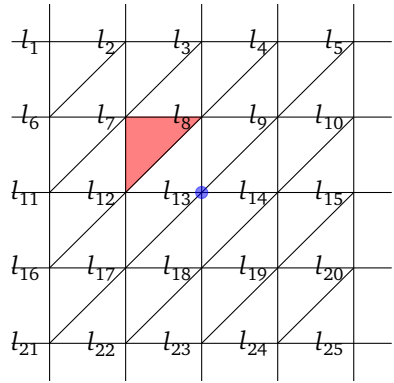

Now we can move on to the twisted smooth boundary which is depicted in Figure \ref{tsbdy}, where we have hidden all the bulk edges. The labels of boundary vertices are chosen to get bigger when we move from right to left and top to bottom on the lattice, e.g. in Figure \ref{tsbdy} $l_1<l_2<...<l_5<l_6<...l_{11}<...<l_{25}$. 

Similarly, for every vertex and plaquette on the boundary we define an operator. The plaquette operators $\tilde{B}_{[i,j,k]}$ are the same as what is in the bulk, $B'_{[i,j,k]}$, while the vertex operators $\tilde{A}_i$ is similar to $A_i^m$ twisted with a 3-cocycle, as depicted in Figure \ref{tsbdy}.
\begin{equation}
\begin{aligned}
&\tilde A_{l_{13}}=\sigma_x^{<l_8,l_{13}>}\sigma_x^{<l_9,l_{13}>}\sigma_x^{<l_{12},l_{13}>}\sigma_x^{<l_{13},l_{14}>}\sigma_x^{<l_{13},l_{17}>}\\
&\times\sigma_x^{<l_{13},l_{18}>}\prod_B\sigma_x^{<l_{13},l_B>}\\
&\times\alpha(<l_8,l_{12}>,<l_{12},l_{13}>\cdot g, g)\\
&\times\alpha(<l_{12},l_{13}>\cdot g,g,<l_{13},l_{17}>)\\
&\times\alpha(g,<l_{13},l_{17}>,<l_{17},l_{18}>)\\
&\times\alpha(<l_8,l_9>,<l_9,l_{13}>\cdot g,g)\\
&\times\alpha(<l_9,l_{13}>\cdot g,g,<l_{13},l_{14}>)\\
&\times\alpha(g,<l_{13},l_{14}>,<l_{14},l_{18}>)\\
&\tilde B_{[l_7,l_8,l_{12}]}=\sigma_z^{<l_7,l_8>}\sigma_z^{<l_7,l_{12}>}\sigma_z^{<l_8,l_{12}>}
\end{aligned}
\end{equation}
where $<l_{13},l_B>$ labels the adjacent bulk edges and $\alpha$ is the non-trivial 3-cocycle of $\mathbb{Z}_2=\{1,g\}$. The only non-trivial component of the twisted 3-cocycle is given by $\alpha(g,g,g) = -1$ . 
Notice that $\alpha^{-1}=\alpha$, hence we do not distinguish them in the above formula. Due to the labeling we choose all the $A_i$ to have the same form. Then the boundary Hamiltonian is defined as
\begin{equation}
H_{bdy}=-\frac{{1}}{{2}}(\sum_i(1+\tilde{A}_i)+\sum_{[i,j,k]}(1+\tilde{B}_{[i,j,k]}))
\end{equation}
All the $\tilde{A}_i$ and $\tilde{B}_{[i,j,k]}$ commute with each other and have eigenvalue $\pm1$. they also commute with all the $A_i$ and $B'_{[i,j,k]}$ in the bulk. Hence the ground state subspace is the common eigenspace of all the $\{A_i,\ B'_{[i,j,k]},\ \tilde{A}_i,\ \tilde{B}_{[i,j,k]}\}$ with +1 eigenvalue.\\

In the case of the ordinary smooth boundary, it corresponds to a trivial 3-cocycle $\alpha$, where it takes value unity when non-vanishing. 

When $\alpha$ corresponds to the non-trivial component of $H^3(\mathbb{Z}_2, U(1))$, the boundary Hamiltonian this is exactly the 2+1D $\mathbb{Z}_2$ twisted quantum double model, which is in fact the well-known double semion model\cite{Levin_2005, Hu_2013}. \\

Let's take a detour to look into this 2+1D model. In the double semion model there are only four simple string operators defined on a loop which commute with the Hamiltonian\cite{Levin_2005}. They correspond to four simple objects of the double semion model, $\{\one, s, \bar{s}, s\bar{s}\}$, respectively.

\begin{figure}[htbp!]
\centering
\begin{tikzpicture}
\foreach \x in {4,5,6,7,8,9,10}
\draw (\x,0.5) -- (\x,6.5);
\foreach \y in {1,2,3,4,5,6}
\draw (3.5,\y) -- (10.5,\y);
\foreach \w in {4,5,6,7,8,9}
\foreach \v in {1,2,3,4,5}
\draw (\w+1,\v+1) -- (\w,\v);
\draw [fill=orange!50!white] (5,1) -- (6,2) -- (5,2) -- (5,1);
\draw [fill=orange!50!white] (5,1) -- (6,2) -- (6,1) -- (5,1);
\draw [fill=orange!50!white] (5,2) -- (6,3) -- (5,3) -- (5,2);
\draw [fill=orange!50!white] (5,3) -- (6,4) -- (5,4) -- (5,3);
\draw [fill=orange!50!white] (6,5) -- (6,4) -- (5,4) -- (6,5);
\draw [fill=orange!50!white] (7,6) -- (7,5) -- (6,5) -- (7,6);
\draw [fill=orange!50!white] (7,6) -- (7,5) -- (8,6) -- (7,6);
\draw [fill=orange!50!white] (8,6) -- (8,5) -- (9,6) -- (8,6);
\draw [fill=orange!50!white] (9,5) -- (8,5) -- (9,6) -- (9,5);
\draw [fill=orange!50!white] (9,5) -- (8,5) -- (8,4) -- (9,5);
\draw [fill=orange!50!white] (8,4) -- (7,4) -- (7,3) -- (8,4);
\draw [fill=orange!50!white] (8,4) -- (7,4) -- (7,3) -- (8,4);
\draw [fill=orange!50!white] (6,2) -- (7,2) -- (7,3) -- (6,2);
\draw [fill=orange!50!white] (6,2) -- (7,2) -- (6,1) -- (6,2);
\draw [fill=yellow!50!white] (5,2) -- (6,2) -- (6,3) -- (5,2);
\draw [fill=yellow!50!white] (5,3) -- (6,3) -- (6,4) -- (5,3);
\draw [fill=yellow!50!white] (7,5) -- (6,5) -- (6,4) -- (7,5);
\draw [fill=yellow!50!white] (7,5) -- (8,5) -- (8,6) -- (7,5);
\draw [fill=yellow!50!white] (7,4) -- (8,4) -- (8,5) -- (7,4);
\draw [fill=yellow!50!white] (6,3) -- (7,3) -- (7,4) -- (6,3);
\draw [fill=yellow!50!white] (6,3) -- (7,3) -- (6,2) -- (6,3);
\draw [red, line width=0.05cm] (5.5,1.7) -- (5.5,4.2) to [out=90, in=180] (5.8,4.5) -- (6.2,4.5) to [out=0, in=270] (6.5,4.8) -- (6.5,5.2) to [out=90, in=180] (6.8,5.5) -- (8.2,5.5) to [out=0, in=90] (8.5,5.2) -- (8.5,4.8) to [out=270, in=0] (8.2,4.5) -- (7.8, 4.5) to [out=180, in=90] (7.5,4.2) -- (7.5,3.8) to [out=270, in=0] (7.2,3.5) -- (6.8,3.5) to [out=180, in=90] (6.5,3.2) -- (6.5,1.8) to [out=270, in=0] (6.2,1.5) -- (5.8,1.5) to [out=180, in=270] (5.5,1.7);
\draw [red, fill=red] (5.4,3.2) -- (5.5,3.4) -- (5.6,3.2) -- (5.4,3.2);
\draw [red, fill=red] (6.4,3.2) -- (6.5,3) -- (6.6,3.2) -- (6.4,3.2);
\draw [orange] (5,1) -- (5,4) -- (7,6) -- (9,6) -- (9,5) -- (7,3) -- (7,2) -- (6,1) -- (5,1);
\draw [yellow] (6,3) -- (6,4) -- (7,5) -- (8,5) -- (6,3);
\draw [line width=0.05cm](4.3,3.5) -- (4.9,3.5);
\draw [fill] (4.7,3.6) -- (4.7,3.4)-- (4.9,3.5) -- (4.7,3.6);
\node [above] at (4.3,3.5) {\footnotesize{$L-edge$}};
\draw [line width=0.05cm] (6.85,3.25) -- (7.45,3.25);
\draw [fill] (6.9,3.35) -- (6.9,3.15) -- (6.7,3.25) -- (6.9,3.35);
\node [right] at (7.5,3.25) {\footnotesize{$R-plaquette$}};
\draw [line width=0.05cm] (6.6,2.25) -- (7.2,2.25);
\draw [fill] (6.75,2.35) -- (6.75,2.15) -- (6.55,2.25) -- (6.75,2.35);
\node [right] at (7.2,2.25) {\footnotesize{$loop\ \tilde{P}$}};
\node [above] at (6.1,3) {\footnotesize{$l$}};
\node [above] at (6.85,3.95) {\footnotesize{$\sm$}};
\node [below] at (7.15,3.05) {\footnotesize{$n$}};
\node [below] at (8.15,4.05) {\footnotesize{$o$}};
\end{tikzpicture}
\caption{The string operator $W(\tilde P)$. The path $\tilde P$ is the red line. The edges and plaquettes $\tilde P$ pass through are in the path $\tilde P$. The edges of these plaquettes on the left (right) side of $\tilde P$ are $L(R)-edges$. The plaquette in the path with $L(R)-edge$ are $L(R)-plaquette$.}\label{loop}
\end{figure}
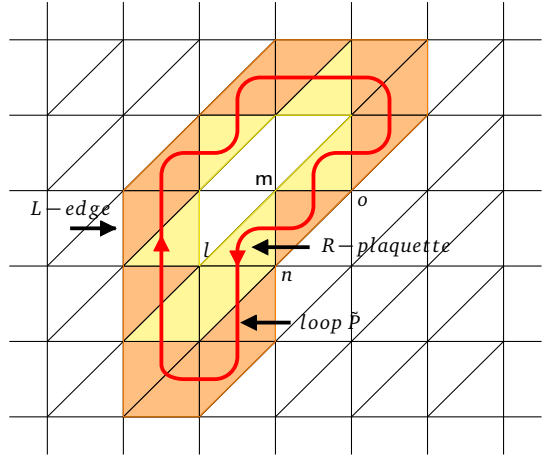

As indicated in Figure \ref{loop}, the loop $\tilde{P}$ is directed and defined on the plaquettes of which $\tilde{P}$ will pass through two adjacent edges, e.g. $<l,n>$ and $<m,n>$ of $[l,m,n]$. We denote these edges crossing the loop $\tilde{P}$ as the edges {\it in} the loop $\tilde{P}$, e.g. $<l,n>, <m,n>,<m,o>\in \tilde{P}$. Then depending on whether the other adjacent edge is on the left or right of $\tilde{P}$, we denote the corresponding plaquette as $L-plaquette$ and $R-plaquette$ respectively and this edge is denoted as the $L-edge$ and $R-edge$ respectively. E.g. $[l,m,n]$ is a $R-plaquette$ and $[m,n,o]$ is a $L-plaquette$ with $<n,o>$ a $L-edge$. Then Four simple string operator is defined as
\begin{equation}
\begin{aligned}
&W_{\one}=1 \\
&W_{s\bar s}=\prod_{L-edge}\sigma_z^{<p,q>}\\
&W_s=\prod_{<j,k>\in \tilde{P}}\sigma_x^{<j,k>}\prod_{L-edge}(i)^{(1-\sigma_z^{<p,q>})/2}\\
&\times\prod_{R-plaquette}(-1)^{s_{[l,m,n]}}\\
&W_{\bar s}=\prod_{<j,k>\in \tilde{P}}\sigma_x^{<j,k>}\prod_{L-edge}(-i)^{(1-\sigma_z^{<p,q>})/2}\\
&\times\prod_{R-plaquette}(-1)^{s_{[l,m,n]}}
\end{aligned}
\end{equation}
where $s_{[l,m,n]}=(1-\sigma_z^{<m,n>})(1+\sigma_z^{<l,n>})/4$. $<m,n>$, $<l,n>$ are two edges in $\tilde{P}$ adjacent to a $R-plaquette$ $[l,m,n]$ and $<m,n>$ is the edge that $\tilde{P}$ passes before $<l,n>$. Note that $W_{{s}\bar{s}}$ is the same loop operator as that of $e$-particle in the bulk. When the string operator is defined on a string $S$ instead of a loop it will create two excitations on two ends of $S$. These open string operators will be denoted as $\tilde W$.\\

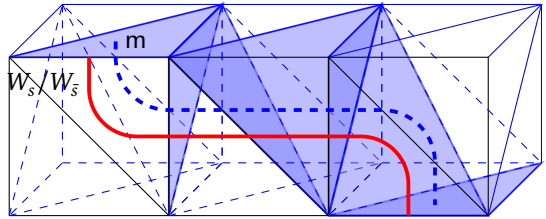
\begin{figure}[htbp!]
\centering
\begin{tikzpicture}[scale=0.7]
\draw[thick, blue, fill=blue!50!white, opacity=0.5] (0,3) -- (3,3) -- (4,4) -- (0,3);
\draw (0,0) -- (0,3) -- (3,3) -- (3,0) -- (0,0);
\draw[blue] (0,3) -- (1,4) -- (4,4) -- (3,3);
\draw[dashed, blue] (4,4) -- (4,1) -- (3,0);
\draw[dashed, blue] (1,4) -- (1,1) -- (0,0);
\draw[dashed, blue] (1,1) -- (4,1);
\draw (0,3) -- (3,0);
\draw[dashed, blue] (4,4) -- (3,0);
\draw[dashed, blue] (0,0) -- (1,4) -- (4,1) -- (0,0);
\draw[dashed, blue] (1,4) -- (3,0);
\draw[blue] (4,4) -- (0,3);

\begin{scope}[xshift=3cm,yshift=0cm]{
\draw[thick, blue, fill=blue!50!white, opacity=0.5] (0,0) -- (0,3) -- (1,4) -- (0,0);
\draw[thick, blue, fill=blue!50!white, opacity=0.5] (3,0) -- (0,3) -- (1,4) -- (3,0);
\draw[thick, blue, fill=blue!50!white, opacity=0.5] (3,0) -- (0,3) -- (4,4) -- (3,0);
\draw (0,0) -- (0,3) -- (3,3) -- (3,0) -- (0,0);
\draw[blue] (0,3) -- (1,4) -- (4,4) -- (3,3);
\draw[dashed, blue] (4,4) -- (4,1) -- (3,0);
\draw[dashed, blue] (1,4) -- (1,1) -- (0,0);
\draw[dashed, blue] (1,1) -- (4,1);
\draw (0,3) -- (3,0);
\draw[dashed, blue] (4,4) -- (3,0);
\draw[dashed, blue] (0,0) -- (1,4) -- (4,1) -- (0,0);
\draw[dashed, blue] (1,4) -- (3,0);
\draw[blue] (4,4) -- (0,3);
}
\end{scope}

\begin{scope}[xshift=6cm,yshift=0cm]{
\draw[thick, blue, fill=blue!50!white, opacity=0.5] (0,0) -- (0,3) -- (1,4) -- (0,0);
\draw[thick, blue, fill=blue!50!white, opacity=0.5] (0,0) -- (3,0) -- (4,1) -- (0,0);
\draw[thick, blue, fill=blue!50!white, opacity=0.5] (0,0) -- (3,0) -- (1,4) -- (0,0);
\draw (0,0) -- (0,3) -- (3,3) -- (3,0) -- (0,0);
\draw[blue] (0,3) -- (1,4) -- (4,4) -- (3,3);
\draw[blue] (4,4) -- (4,1) -- (3,0);
\draw[dashed, blue] (1,4) -- (1,1) -- (0,0);
\draw[dashed, blue] (1,1) -- (4,1);
\draw (0,3) -- (3,0);
\draw[blue] (4,4) -- (3,0);
\draw[dashed, blue] (0,0) -- (1,4) -- (4,1) -- (0,0);
\draw[dashed, blue] (1,4) -- (3,0);
\draw[blue] (4,4) -- (0,3);
}
\end{scope}
\draw[red, line width=0.05cm] (1.5,3) -- (1.5,2.4) to [out=270, in=180] (2.4,1.5) -- (6.6, 1.5) to [out=0, in=90] (7.5,0.6) -- (7.5,0);
\draw[line width=0.05cm, dashed, blue] (2,3.3) -- (2,3) to [out=270, in=180] (3,2) -- (7,2) to [out=0, in=90] (8,1) -- (8,0.2);
\node[right] at (2,3.3) {$\sm$};
\node[left] at (1.5,2.5) {$W_s/W_{\bar{s}}$};
\end{tikzpicture}
\caption{Action of the $W_s$ or $W_{\bar{s}}$ loop operator with the path denoted as the red line partly. They create a $\sm$-string in the bulk denoted as the blue dashed line and blue plaquettes.}\label{actionws}
\end{figure}

Now we can come back to the boundary. Note that these four string operators commute with $H_{bdy}$. $W_{\one}$ and $W_{s\bar{s}}$ also commute with the bulk Hamiltonian $H_{DW}$ while $W_s$ and $W_{\bar{s}}$ are not. Obviously they commute with the vertex operators in the bulk, $A_i$. However when the plaquette operators in the bulk, $B'_{[i,j,k]}$, has one edge on the boundary and in the loop $\tilde{P}$, the $\prod_{<j,k>\in \tilde{P}}\sigma_x^{<j,k>}$ part in $W_s$ and $W_{\bar{s}}$ anti-commute with them. Namely the action of $W_s$ and $W_{\bar{s}}$ on the boundary will create a $\sm$-string in the bulk just beside the boundary as partly shown in Figure \ref{actionws}.
\\
In other words when a $\sm$-string is pulled towards the boundary, it can be moved onto the boundary and vanishes by acting $W_s$, $W_{\bar{s}}$ or even their linear combinations with a suitable loop $\tilde{P}$, which indicates that the $\sm$-string is condensed on the twisted smooth boundary. Now we can denote the trivial string on the boundary as $\one\oplus \sm$.\\
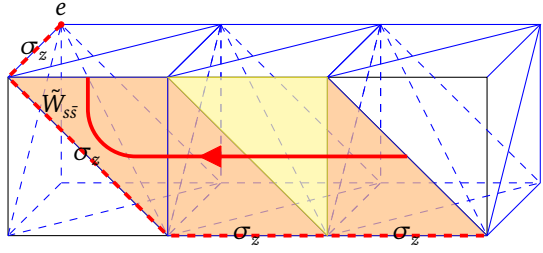
\begin{figure}[htbp!]
\centering
\begin{tikzpicture}[scale=0.7]
\draw (0,0) -- (0,3) -- (3,3) -- (3,0) -- (0,0);
\draw[blue] (0,3) -- (1,4) -- (4,4) -- (3,3);
\draw[dashed, blue] (4,4) -- (4,1) -- (3,0);
\draw[dashed, blue] (1,4) -- (1,1) -- (0,0);
\draw[dashed, blue] (1,1) -- (4,1);
\draw (0,3) -- (3,0);
\draw[dashed, blue] (4,4) -- (3,0);
\draw[dashed, blue] (0,0) -- (1,4) -- (4,1) -- (0,0);
\draw[dashed, blue] (1,4) -- (3,0);
\draw[blue] (4,4) -- (0,3);

\begin{scope}[xshift=3cm,yshift=0cm]{
\draw (0,0) -- (0,3) -- (3,3) -- (3,0) -- (0,0);
\draw[blue] (0,3) -- (1,4) -- (4,4) -- (3,3);
\draw[dashed, blue] (4,4) -- (4,1) -- (3,0);
\draw[dashed, blue] (1,4) -- (1,1) -- (0,0);
\draw[dashed, blue] (1,1) -- (4,1);
\draw (0,3) -- (3,0);
\draw[dashed, blue] (4,4) -- (3,0);
\draw[dashed, blue] (0,0) -- (1,4) -- (4,1) -- (0,0);
\draw[dashed, blue] (1,4) -- (3,0);
\draw[blue] (4,4) -- (0,3);
}
\end{scope}

\begin{scope}[xshift=6cm,yshift=0cm]{
\draw (0,0) -- (0,3) -- (3,3) -- (3,0) -- (0,0);
\draw[blue] (0,3) -- (1,4) -- (4,4) -- (3,3);
\draw[blue] (4,4) -- (4,1) -- (3,0);
\draw[dashed, blue] (1,4) -- (1,1) -- (0,0);
\draw[dashed, blue] (1,1) -- (4,1);
\draw (0,3) -- (3,0);
\draw[ blue] (4,4) -- (3,0);
\draw[dashed, blue] (0,0) -- (1,4) -- (4,1) -- (0,0);
\draw[dashed, blue] (1,4) -- (3,0);
\draw[blue] (4,4) -- (0,3);
}
\end{scope}
\draw[blue, fill=orange!50!white, opacity=0.7] (0,3) -- (3,0) -- (3,3) -- (0,3);
\draw[blue, fill=orange!50!white, opacity=0.7] (6,0) -- (3,0) -- (3,3) -- (6,0);
\draw[blue, fill=orange!50!white, opacity=0.7] (9,0) -- (6,0) -- (6,3) -- (9,0);
\draw[yellow, fill=yellow!50!white, opacity=0.7] (6,0) -- (6,3) -- (3,3) -- (6,0);
\draw[red, line width=0.05cm] (1.5,3) -- (1.5,2.4) to [out=270, in=180] (2.4,1.5) -- (7.5, 1.5);
\draw[red, fill=red] (4,1.7) -- (4,1.3) -- (3.6,1.5) -- (4,1.7);
\node[left] at (1.5,2.5) {$\tilde{W}_{s\bar{s}}$};
\draw[red, fill=red] (1,4) circle [radius=0.05cm];
\node[above] at (1,4) {$e$};
\draw[red, dashed, line width=0.05cm] (1,4) -- (0,3) -- (3,0) -- (9,0);
\node at (0.5,3.5) {$\sigma_z$};
\node at (1.5,1.5) {$\sigma_z$};
\node at (4.5,0) {$\sigma_z$};
\node at (7.5,0) {$\sigma_z$};
\end{tikzpicture}
\caption{The $e$-particle on the boundary and its moving path is denoted as the red dashed line. The string operator $\tilde W_{s\bar s}$ is defined on the red string.}\label{eandssbar}
\end{figure}
What happened to the $e$-particle? Note that if we have a string operator of $e$-particle which has an end point $p$ on the boundary, it will commute with all the $A_i$, $B'_{[i,j,k]}$, $\tilde{A}_i$ and $\tilde{B}_{[i,j,k]}$ except $\tilde{A}_p$ at vertex $p$. It anti-commutes with $\tilde{A}_p$ indicating that the $e$-particle survives on the boundary. Further on the boundary we can move it by $\tilde W_{s\bar{s}}$ as shown in Figure \ref{eandssbar}.
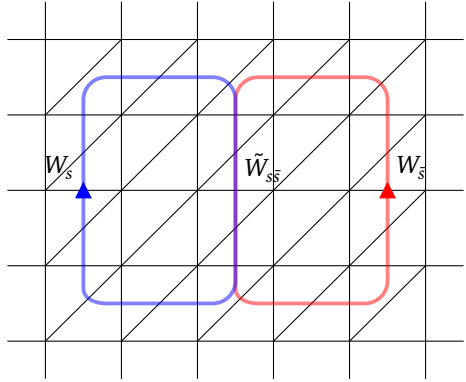
\begin{figure}[htbp!]
\centering
\begin{tikzpicture}
\foreach \x in {5,6,7,8,9,10}
\draw (\x,0.5) -- (\x,5.5);
\foreach \y in {1,2,3,4,5}
\draw (4.5,\y) -- (10.5,\y);
\foreach \w in {5,6,7,8,9}
\foreach \v in {1,2,3,4}
\draw (\w+1,\v+1) -- (\w,\v);
\begin{scope}[xshift=2cm, yshift=0]{
\draw [red, line width=0.05cm ,opacity=0.5] (5.5,1.7) -- (5.5,4.2) to [out=90, in=180] (5.8,4.5) -- (7.2, 4.5) to [out=0, in=90] (7.5,4.2) -- (7.5,1.8) to [out=270, in=0] (7.2,1.5) -- (5.8,1.5) to [out=180, in=270] (5.5,1.7);
\draw [red, fill=red] (7.5,3.1) -- (7.4,2.9) -- (7.6,2.9) -- (7.5,3.1);
}
\end{scope}
\draw [blue, line width=0.05cm, opacity=0.5] (5.5,1.7) -- (5.5,4.2) to [out=90, in=180] (5.8,4.5) -- (7.2, 4.5) to [out=0, in=90] (7.5,4.2) -- (7.5,1.8) to [out=270, in=0] (7.2,1.5) -- (5.8,1.5) to [out=180, in=270] (5.5,1.7);
\draw [blue, fill=blue] (5.4, 2.9) -- (5.6,2.9) -- (5.5, 3.1) -- (5.4,2.9);
\node [left] at (5.5,3.3) {$W_s$};
\node [right] at (9.5,3.3) {$W_{\bar{s}}$};
\node [right] at (7.5,3.3) {$\tilde{W}_{s\bar{s}}$};
\end{tikzpicture}
\caption{The overlap of $W_s$ and $W_{\bar s}$ is $\tilde W_{s\bar s}$}\label{mofm}
\end{figure}\\
Compared to the ordinary smooth boundary on which the $\sm$-string can only be moved onto the boundary and annihilated by only one kind of loop operator, i.e., $\prod\sigma_x$, here it can be done by two kinds of loop operator. 
Note that we can use the open string operators $\tilde W_s$, $\tilde W_{\bar s}$ and their linear combinations to pull half of the $\sm$-string onto the boundary and annihilate them, while leaving two excitations on two ends of the $\sm$-string on the boundary. From the point of view of the double semion model, these excitations correspond to the semions $s$ and $\bar s$ respectively therefore two types of end points can change into each other by fusing with the $e$-particle on the boundary. Further if we only have a pair of end points then they must be of the same type.
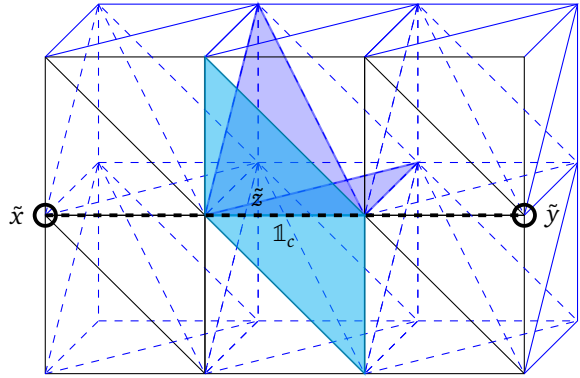
\begin{figure}[htbp!]
\centering
\begin{tikzpicture}[scale=0.7]
\draw[thick, blue, fill=blue!50!white, opacity=0.5] (3,0) -- (6,0) -- (4,4) -- (3,0);
\draw[thick, blue, fill=blue!50!white, opacity=0.5] (3,0) -- (6,0) -- (7,1) -- (3,0);
\draw (0,0) -- (0,3) -- (3,3) -- (3,0) -- (0,0);
\draw[blue] (0,3) -- (1,4) -- (4,4) -- (3,3);
\draw[dashed, blue] (4,4) -- (4,1) -- (3,0);
\draw[dashed, blue] (1,4) -- (1,1) -- (0,0);
\draw[dashed, blue] (1,1) -- (4,1);
\draw (0,3) -- (3,0);
\draw[dashed, blue] (4,4) -- (3,0);
\draw[dashed, blue] (0,0) -- (1,4) -- (4,1) -- (0,0);
\draw[dashed, blue] (1,4) -- (3,0);
\draw[blue] (4,4) -- (0,3);

\begin{scope}[xshift=3cm,yshift=0cm]{
\draw (0,0) -- (0,3) -- (3,3) -- (3,0) -- (0,0);
\draw[blue] (0,3) -- (1,4) -- (4,4) -- (3,3);
\draw[dashed, blue] (4,4) -- (4,1) -- (3,0);
\draw[dashed, blue] (1,4) -- (1,1) -- (0,0);
\draw[dashed, blue] (1,1) -- (4,1);
\draw (0,3) -- (3,0);
\draw[dashed, blue] (4,4) -- (3,0);
\draw[dashed, blue] (0,0) -- (1,4) -- (4,1) -- (0,0);
\draw[dashed, blue] (1,4) -- (3,0);
\draw[blue] (4,4) -- (0,3);
}
\end{scope}

\begin{scope}[xshift=6cm,yshift=0cm]{
\draw (0,0) -- (0,3) -- (3,3) -- (3,0) -- (0,0);
\draw[blue] (0,3) -- (1,4) -- (4,4) -- (3,3);
\draw[blue] (4,4) -- (4,1) -- (3,0);
\draw[dashed, blue] (1,4) -- (1,1) -- (0,0);
\draw[dashed, blue] (1,1) -- (4,1);
\draw (0,3) -- (3,0);
\draw[blue] (4,4) -- (3,0);
\draw[dashed, blue] (0,0) -- (1,4) -- (4,1) -- (0,0);
\draw[dashed, blue] (1,4) -- (3,0);
\draw[blue] (4,4) -- (0,3);
}
\end{scope}

\begin{scope}[xshift=0cm, yshift=-3cm]{
\draw (0,0) -- (0,3) -- (3,3) -- (3,0) -- (0,0);
\draw[dashed, blue] (4,4) -- (4,1) -- (3,0);
\draw[dashed, blue] (1,4) -- (1,1) -- (0,0);
\draw[dashed, blue] (1,1) -- (4,1);
\draw (0,3) -- (3,0);
\draw[dashed, blue] (4,4) -- (3,0);
\draw[dashed, blue] (0,0) -- (1,4) -- (4,1) -- (0,0);
\draw[dashed, blue] (1,4) -- (3,0);

\begin{scope}[xshift=3cm,yshift=0cm]{
\draw (0,0) -- (0,3) -- (3,3) -- (3,0) -- (0,0);
\draw[dashed, blue] (4,4) -- (4,1) -- (3,0);
\draw[dashed, blue] (1,4) -- (1,1) -- (0,0);
\draw[dashed, blue] (1,1) -- (4,1);
\draw (0,3) -- (3,0);
\draw[dashed, blue] (4,4) -- (3,0);
\draw[dashed, blue] (0,0) -- (1,4) -- (4,1) -- (0,0);
\draw[dashed, blue] (1,4) -- (3,0);
}
\end{scope}

\begin{scope}[xshift=6cm,yshift=0cm]{
\draw (0,0) -- (0,3) -- (3,3) -- (3,0) -- (0,0);
\draw[blue] (4,4) -- (4,1) -- (3,0);
\draw[dashed, blue] (1,4) -- (1,1) -- (0,0);
\draw[dashed, blue] (1,1) -- (4,1);
\draw (0,3) -- (3,0);
\draw[blue] (4,4) -- (3,0);
\draw[dashed, blue] (0,0) -- (1,4) -- (4,1) -- (0,0);
\draw[dashed, blue] (1,4) -- (3,0);
}
\end{scope}
}
\end{scope}
\draw[thick, cyan, fill=cyan, opacity=0.5] (3,0) -- (6,0) -- (3,3) -- (3,0);
\draw[thick, cyan, fill=cyan, opacity=0.5] (3,0) -- (6,0) -- (6,-3) -- (3,0);
\draw[dashed, line width=0.05cm] (0,0) -- (9,0);
\draw[line width=0.05cm] (0,0) circle [radius=0.2cm];
\draw[line width=0.05cm] (9,0) circle [radius=0.2cm];
\node[below] at (4.5,0) {$\one_c$};
\node[left] at (-0.2,0) {$\tilde x$};
\node[right] at (9.2,0) {$\tilde y$};
\node[above] at (4,0) {$\tilde z$};
\end{tikzpicture}
\caption{$\one_c$-sting on the boundary is also denoted as the black dashed line. The 0d excitations between $\one_c$ and $\one$ strings on the boundary: $\tilde x$ and $\tilde y$ are denoted as circles. The 0d excitations between $\one_c$-strings: $\tilde z$ is a half of $\sm$-string winding around the $\one_c$-string on the boundary.}\label{1_cbdy}
\end{figure}\\

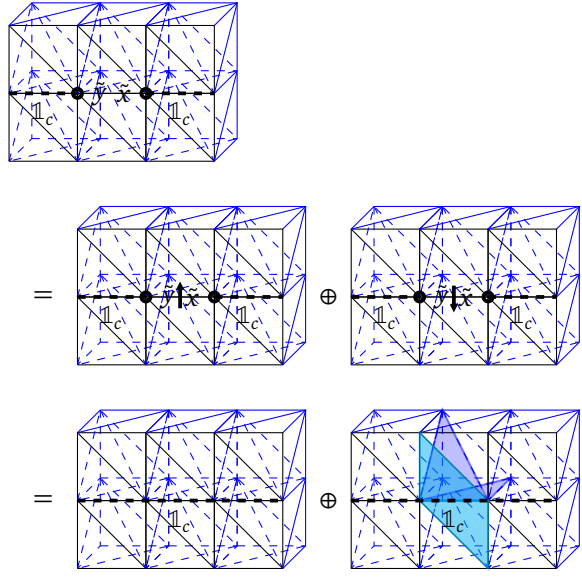
\begin{figure}[htbp!]
\centering
\begin{tikzpicture}[scale=0.3]
\draw (0,0) -- (0,3) -- (3,3) -- (3,0) -- (0,0);
\draw[blue] (0,3) -- (1,4) -- (4,4) -- (3,3);
\draw[dashed, blue] (4,4) -- (4,1) -- (3,0);
\draw[dashed, blue] (1,4) -- (1,1) -- (0,0);
\draw[dashed, blue] (1,1) -- (4,1);
\draw (0,3) -- (3,0);
\draw[dashed, blue] (4,4) -- (3,0);
\draw[dashed, blue] (0,0) -- (1,4) -- (4,1) -- (0,0);
\draw[dashed, blue] (1,4) -- (3,0);
\draw[blue] (4,4) -- (0,3);

\begin{scope}[xshift=3cm,yshift=0cm]{
\draw (0,0) -- (0,3) -- (3,3) -- (3,0) -- (0,0);
\draw[blue] (0,3) -- (1,4) -- (4,4) -- (3,3);
\draw[dashed, blue] (4,4) -- (4,1) -- (3,0);
\draw[dashed, blue] (1,4) -- (1,1) -- (0,0);
\draw[dashed, blue] (1,1) -- (4,1);
\draw (0,3) -- (3,0);
\draw[dashed, blue] (4,4) -- (3,0);
\draw[dashed, blue] (0,0) -- (1,4) -- (4,1) -- (0,0);
\draw[dashed, blue] (1,4) -- (3,0);
\draw[blue] (4,4) -- (0,3);
}
\end{scope}

\begin{scope}[xshift=6cm,yshift=0cm]{
\draw (0,0) -- (0,3) -- (3,3) -- (3,0) -- (0,0);
\draw[blue] (0,3) -- (1,4) -- (4,4) -- (3,3);
\draw[blue] (4,4) -- (4,1) -- (3,0);
\draw[dashed, blue] (1,4) -- (1,1) -- (0,0);
\draw[dashed, blue] (1,1) -- (4,1);
\draw (0,3) -- (3,0);
\draw[blue] (4,4) -- (3,0);
\draw[dashed, blue] (0,0) -- (1,4) -- (4,1) -- (0,0);
\draw[dashed, blue] (1,4) -- (3,0);
\draw[blue] (4,4) -- (0,3);
}
\end{scope}

\begin{scope}[xshift=0cm, yshift=-3cm]{
\draw (0,0) -- (0,3) -- (3,3) -- (3,0) -- (0,0);
\draw[dashed, blue] (4,4) -- (4,1) -- (3,0);
\draw[dashed, blue] (1,4) -- (1,1) -- (0,0);
\draw[dashed, blue] (1,1) -- (4,1);
\draw (0,3) -- (3,0);
\draw[dashed, blue] (4,4) -- (3,0);
\draw[dashed, blue] (0,0) -- (1,4) -- (4,1) -- (0,0);
\draw[dashed, blue] (1,4) -- (3,0);

\begin{scope}[xshift=3cm,yshift=0cm]{
\draw (0,0) -- (0,3) -- (3,3) -- (3,0) -- (0,0);
\draw[dashed, blue] (4,4) -- (4,1) -- (3,0);
\draw[dashed, blue] (1,4) -- (1,1) -- (0,0);
\draw[dashed, blue] (1,1) -- (4,1);
\draw (0,3) -- (3,0);
\draw[dashed, blue] (4,4) -- (3,0);
\draw[dashed, blue] (0,0) -- (1,4) -- (4,1) -- (0,0);
\draw[dashed, blue] (1,4) -- (3,0);
}
\end{scope}

\begin{scope}[xshift=6cm,yshift=0cm]{
\draw (0,0) -- (0,3) -- (3,3) -- (3,0) -- (0,0);
\draw[blue] (4,4) -- (4,1) -- (3,0);
\draw[dashed, blue] (1,4) -- (1,1) -- (0,0);
\draw[dashed, blue] (1,1) -- (4,1);
\draw (0,3) -- (3,0);
\draw[blue] (4,4) -- (3,0);
\draw[dashed, blue] (0,0) -- (1,4) -- (4,1) -- (0,0);
\draw[dashed, blue] (1,4) -- (3,0);
}
\end{scope}
}
\end{scope}
\draw[dashed, line width=0.05cm] (0,0) -- (3,0);
\draw[dashed, line width=0.05cm] (6,0) -- (9,0);
\draw[line width=0.05cm] (3,0) circle [radius=0.2cm];
\draw[line width=0.05cm] (6,0) circle [radius=0.2cm];
\node[below] at (1.5,0) {$\one_c$};
\node[below] at (7.5,0) {$\one_c$};
\node[right] at (3.2,0) {$\tilde y$};
\node[left] at (5.8,0) {$\tilde x$};

\begin{scope}[xshift=3cm, yshift=-9cm]{
\draw (0,0) -- (0,3) -- (3,3) -- (3,0) -- (0,0);
\draw[blue] (0,3) -- (1,4) -- (4,4) -- (3,3);
\draw[dashed, blue] (4,4) -- (4,1) -- (3,0);
\draw[dashed, blue] (1,4) -- (1,1) -- (0,0);
\draw[dashed, blue] (1,1) -- (4,1);
\draw (0,3) -- (3,0);
\draw[dashed, blue] (4,4) -- (3,0);
\draw[dashed, blue] (0,0) -- (1,4) -- (4,1) -- (0,0);
\draw[dashed, blue] (1,4) -- (3,0);
\draw[blue] (4,4) -- (0,3);

\begin{scope}[xshift=3cm,yshift=0cm]{
\draw (0,0) -- (0,3) -- (3,3) -- (3,0) -- (0,0);
\draw[blue] (0,3) -- (1,4) -- (4,4) -- (3,3);
\draw[dashed, blue] (4,4) -- (4,1) -- (3,0);
\draw[dashed, blue] (1,4) -- (1,1) -- (0,0);
\draw[dashed, blue] (1,1) -- (4,1);
\draw (0,3) -- (3,0);
\draw[dashed, blue] (4,4) -- (3,0);
\draw[dashed, blue] (0,0) -- (1,4) -- (4,1) -- (0,0);
\draw[dashed, blue] (1,4) -- (3,0);
\draw[blue] (4,4) -- (0,3);
}
\end{scope}

\begin{scope}[xshift=6cm,yshift=0cm]{
\draw (0,0) -- (0,3) -- (3,3) -- (3,0) -- (0,0);
\draw[blue] (0,3) -- (1,4) -- (4,4) -- (3,3);
\draw[blue] (4,4) -- (4,1) -- (3,0);
\draw[dashed, blue] (1,4) -- (1,1) -- (0,0);
\draw[dashed, blue] (1,1) -- (4,1);
\draw (0,3) -- (3,0);
\draw[blue] (4,4) -- (3,0);
\draw[dashed, blue] (0,0) -- (1,4) -- (4,1) -- (0,0);
\draw[dashed, blue] (1,4) -- (3,0);
\draw[blue] (4,4) -- (0,3);
}
\end{scope}

\begin{scope}[xshift=0cm, yshift=-3cm]{
\draw (0,0) -- (0,3) -- (3,3) -- (3,0) -- (0,0);
\draw[dashed, blue] (4,4) -- (4,1) -- (3,0);
\draw[dashed, blue] (1,4) -- (1,1) -- (0,0);
\draw[dashed, blue] (1,1) -- (4,1);
\draw (0,3) -- (3,0);
\draw[dashed, blue] (4,4) -- (3,0);
\draw[dashed, blue] (0,0) -- (1,4) -- (4,1) -- (0,0);
\draw[dashed, blue] (1,4) -- (3,0);

\begin{scope}[xshift=3cm,yshift=0cm]{
\draw (0,0) -- (0,3) -- (3,3) -- (3,0) -- (0,0);
\draw[dashed, blue] (4,4) -- (4,1) -- (3,0);
\draw[dashed, blue] (1,4) -- (1,1) -- (0,0);
\draw[dashed, blue] (1,1) -- (4,1);
\draw (0,3) -- (3,0);
\draw[dashed, blue] (4,4) -- (3,0);
\draw[dashed, blue] (0,0) -- (1,4) -- (4,1) -- (0,0);
\draw[dashed, blue] (1,4) -- (3,0);
}
\end{scope}

\begin{scope}[xshift=6cm,yshift=0cm]{
\draw (0,0) -- (0,3) -- (3,3) -- (3,0) -- (0,0);
\draw[blue] (4,4) -- (4,1) -- (3,0);
\draw[dashed, blue] (1,4) -- (1,1) -- (0,0);
\draw[dashed, blue] (1,1) -- (4,1);
\draw (0,3) -- (3,0);
\draw[blue] (4,4) -- (3,0);
\draw[dashed, blue] (0,0) -- (1,4) -- (4,1) -- (0,0);
\draw[dashed, blue] (1,4) -- (3,0);
}
\end{scope}
}
\end{scope}
\draw[dashed, line width=0.05cm] (0,0) -- (3,0);
\draw[dashed, line width=0.05cm] (6,0) -- (9,0);
\draw[line width=0.05cm] (3,0) circle [radius=0.2cm];
\draw[line width=0.05cm] (6,0) circle [radius=0.2cm];
\node[below] at (1.5,0) {$\one_c$};
\node[below] at (7.5,0) {$\one_c$};
\node[right] at (3.2,0) {$\tilde y$};
\node[left] at (5.8,0) {$\tilde x$};
\draw[line width=0.05cm] (4.5,-0.5) -- (4.5,0.5);
\draw[line width=0.05cm] (4.4,0.3) -- (4.5,0.5) -- (4.6,0.3);
}
\end{scope}

\begin{scope}[xshift=15cm, yshift=-9cm]{
\draw (0,0) -- (0,3) -- (3,3) -- (3,0) -- (0,0);
\draw[blue] (0,3) -- (1,4) -- (4,4) -- (3,3);
\draw[dashed, blue] (4,4) -- (4,1) -- (3,0);
\draw[dashed, blue] (1,4) -- (1,1) -- (0,0);
\draw[dashed, blue] (1,1) -- (4,1);
\draw (0,3) -- (3,0);
\draw[dashed, blue] (4,4) -- (3,0);
\draw[dashed, blue] (0,0) -- (1,4) -- (4,1) -- (0,0);
\draw[dashed, blue] (1,4) -- (3,0);
\draw[blue] (4,4) -- (0,3);

\begin{scope}[xshift=3cm,yshift=0cm]{
\draw (0,0) -- (0,3) -- (3,3) -- (3,0) -- (0,0);
\draw[blue] (0,3) -- (1,4) -- (4,4) -- (3,3);
\draw[dashed, blue] (4,4) -- (4,1) -- (3,0);
\draw[dashed, blue] (1,4) -- (1,1) -- (0,0);
\draw[dashed, blue] (1,1) -- (4,1);
\draw (0,3) -- (3,0);
\draw[dashed, blue] (4,4) -- (3,0);
\draw[dashed, blue] (0,0) -- (1,4) -- (4,1) -- (0,0);
\draw[dashed, blue] (1,4) -- (3,0);
\draw[blue] (4,4) -- (0,3);
}
\end{scope}

\begin{scope}[xshift=6cm,yshift=0cm]{
\draw (0,0) -- (0,3) -- (3,3) -- (3,0) -- (0,0);
\draw[blue] (0,3) -- (1,4) -- (4,4) -- (3,3);
\draw[blue] (4,4) -- (4,1) -- (3,0);
\draw[dashed, blue] (1,4) -- (1,1) -- (0,0);
\draw[dashed, blue] (1,1) -- (4,1);
\draw (0,3) -- (3,0);
\draw[blue] (4,4) -- (3,0);
\draw[dashed, blue] (0,0) -- (1,4) -- (4,1) -- (0,0);
\draw[dashed, blue] (1,4) -- (3,0);
\draw[blue] (4,4) -- (0,3);
}
\end{scope}

\begin{scope}[xshift=0cm, yshift=-3cm]{
\draw (0,0) -- (0,3) -- (3,3) -- (3,0) -- (0,0);
\draw[dashed, blue] (4,4) -- (4,1) -- (3,0);
\draw[dashed, blue] (1,4) -- (1,1) -- (0,0);
\draw[dashed, blue] (1,1) -- (4,1);
\draw (0,3) -- (3,0);
\draw[dashed, blue] (4,4) -- (3,0);
\draw[dashed, blue] (0,0) -- (1,4) -- (4,1) -- (0,0);
\draw[dashed, blue] (1,4) -- (3,0);

\begin{scope}[xshift=3cm,yshift=0cm]{
\draw (0,0) -- (0,3) -- (3,3) -- (3,0) -- (0,0);
\draw[dashed, blue] (4,4) -- (4,1) -- (3,0);
\draw[dashed, blue] (1,4) -- (1,1) -- (0,0);
\draw[dashed, blue] (1,1) -- (4,1);
\draw (0,3) -- (3,0);
\draw[dashed, blue] (4,4) -- (3,0);
\draw[dashed, blue] (0,0) -- (1,4) -- (4,1) -- (0,0);
\draw[dashed, blue] (1,4) -- (3,0);
}
\end{scope}

\begin{scope}[xshift=6cm,yshift=0cm]{
\draw (0,0) -- (0,3) -- (3,3) -- (3,0) -- (0,0);
\draw[blue] (4,4) -- (4,1) -- (3,0);
\draw[dashed, blue] (1,4) -- (1,1) -- (0,0);
\draw[dashed, blue] (1,1) -- (4,1);
\draw (0,3) -- (3,0);
\draw[blue] (4,4) -- (3,0);
\draw[dashed, blue] (0,0) -- (1,4) -- (4,1) -- (0,0);
\draw[dashed, blue] (1,4) -- (3,0);
}
\end{scope}
}
\end{scope}
\draw[dashed, line width=0.05cm] (0,0) -- (3,0);
\draw[dashed, line width=0.05cm] (6,0) -- (9,0);
\draw[line width=0.05cm] (3,0) circle [radius=0.2cm];
\draw[line width=0.05cm] (6,0) circle [radius=0.2cm];
\node[below] at (1.5,0) {$\one_c$};
\node[below] at (7.5,0) {$\one_c$};
\node[right] at (3.2,0) {$\tilde y$};
\node[left] at (5.8,0) {$\tilde x$};
\draw[line width=0.05cm] (4.5,-0.5) -- (4.5,0.5);
\draw[line width=0.05cm] (4.4,-0.3) -- (4.5,-0.5) -- (4.6,-0.3);
}
\end{scope}
\node at (1.5,-9) {$=$};
\node at (14,-9) {$\oplus$};

\begin{scope}[xshift=3cm, yshift=-18cm]{
\draw (0,0) -- (0,3) -- (3,3) -- (3,0) -- (0,0);
\draw[blue] (0,3) -- (1,4) -- (4,4) -- (3,3);
\draw[dashed, blue] (4,4) -- (4,1) -- (3,0);
\draw[dashed, blue] (1,4) -- (1,1) -- (0,0);
\draw[dashed, blue] (1,1) -- (4,1);
\draw (0,3) -- (3,0);
\draw[dashed, blue] (4,4) -- (3,0);
\draw[dashed, blue] (0,0) -- (1,4) -- (4,1) -- (0,0);
\draw[dashed, blue] (1,4) -- (3,0);
\draw[blue] (4,4) -- (0,3);

\begin{scope}[xshift=3cm,yshift=0cm]{
\draw (0,0) -- (0,3) -- (3,3) -- (3,0) -- (0,0);
\draw[blue] (0,3) -- (1,4) -- (4,4) -- (3,3);
\draw[dashed, blue] (4,4) -- (4,1) -- (3,0);
\draw[dashed, blue] (1,4) -- (1,1) -- (0,0);
\draw[dashed, blue] (1,1) -- (4,1);
\draw (0,3) -- (3,0);
\draw[dashed, blue] (4,4) -- (3,0);
\draw[dashed, blue] (0,0) -- (1,4) -- (4,1) -- (0,0);
\draw[dashed, blue] (1,4) -- (3,0);
\draw[blue] (4,4) -- (0,3);
}
\end{scope}

\begin{scope}[xshift=6cm,yshift=0cm]{
\draw (0,0) -- (0,3) -- (3,3) -- (3,0) -- (0,0);
\draw[blue] (0,3) -- (1,4) -- (4,4) -- (3,3);
\draw[blue] (4,4) -- (4,1) -- (3,0);
\draw[dashed, blue] (1,4) -- (1,1) -- (0,0);
\draw[dashed, blue] (1,1) -- (4,1);
\draw (0,3) -- (3,0);
\draw[blue] (4,4) -- (3,0);
\draw[dashed, blue] (0,0) -- (1,4) -- (4,1) -- (0,0);
\draw[dashed, blue] (1,4) -- (3,0);
\draw[blue] (4,4) -- (0,3);
}
\end{scope}

\begin{scope}[xshift=0cm, yshift=-3cm]{
\draw (0,0) -- (0,3) -- (3,3) -- (3,0) -- (0,0);
\draw[dashed, blue] (4,4) -- (4,1) -- (3,0);
\draw[dashed, blue] (1,4) -- (1,1) -- (0,0);
\draw[dashed, blue] (1,1) -- (4,1);
\draw (0,3) -- (3,0);
\draw[dashed, blue] (4,4) -- (3,0);
\draw[dashed, blue] (0,0) -- (1,4) -- (4,1) -- (0,0);
\draw[dashed, blue] (1,4) -- (3,0);

\begin{scope}[xshift=3cm,yshift=0cm]{
\draw (0,0) -- (0,3) -- (3,3) -- (3,0) -- (0,0);
\draw[dashed, blue] (4,4) -- (4,1) -- (3,0);
\draw[dashed, blue] (1,4) -- (1,1) -- (0,0);
\draw[dashed, blue] (1,1) -- (4,1);
\draw (0,3) -- (3,0);
\draw[dashed, blue] (4,4) -- (3,0);
\draw[dashed, blue] (0,0) -- (1,4) -- (4,1) -- (0,0);
\draw[dashed, blue] (1,4) -- (3,0);
}
\end{scope}

\begin{scope}[xshift=6cm,yshift=0cm]{
\draw (0,0) -- (0,3) -- (3,3) -- (3,0) -- (0,0);
\draw[blue] (4,4) -- (4,1) -- (3,0);
\draw[dashed, blue] (1,4) -- (1,1) -- (0,0);
\draw[dashed, blue] (1,1) -- (4,1);
\draw (0,3) -- (3,0);
\draw[blue] (4,4) -- (3,0);
\draw[dashed, blue] (0,0) -- (1,4) -- (4,1) -- (0,0);
\draw[dashed, blue] (1,4) -- (3,0);
}
\end{scope}
}
\end{scope}
\draw[dashed, line width=0.05cm] (0,0) -- (9,0);
\node[below] at (4.5,0) {$\one_c$};
}
\end{scope}

\begin{scope}[xshift=15cm, yshift=-18cm]{
\draw[thick, blue, fill=blue!50!white, opacity=0.5] (3,0) -- (6,0) -- (4,4) -- (3,0);
\draw[thick, blue, fill=blue!50!white, opacity=0.5] (3,0) -- (6,0) -- (7,1) -- (3,0);
\draw (0,0) -- (0,3) -- (3,3) -- (3,0) -- (0,0);
\draw[blue] (0,3) -- (1,4) -- (4,4) -- (3,3);
\draw[dashed, blue] (4,4) -- (4,1) -- (3,0);
\draw[dashed, blue] (1,4) -- (1,1) -- (0,0);
\draw[dashed, blue] (1,1) -- (4,1);
\draw (0,3) -- (3,0);
\draw[dashed, blue] (4,4) -- (3,0);
\draw[dashed, blue] (0,0) -- (1,4) -- (4,1) -- (0,0);
\draw[dashed, blue] (1,4) -- (3,0);
\draw[blue] (4,4) -- (0,3);

\begin{scope}[xshift=3cm,yshift=0cm]{
\draw (0,0) -- (0,3) -- (3,3) -- (3,0) -- (0,0);
\draw[blue] (0,3) -- (1,4) -- (4,4) -- (3,3);
\draw[dashed, blue] (4,4) -- (4,1) -- (3,0);
\draw[dashed, blue] (1,4) -- (1,1) -- (0,0);
\draw[dashed, blue] (1,1) -- (4,1);
\draw (0,3) -- (3,0);
\draw[dashed, blue] (4,4) -- (3,0);
\draw[dashed, blue] (0,0) -- (1,4) -- (4,1) -- (0,0);
\draw[dashed, blue] (1,4) -- (3,0);
\draw[blue] (4,4) -- (0,3);
}
\end{scope}

\begin{scope}[xshift=6cm,yshift=0cm]{
\draw (0,0) -- (0,3) -- (3,3) -- (3,0) -- (0,0);
\draw[blue] (0,3) -- (1,4) -- (4,4) -- (3,3);
\draw[blue] (4,4) -- (4,1) -- (3,0);
\draw[dashed, blue] (1,4) -- (1,1) -- (0,0);
\draw[dashed, blue] (1,1) -- (4,1);
\draw (0,3) -- (3,0);
\draw[blue] (4,4) -- (3,0);
\draw[dashed, blue] (0,0) -- (1,4) -- (4,1) -- (0,0);
\draw[dashed, blue] (1,4) -- (3,0);
\draw[blue] (4,4) -- (0,3);
}
\end{scope}

\begin{scope}[xshift=0cm, yshift=-3cm]{
\draw (0,0) -- (0,3) -- (3,3) -- (3,0) -- (0,0);
\draw[dashed, blue] (4,4) -- (4,1) -- (3,0);
\draw[dashed, blue] (1,4) -- (1,1) -- (0,0);
\draw[dashed, blue] (1,1) -- (4,1);
\draw (0,3) -- (3,0);
\draw[dashed, blue] (4,4) -- (3,0);
\draw[dashed, blue] (0,0) -- (1,4) -- (4,1) -- (0,0);
\draw[dashed, blue] (1,4) -- (3,0);

\begin{scope}[xshift=3cm,yshift=0cm]{
\draw (0,0) -- (0,3) -- (3,3) -- (3,0) -- (0,0);
\draw[dashed, blue] (4,4) -- (4,1) -- (3,0);
\draw[dashed, blue] (1,4) -- (1,1) -- (0,0);
\draw[dashed, blue] (1,1) -- (4,1);
\draw (0,3) -- (3,0);
\draw[dashed, blue] (4,4) -- (3,0);
\draw[dashed, blue] (0,0) -- (1,4) -- (4,1) -- (0,0);
\draw[dashed, blue] (1,4) -- (3,0);
}
\end{scope}

\begin{scope}[xshift=6cm,yshift=0cm]{
\draw (0,0) -- (0,3) -- (3,3) -- (3,0) -- (0,0);
\draw[blue] (4,4) -- (4,1) -- (3,0);
\draw[dashed, blue] (1,4) -- (1,1) -- (0,0);
\draw[dashed, blue] (1,1) -- (4,1);
\draw (0,3) -- (3,0);
\draw[blue] (4,4) -- (3,0);
\draw[dashed, blue] (0,0) -- (1,4) -- (4,1) -- (0,0);
\draw[dashed, blue] (1,4) -- (3,0);
}
\end{scope}
}
\end{scope}
\draw[thick, cyan, fill=cyan, opacity=0.5] (3,0) -- (6,0) -- (3,3) -- (3,0);
\draw[thick, cyan, fill=cyan, opacity=0.5] (3,0) -- (6,0) -- (6,-3) -- (3,0);
\draw[dashed, line width=0.05cm] (0,0) -- (9,0);
\node[below] at (4.5,0) {$\one_c$};
}
\end{scope}
\node at (1.5,-18) {$=$};
\node at (14,-18) {$\oplus$};
\end{tikzpicture}
\caption{The fusion of $\tilde y$ and $\tilde x$ is $1_{\one\oplus \sm}\oplus e$}\label{yfx}
\end{figure}

Then what happens to the condensation descendants? When the $\one_c$-string is moved onto the boundary it survives as the Figure \ref{1_cbdy} shows. The creation of $\one_c$ on the boundary is similar to the case in the bulk. Simply add $\tilde A_i/2$ and $(\sigma_z+1)/2$ for the vertices and edges respectively in the path of $\one_c$ to the boundary Hamiltonian. Due to the condensation of the $\sm$-string, $\sm_c$ and $\one_c$ strings are now indinguishable. They are identified. We denote them as $\one_c\oplus \sm_c$. Still it can have end points. The one from $\one\oplus \sm$ to $\one_c\oplus \sm_c$ is denoted as $\tilde x \colon \one\oplus \sm\rightarrow\one_c\oplus \sm_c$ and the adjoint is denoted as $\tilde y \colon \one_c\oplus \sm_c\rightarrow\one\oplus \sm$. They are the corresponding 0d excitations of $x$ and $y$ pulled onto the boundary. Further there is still one non-trivial 0d excitation, $\tilde z$, on $\one_c\oplus \sm_c$ corresponding to $z$ pulled onto the boundary. However now on the boundary $\tilde z$ is half of the $\sm$-string winding around the $\one_c\oplus \sm_c$ as depicted in Figure \ref{1_cbdy}. The end points of the $\sm$-string can either be $s$, $\bar s$ or their linear combination. As the $e$-particle condenses on the $\one_c$-string, the $\sm$-strings winding around the $\one_c$-string with different types of end points are identified.\\
The fusion of $\tilde y$ and $\tilde x$ is depicted as Figure \ref{yfx}. 
\\
which gives
\begin{equation}
\tilde y\circ_{\one\oplus \sm}\tilde x=1_{\one_c\oplus \sm_c}\oplus\tilde z
\end{equation}
where $1_{\one_c\oplus \sm_c}$ is the trivial 0d excitation on $\one_c\oplus \sm_c$. While for the fusion of $\tilde x$ and $\tilde y$, it can be viewed as a 0d $\one_c\oplus \sm_c$. Namely only add to the Hamiltonian $\tilde A_i$ at vertex $i$. Hence it will be a superposition of states with $\tilde A_i=\pm1$, which means
\begin{equation}
\tilde x\circ_{\one\oplus \sm}\tilde y=1_{\one\oplus \sm}\oplus e
\end{equation}
And for the fusion of $\tilde z$ and itself. As depicted in Figure \ref{zfz}, there are four end points.
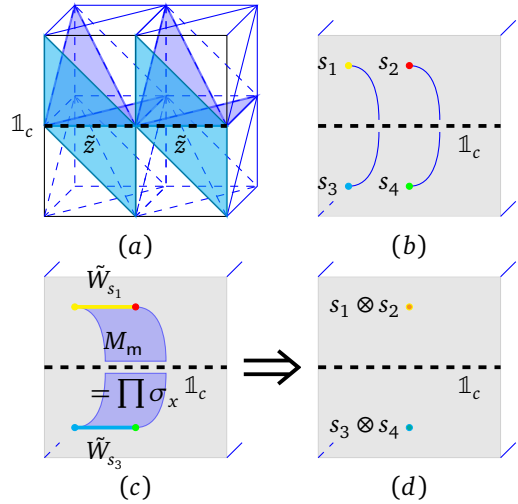
\begin{figure}[htbp!]
\centering
\begin{tikzpicture}[scale=0.4]
\begin{scope}[xshift=0cm, yshift=-3cm]{
\begin{scope}[xshift=3cm,yshift=0cm]{
\draw (0,0) -- (0,3) -- (3,3) -- (3,0) -- (0,0);
\draw[dashed, blue] (4,4) -- (4,1) -- (3,0);
\draw[dashed, blue] (1,4) -- (1,1) -- (0,0);
\draw[dashed, blue] (1,1) -- (4,1);
\draw (0,3) -- (3,0);
\draw[dashed, blue] (4,4) -- (3,0);
\draw[dashed, blue] (0,0) -- (1,4) -- (4,1) -- (0,0);
\draw[dashed, blue] (1,4) -- (3,0);
}
\end{scope}

\begin{scope}[xshift=6cm,yshift=0cm]{
\draw (0,0) -- (0,3) -- (3,3) -- (3,0) -- (0,0);
\draw[blue] (4,4) -- (4,1) -- (3,0);
\draw[dashed, blue] (1,4) -- (1,1) -- (0,0);
\draw[dashed, blue] (1,1) -- (4,1);
\draw (0,3) -- (3,0);
\draw[blue] (4,4) -- (3,0);
\draw[dashed, blue] (0,0) -- (1,4) -- (4,1) -- (0,0);
\draw[dashed, blue] (1,4) -- (3,0);
}
\end{scope}
}
\end{scope}

\begin{scope}[xshift=3cm,yshift=0cm]{
\draw[thick, blue, fill=blue!50!white, opacity=0.5] (0,0) -- (3,0) -- (1,4) -- (0,0);
\draw[thick, blue, fill=blue!50!white, opacity=0.5] (0,0) -- (3,0) -- (4,1) -- (0,0);
\draw (0,0) -- (0,3) -- (3,3) -- (3,0) -- (0,0);
\draw[blue] (0,3) -- (1,4) -- (4,4) -- (3,3);
\draw[dashed, blue] (4,4) -- (4,1) -- (3,0);
\draw[dashed, blue] (1,4) -- (1,1) -- (0,0);
\draw[dashed, blue] (1,1) -- (4,1);
\draw (0,3) -- (3,0);
\draw[dashed, blue] (4,4) -- (3,0);
\draw[dashed, blue] (0,0) -- (1,4) -- (4,1) -- (0,0);
\draw[dashed, blue] (1,4) -- (3,0);
\draw[blue] (4,4) -- (0,3);
\draw[thick, cyan, fill=cyan, opacity=0.5] (0,0) -- (3,0) -- (0,3) -- (0,0);
\draw[thick, cyan, fill=cyan, opacity=0.5] (0,0) -- (3,0) -- (3,-3) -- (0,0);
}
\end{scope}

\begin{scope}[xshift=6cm,yshift=0cm]{
\draw[thick, blue, fill=blue!50!white, opacity=0.5] (0,0) -- (3,0) -- (1,4) -- (0,0);
\draw[thick, blue, fill=blue!50!white, opacity=0.5] (0,0) -- (3,0) -- (4,1) -- (0,0);
\draw (0,0) -- (0,3) -- (3,3) -- (3,0) -- (0,0);
\draw[blue] (0,3) -- (1,4) -- (4,4) -- (3,3);
\draw[blue] (4,4) -- (4,1) -- (3,0);
\draw[dashed, blue] (1,4) -- (1,1) -- (0,0);
\draw[dashed, blue] (1,1) -- (4,1);
\draw (0,3) -- (3,0);
\draw[blue] (4,4) -- (3,0);
\draw[dashed, blue] (0,0) -- (1,4) -- (4,1) -- (0,0);
\draw[dashed, blue] (1,4) -- (3,0);
\draw[blue] (4,4) -- (0,3);
\draw[thick, cyan, fill=cyan, opacity=0.5] (0,0) -- (3,0) -- (0,3) -- (0,0);
\draw[thick, cyan, fill=cyan, opacity=0.5] (0,0) -- (3,0) -- (3,-3) -- (0,0);
}
\end{scope}
\node at (6,-4) {$(a)$};
\node at (15,-4) {$(b)$};
\draw[dashed, line width=0.05cm] (3,0) -- (9,0);
\node[left] at (3,0) {$\one_c$};
\node[below] at (4.5,0) {$\tilde z$};
\node[below] at (7.5,0) {$\tilde z$};
\node[below] at (17,0) {$\one_c$};
\draw[dashed, line width=0.05cm] (12,0) -- (18,0);
\draw[fill=black, opacity=0.1] (12,-3) -- (12,3) -- (18,3) -- (18,-3) -- (12,-3);
\draw[blue, dashed] (12,-3) -- (12.5,-2.5);
\draw[blue] (12,3) -- (12.5,3.5);
\draw[blue] (18,3) -- (18.5,3.5);
\draw[blue] (18,-3) -- (18.5,-2.5);
\draw[blue] (13,2) to [out=0, in=90] (14,0.2);
\draw[blue] (13,-2) to [out=0, in=-90] (14,-0.2);
\draw[blue] (15,2) to [out=0, in=90] (16,0.2);
\draw[blue] (15,-2) to [out=0, in=-90] (16,-0.2);
\draw[yellow, fill=yellow] (13,2) circle [radius=0.1cm];
\draw[red,fill=red] (15,2) circle [radius=0.1cm];
\draw[cyan, fill=cyan] (13,-2) circle [radius=0.1cm];
\draw[green,fill=green] (15,-2) circle [radius=0.1cm];
\node[left] at (13,2) {$s_1$};
\node[left] at (15,2) {$s_2$};
\node[left] at (13,-2) {$s_3$};
\node[left] at (15,-2) {$s_4$};
\begin{scope}[xshift=-9cm, yshift=-8cm]{
\node at (15,-4) {$(c)$};
\node[below] at (17,0) {$\one_c$};
\draw[dashed, line width=0.05cm] (12,0) -- (18,0);
\draw[fill=black, opacity=0.1] (12,-3) -- (12,3) -- (18,3) -- (18,-3) -- (12,-3);
\draw[blue, dashed] (12,-3) -- (12.5,-2.5);
\draw[blue] (12,3) -- (12.5,3.5);
\draw[blue] (18,3) -- (18.5,3.5);
\draw[blue] (18,-3) -- (18.5,-2.5);
\draw[blue, fill=blue!50!white, opacity=0.5] (13,2) to [out=0, in=90] (14,0.2) -- (16,0.2) to [out=90, in=0] (15,2) -- (13,2);
\draw[blue, fill=blue!50!white, opacity=0.5] (13,-2) to [out=0, in=-90] (14,-0.2) -- (16,-0.2) to [out=-90, in=0] (15,-2) -- (13,-2);
\draw[yellow, line width=0.05cm] (13,2) -- (15,2);
\draw[cyan, line width=0.05cm] (13,-2) -- (15,-2);
\draw[yellow, fill=yellow] (13,2) circle [radius=0.1cm];
\draw[red,fill=red] (15,2) circle [radius=0.1cm];
\draw[cyan, fill=cyan] (13,-2) circle [radius=0.1cm];
\draw[green,fill=green] (15,-2) circle [radius=0.1cm];
\node[above] at (14,2) {$\tilde W_{s_1}$};
\node[below] at (14,-2) {$\tilde W_{s_3}$};
\node[above] at (14.6,0.1) {$M_\sm$}; 
\node[below] at (15,-0.1) {$=\prod\sigma_x$};
}
\end{scope}
\begin{scope}[xshift=0cm, yshift=-8cm]{
\node at (15,-4) {$(d)$};
\node[below] at (17,0) {$\one_c$};
\draw[dashed, line width=0.05cm] (12,0) -- (18,0);
\draw[fill=black, opacity=0.1] (12,-3) -- (12,3) -- (18,3) -- (18,-3) -- (12,-3);
\draw[blue, dashed] (12,-3) -- (12.5,-2.5);
\draw[blue] (12,3) -- (12.5,3.5);
\draw[blue] (18,3) -- (18.5,3.5);
\draw[blue] (18,-3) -- (18.5,-2.5);
\draw[yellow, fill=yellow!50!red] (15,2) circle [radius=0.1cm];
\draw[cyan, fill=cyan!50!green] (15,-2) circle [radius=0.1cm];
\node[left] at (15,2) {$s_1\otimes s_2$};
\node[left] at (15,-2) {$s_3\otimes s_4$};
\node at (10.5,0) {\Huge{$\Rightarrow$}};
}
\end{scope}
\end{tikzpicture}
\caption{The fusion of $\tilde z$ and $\tilde z$ is $1_{\one_c\oplus \sm_c}$. $(a)$ and $(b)$: two $\tilde z$ on the $\one_c$-string with $s_1-s_4$ types of end points on the boundary. $(c)$: move the left $\tilde z$ towards the right one by string operators $\tilde W_{s_1}$ and $\tilde W_{s_3}$ on the boundary together with the membrane operator $M_m$ in the bulk. $(d)$: The remnant of the fusion is $s_1\otimes s_2$ and $s_3\otimes s_4$ which can be either the vacuum or $e$-particle on the boundary.}\label{zfz}
\end{figure}\\
No matter what types four end points are, we can pull them together with proper open string operators on the boundary and membrane operator in the bulk, e.g. $\tilde W_{s_1}$, $\tilde W_{s_3}$ and $M_m$ in Figure \ref{zfz}. Then what left are $s_1\otimes s_2$ and $s_3\otimes s_4$ on the boundary which are either the $e$-particle or vacuum. If there are $e$-particles left we can further pull them onto the $\one_c$-string and condense on it. Therefore this fusion is given by
\begin{equation}
\tilde z\circ_{\one\oplus \sm}\tilde z=1_{\one_c\oplus \sm_c}
\end{equation}
The last fusion rule we need to consider is between $\one_c\oplus \sm_c$ and itself, which is depicted in Figure \ref{1_cf1_c}, where the $\sm$-string in the bulk just beside the thicken $\one_c\oplus \sm_c$ can be move onto the boundary and disappears. Hence
\begin{equation}
(\one_c\oplus \sm_c)\otimes_{\one\oplus \sm}(\one_c\oplus \sm_c)=(\one_c\oplus \sm_c)\oplus(\one_c\oplus \sm_c)
\end{equation}

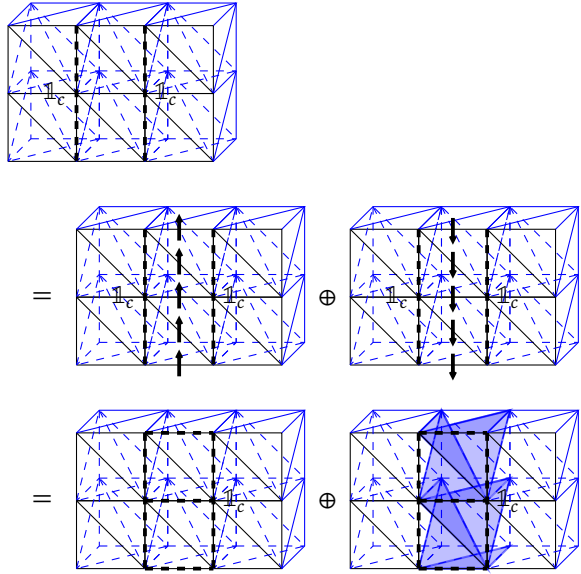
\begin{figure}[htbp!]
\centering
\begin{tikzpicture}[scale=0.3]
\draw (0,0) -- (0,3) -- (3,3) -- (3,0) -- (0,0);
\draw[blue] (0,3) -- (1,4) -- (4,4) -- (3,3);
\draw[dashed, blue] (4,4) -- (4,1) -- (3,0);
\draw[dashed, blue] (1,4) -- (1,1) -- (0,0);
\draw[dashed, blue] (1,1) -- (4,1);
\draw (0,3) -- (3,0);
\draw[dashed, blue] (4,4) -- (3,0);
\draw[dashed, blue] (0,0) -- (1,4) -- (4,1) -- (0,0);
\draw[dashed, blue] (1,4) -- (3,0);
\draw[blue] (4,4) -- (0,3);

\begin{scope}[xshift=3cm,yshift=0cm]{
\draw (0,0) -- (0,3) -- (3,3) -- (3,0) -- (0,0);
\draw[blue] (0,3) -- (1,4) -- (4,4) -- (3,3);
\draw[dashed, blue] (4,4) -- (4,1) -- (3,0);
\draw[dashed, blue] (1,4) -- (1,1) -- (0,0);
\draw[dashed, blue] (1,1) -- (4,1);
\draw (0,3) -- (3,0);
\draw[dashed, blue] (4,4) -- (3,0);
\draw[dashed, blue] (0,0) -- (1,4) -- (4,1) -- (0,0);
\draw[dashed, blue] (1,4) -- (3,0);
\draw[blue] (4,4) -- (0,3);
}
\end{scope}

\begin{scope}[xshift=6cm,yshift=0cm]{
\draw (0,0) -- (0,3) -- (3,3) -- (3,0) -- (0,0);
\draw[blue] (0,3) -- (1,4) -- (4,4) -- (3,3);
\draw[blue] (4,4) -- (4,1) -- (3,0);
\draw[dashed, blue] (1,4) -- (1,1) -- (0,0);
\draw[dashed, blue] (1,1) -- (4,1);
\draw (0,3) -- (3,0);
\draw[blue] (4,4) -- (3,0);
\draw[dashed, blue] (0,0) -- (1,4) -- (4,1) -- (0,0);
\draw[dashed, blue] (1,4) -- (3,0);
\draw[blue] (4,4) -- (0,3);
}
\end{scope}

\begin{scope}[xshift=0cm, yshift=-3cm]{
\draw (0,0) -- (0,3) -- (3,3) -- (3,0) -- (0,0);
\draw[dashed, blue] (4,4) -- (4,1) -- (3,0);
\draw[dashed, blue] (1,4) -- (1,1) -- (0,0);
\draw[dashed, blue] (1,1) -- (4,1);
\draw (0,3) -- (3,0);
\draw[dashed, blue] (4,4) -- (3,0);
\draw[dashed, blue] (0,0) -- (1,4) -- (4,1) -- (0,0);
\draw[dashed, blue] (1,4) -- (3,0);

\begin{scope}[xshift=3cm,yshift=0cm]{
\draw (0,0) -- (0,3) -- (3,3) -- (3,0) -- (0,0);
\draw[dashed, blue] (4,4) -- (4,1) -- (3,0);
\draw[dashed, blue] (1,4) -- (1,1) -- (0,0);
\draw[dashed, blue] (1,1) -- (4,1);
\draw (0,3) -- (3,0);
\draw[dashed, blue] (4,4) -- (3,0);
\draw[dashed, blue] (0,0) -- (1,4) -- (4,1) -- (0,0);
\draw[dashed, blue] (1,4) -- (3,0);
}
\end{scope}

\begin{scope}[xshift=6cm,yshift=0cm]{
\draw (0,0) -- (0,3) -- (3,3) -- (3,0) -- (0,0);
\draw[blue] (4,4) -- (4,1) -- (3,0);
\draw[dashed, blue] (1,4) -- (1,1) -- (0,0);
\draw[dashed, blue] (1,1) -- (4,1);
\draw (0,3) -- (3,0);
\draw[blue] (4,4) -- (3,0);
\draw[dashed, blue] (0,0) -- (1,4) -- (4,1) -- (0,0);
\draw[dashed, blue] (1,4) -- (3,0);
}
\end{scope}
}
\end{scope}
\draw[dashed, line width=0.05cm] (3,3) -- (3,-3);
\draw[dashed, line width=0.05cm] (6,3) -- (6,-3);
\node[right] at (6,0) {$\one_c$};
\node[left] at (3,0) {$\one_c$};

\begin{scope}[xshift=3cm, yshift=-9cm]{
\draw (0,0) -- (0,3) -- (3,3) -- (3,0) -- (0,0);
\draw[blue] (0,3) -- (1,4) -- (4,4) -- (3,3);
\draw[dashed, blue] (4,4) -- (4,1) -- (3,0);
\draw[dashed, blue] (1,4) -- (1,1) -- (0,0);
\draw[dashed, blue] (1,1) -- (4,1);
\draw (0,3) -- (3,0);
\draw[dashed, blue] (4,4) -- (3,0);
\draw[dashed, blue] (0,0) -- (1,4) -- (4,1) -- (0,0);
\draw[dashed, blue] (1,4) -- (3,0);
\draw[blue] (4,4) -- (0,3);

\begin{scope}[xshift=3cm,yshift=0cm]{
\draw (0,0) -- (0,3) -- (3,3) -- (3,0) -- (0,0);
\draw[blue] (0,3) -- (1,4) -- (4,4) -- (3,3);
\draw[dashed, blue] (4,4) -- (4,1) -- (3,0);
\draw[dashed, blue] (1,4) -- (1,1) -- (0,0);
\draw[dashed, blue] (1,1) -- (4,1);
\draw (0,3) -- (3,0);
\draw[dashed, blue] (4,4) -- (3,0);
\draw[dashed, blue] (0,0) -- (1,4) -- (4,1) -- (0,0);
\draw[dashed, blue] (1,4) -- (3,0);
\draw[blue] (4,4) -- (0,3);
}
\end{scope}

\begin{scope}[xshift=6cm,yshift=0cm]{
\draw (0,0) -- (0,3) -- (3,3) -- (3,0) -- (0,0);
\draw[blue] (0,3) -- (1,4) -- (4,4) -- (3,3);
\draw[blue] (4,4) -- (4,1) -- (3,0);
\draw[dashed, blue] (1,4) -- (1,1) -- (0,0);
\draw[dashed, blue] (1,1) -- (4,1);
\draw (0,3) -- (3,0);
\draw[blue] (4,4) -- (3,0);
\draw[dashed, blue] (0,0) -- (1,4) -- (4,1) -- (0,0);
\draw[dashed, blue] (1,4) -- (3,0);
\draw[blue] (4,4) -- (0,3);
}
\end{scope}

\begin{scope}[xshift=0cm, yshift=-3cm]{
\draw (0,0) -- (0,3) -- (3,3) -- (3,0) -- (0,0);
\draw[dashed, blue] (4,4) -- (4,1) -- (3,0);
\draw[dashed, blue] (1,4) -- (1,1) -- (0,0);
\draw[dashed, blue] (1,1) -- (4,1);
\draw (0,3) -- (3,0);
\draw[dashed, blue] (4,4) -- (3,0);
\draw[dashed, blue] (0,0) -- (1,4) -- (4,1) -- (0,0);
\draw[dashed, blue] (1,4) -- (3,0);

\begin{scope}[xshift=3cm,yshift=0cm]{
\draw (0,0) -- (0,3) -- (3,3) -- (3,0) -- (0,0);
\draw[dashed, blue] (4,4) -- (4,1) -- (3,0);
\draw[dashed, blue] (1,4) -- (1,1) -- (0,0);
\draw[dashed, blue] (1,1) -- (4,1);
\draw (0,3) -- (3,0);
\draw[dashed, blue] (4,4) -- (3,0);
\draw[dashed, blue] (0,0) -- (1,4) -- (4,1) -- (0,0);
\draw[dashed, blue] (1,4) -- (3,0);
}
\end{scope}

\begin{scope}[xshift=6cm,yshift=0cm]{
\draw (0,0) -- (0,3) -- (3,3) -- (3,0) -- (0,0);
\draw[blue] (4,4) -- (4,1) -- (3,0);
\draw[dashed, blue] (1,4) -- (1,1) -- (0,0);
\draw[dashed, blue] (1,1) -- (4,1);
\draw (0,3) -- (3,0);
\draw[blue] (4,4) -- (3,0);
\draw[dashed, blue] (0,0) -- (1,4) -- (4,1) -- (0,0);
\draw[dashed, blue] (1,4) -- (3,0);
}
\end{scope}
}
\end{scope}
\draw[dashed, line width=0.05cm] (3,3) -- (3,-3);
\draw[dashed, line width=0.05cm] (6,3) -- (6,-3);
\node[right] at (6,0) {$\one_c$};
\node[left] at (3,0) {$\one_c$};
\draw[line width=0.05cm] (4.5,-0.5) -- (4.5,0.5);
\draw[line width=0.05cm] (4.4,0.3) -- (4.5,0.5) -- (4.6,0.3);
\draw[line width=0.05cm] (4.5,2.5) -- (4.5,3.5);
\draw[line width=0.05cm] (4.4,3.3) -- (4.5,3.5) -- (4.6,3.3);
\draw[line width=0.05cm] (4.5,-3.5) -- (4.5,-2.5);
\draw[line width=0.05cm] (4.4,-2.7) -- (4.5,-2.5) -- (4.6,-2.7);
\draw[line width=0.05cm] (4.5,1) -- (4.5,2);
\draw[line width=0.05cm] (4.4,1.8) -- (4.5,2) -- (4.6,1.8);
\draw[line width=0.05cm] (4.5,-2) -- (4.5,-1);
\draw[line width=0.05cm] (4.4,-1.2) -- (4.5,-1) -- (4.6,-1.2);
}
\end{scope}

\begin{scope}[xshift=15cm, yshift=-9cm]{
\draw (0,0) -- (0,3) -- (3,3) -- (3,0) -- (0,0);
\draw[blue] (0,3) -- (1,4) -- (4,4) -- (3,3);
\draw[dashed, blue] (4,4) -- (4,1) -- (3,0);
\draw[dashed, blue] (1,4) -- (1,1) -- (0,0);
\draw[dashed, blue] (1,1) -- (4,1);
\draw (0,3) -- (3,0);
\draw[dashed, blue] (4,4) -- (3,0);
\draw[dashed, blue] (0,0) -- (1,4) -- (4,1) -- (0,0);
\draw[dashed, blue] (1,4) -- (3,0);
\draw[blue] (4,4) -- (0,3);

\begin{scope}[xshift=3cm,yshift=0cm]{
\draw (0,0) -- (0,3) -- (3,3) -- (3,0) -- (0,0);
\draw[blue] (0,3) -- (1,4) -- (4,4) -- (3,3);
\draw[dashed, blue] (4,4) -- (4,1) -- (3,0);
\draw[dashed, blue] (1,4) -- (1,1) -- (0,0);
\draw[dashed, blue] (1,1) -- (4,1);
\draw (0,3) -- (3,0);
\draw[dashed, blue] (4,4) -- (3,0);
\draw[dashed, blue] (0,0) -- (1,4) -- (4,1) -- (0,0);
\draw[dashed, blue] (1,4) -- (3,0);
\draw[blue] (4,4) -- (0,3);
}
\end{scope}

\begin{scope}[xshift=6cm,yshift=0cm]{
\draw (0,0) -- (0,3) -- (3,3) -- (3,0) -- (0,0);
\draw[blue] (0,3) -- (1,4) -- (4,4) -- (3,3);
\draw[blue] (4,4) -- (4,1) -- (3,0);
\draw[dashed, blue] (1,4) -- (1,1) -- (0,0);
\draw[dashed, blue] (1,1) -- (4,1);
\draw (0,3) -- (3,0);
\draw[blue] (4,4) -- (3,0);
\draw[dashed, blue] (0,0) -- (1,4) -- (4,1) -- (0,0);
\draw[dashed, blue] (1,4) -- (3,0);
\draw[blue] (4,4) -- (0,3);
}
\end{scope}

\begin{scope}[xshift=0cm, yshift=-3cm]{
\draw (0,0) -- (0,3) -- (3,3) -- (3,0) -- (0,0);
\draw[dashed, blue] (4,4) -- (4,1) -- (3,0);
\draw[dashed, blue] (1,4) -- (1,1) -- (0,0);
\draw[dashed, blue] (1,1) -- (4,1);
\draw (0,3) -- (3,0);
\draw[dashed, blue] (4,4) -- (3,0);
\draw[dashed, blue] (0,0) -- (1,4) -- (4,1) -- (0,0);
\draw[dashed, blue] (1,4) -- (3,0);

\begin{scope}[xshift=3cm,yshift=0cm]{
\draw (0,0) -- (0,3) -- (3,3) -- (3,0) -- (0,0);
\draw[dashed, blue] (4,4) -- (4,1) -- (3,0);
\draw[dashed, blue] (1,4) -- (1,1) -- (0,0);
\draw[dashed, blue] (1,1) -- (4,1);
\draw (0,3) -- (3,0);
\draw[dashed, blue] (4,4) -- (3,0);
\draw[dashed, blue] (0,0) -- (1,4) -- (4,1) -- (0,0);
\draw[dashed, blue] (1,4) -- (3,0);
}
\end{scope}

\begin{scope}[xshift=6cm,yshift=0cm]{
\draw (0,0) -- (0,3) -- (3,3) -- (3,0) -- (0,0);
\draw[blue] (4,4) -- (4,1) -- (3,0);
\draw[dashed, blue] (1,4) -- (1,1) -- (0,0);
\draw[dashed, blue] (1,1) -- (4,1);
\draw (0,3) -- (3,0);
\draw[blue] (4,4) -- (3,0);
\draw[dashed, blue] (0,0) -- (1,4) -- (4,1) -- (0,0);
\draw[dashed, blue] (1,4) -- (3,0);
}
\end{scope}
}
\end{scope}
\draw[dashed, line width=0.05cm] (3,3) -- (3,-3);
\draw[dashed, line width=0.05cm] (6,3) -- (6,-3);
\node[right] at (6,0) {$\one_c$};
\node[left] at (3,0) {$\one_c$};
\draw[line width=0.05cm] (4.5,-0.5) -- (4.5,0.5);
\draw[line width=0.05cm] (4.4,-0.3) -- (4.5,-0.5) -- (4.6,-0.3);
\draw[line width=0.05cm] (4.5,2.5) -- (4.5,3.5);
\draw[line width=0.05cm] (4.4,2.7) -- (4.5,2.5) -- (4.6,2.7);
\draw[line width=0.05cm] (4.5,-3.5) -- (4.5,-2.5);
\draw[line width=0.05cm] (4.4,-3.3) -- (4.5,-3.5) -- (4.6,-3.3);
\draw[line width=0.05cm] (4.5,1) -- (4.5,2);
\draw[line width=0.05cm] (4.4,1.2) -- (4.5,1) -- (4.6,1.2);
\draw[line width=0.05cm] (4.5,-2) -- (4.5,-1);
\draw[line width=0.05cm] (4.4,-1.8) -- (4.5,-2) -- (4.6,-1.8);
}
\end{scope}
\node at (1.5,-9) {$=$};
\node at (14,-9) {$\oplus$};

\begin{scope}[xshift=3cm, yshift=-18cm]{
\draw (0,0) -- (0,3) -- (3,3) -- (3,0) -- (0,0);
\draw[blue] (0,3) -- (1,4) -- (4,4) -- (3,3);
\draw[dashed, blue] (4,4) -- (4,1) -- (3,0);
\draw[dashed, blue] (1,4) -- (1,1) -- (0,0);
\draw[dashed, blue] (1,1) -- (4,1);
\draw (0,3) -- (3,0);
\draw[dashed, blue] (4,4) -- (3,0);
\draw[dashed, blue] (0,0) -- (1,4) -- (4,1) -- (0,0);
\draw[dashed, blue] (1,4) -- (3,0);
\draw[blue] (4,4) -- (0,3);

\begin{scope}[xshift=3cm,yshift=0cm]{
\draw (0,0) -- (0,3) -- (3,3) -- (3,0) -- (0,0);
\draw[blue] (0,3) -- (1,4) -- (4,4) -- (3,3);
\draw[dashed, blue] (4,4) -- (4,1) -- (3,0);
\draw[dashed, blue] (1,4) -- (1,1) -- (0,0);
\draw[dashed, blue] (1,1) -- (4,1);
\draw (0,3) -- (3,0);
\draw[dashed, blue] (4,4) -- (3,0);
\draw[dashed, blue] (0,0) -- (1,4) -- (4,1) -- (0,0);
\draw[dashed, blue] (1,4) -- (3,0);
\draw[blue] (4,4) -- (0,3);
}
\end{scope}

\begin{scope}[xshift=6cm,yshift=0cm]{
\draw (0,0) -- (0,3) -- (3,3) -- (3,0) -- (0,0);
\draw[blue] (0,3) -- (1,4) -- (4,4) -- (3,3);
\draw[blue] (4,4) -- (4,1) -- (3,0);
\draw[dashed, blue] (1,4) -- (1,1) -- (0,0);
\draw[dashed, blue] (1,1) -- (4,1);
\draw (0,3) -- (3,0);
\draw[blue] (4,4) -- (3,0);
\draw[dashed, blue] (0,0) -- (1,4) -- (4,1) -- (0,0);
\draw[dashed, blue] (1,4) -- (3,0);
\draw[blue] (4,4) -- (0,3);
}
\end{scope}

\begin{scope}[xshift=0cm, yshift=-3cm]{
\draw (0,0) -- (0,3) -- (3,3) -- (3,0) -- (0,0);
\draw[dashed, blue] (4,4) -- (4,1) -- (3,0);
\draw[dashed, blue] (1,4) -- (1,1) -- (0,0);
\draw[dashed, blue] (1,1) -- (4,1);
\draw (0,3) -- (3,0);
\draw[dashed, blue] (4,4) -- (3,0);
\draw[dashed, blue] (0,0) -- (1,4) -- (4,1) -- (0,0);
\draw[dashed, blue] (1,4) -- (3,0);

\begin{scope}[xshift=3cm,yshift=0cm]{
\draw (0,0) -- (0,3) -- (3,3) -- (3,0) -- (0,0);
\draw[dashed, blue] (4,4) -- (4,1) -- (3,0);
\draw[dashed, blue] (1,4) -- (1,1) -- (0,0);
\draw[dashed, blue] (1,1) -- (4,1);
\draw (0,3) -- (3,0);
\draw[dashed, blue] (4,4) -- (3,0);
\draw[dashed, blue] (0,0) -- (1,4) -- (4,1) -- (0,0);
\draw[dashed, blue] (1,4) -- (3,0);
}
\end{scope}

\begin{scope}[xshift=6cm,yshift=0cm]{
\draw (0,0) -- (0,3) -- (3,3) -- (3,0) -- (0,0);
\draw[blue] (4,4) -- (4,1) -- (3,0);
\draw[dashed, blue] (1,4) -- (1,1) -- (0,0);
\draw[dashed, blue] (1,1) -- (4,1);
\draw (0,3) -- (3,0);
\draw[blue] (4,4) -- (3,0);
\draw[dashed, blue] (0,0) -- (1,4) -- (4,1) -- (0,0);
\draw[dashed, blue] (1,4) -- (3,0);
}
\end{scope}
}
\end{scope}
\draw[dashed, line width=0.05cm] (3,3) -- (3,-3);
\draw[dashed, line width=0.05cm] (6,3) -- (6,-3);
\draw[dashed, line width=0.05cm] (3,0) -- (6,0);
\draw[dashed, line width=0.05cm] (3,3) -- (6,3);
\draw[dashed, line width=0.05cm] (3,-3) -- (6,-3);
\node[right] at (6,0) {$\one_c$};
}
\end{scope}

\begin{scope}[xshift=15cm, yshift=-18cm]{
\draw[blue, thick, fill=blue!50!white, opacity=0.5] (3,3) -- (6,3) -- (7,4) -- (3,3);
\draw[blue, thick, fill=blue!50!white, opacity=0.5] (3,3) -- (6,0) -- (7,4) -- (3,3);
\draw[blue, thick, fill=blue!50!white, opacity=0.5] (3,3) -- (6,0) -- (4,4) -- (3,3);
\draw[blue, thick, fill=blue!50!white, opacity=0.5] (3,0) -- (6,0) -- (4,4) -- (3,0);
\draw[blue, thick, fill=blue!50!white, opacity=0.5] (3,0) -- (6,0) -- (7,1) -- (3,0);
\draw[blue, thick, fill=blue!50!white, opacity=0.5] (3,0) -- (6,-3) -- (7,1) -- (3,0);
\draw[blue, thick, fill=blue!50!white, opacity=0.5] (3,0) -- (6,-3) -- (4,1) -- (3,0);
\draw[blue, thick, fill=blue!50!white, opacity=0.5] (3,-3) -- (6,-3) -- (4,1) -- (3,-3);
\draw[blue, thick, fill=blue!50!white, opacity=0.5] (3,-3) -- (6,-3) -- (7,-2) -- (3,-3);
\draw (0,0) -- (0,3) -- (3,3) -- (3,0) -- (0,0);
\draw[blue] (0,3) -- (1,4) -- (4,4) -- (3,3);
\draw[dashed, blue] (4,4) -- (4,1) -- (3,0);
\draw[dashed, blue] (1,4) -- (1,1) -- (0,0);
\draw[dashed, blue] (1,1) -- (4,1);
\draw (0,3) -- (3,0);
\draw[dashed, blue] (4,4) -- (3,0);
\draw[dashed, blue] (0,0) -- (1,4) -- (4,1) -- (0,0);
\draw[dashed, blue] (1,4) -- (3,0);
\draw[blue] (4,4) -- (0,3);

\begin{scope}[xshift=3cm,yshift=0cm]{
\draw (0,0) -- (0,3) -- (3,3) -- (3,0) -- (0,0);
\draw[blue] (0,3) -- (1,4) -- (4,4) -- (3,3);
\draw[dashed, blue] (4,4) -- (4,1) -- (3,0);
\draw[dashed, blue] (1,4) -- (1,1) -- (0,0);
\draw[dashed, blue] (1,1) -- (4,1);
\draw (0,3) -- (3,0);
\draw[dashed, blue] (4,4) -- (3,0);
\draw[dashed, blue] (0,0) -- (1,4) -- (4,1) -- (0,0);
\draw[dashed, blue] (1,4) -- (3,0);
\draw[blue] (4,4) -- (0,3);
}
\end{scope}

\begin{scope}[xshift=6cm,yshift=0cm]{
\draw (0,0) -- (0,3) -- (3,3) -- (3,0) -- (0,0);
\draw[blue] (0,3) -- (1,4) -- (4,4) -- (3,3);
\draw[blue] (4,4) -- (4,1) -- (3,0);
\draw[dashed, blue] (1,4) -- (1,1) -- (0,0);
\draw[dashed, blue] (1,1) -- (4,1);
\draw (0,3) -- (3,0);
\draw[blue] (4,4) -- (3,0);
\draw[dashed, blue] (0,0) -- (1,4) -- (4,1) -- (0,0);
\draw[dashed, blue] (1,4) -- (3,0);
\draw[blue] (4,4) -- (0,3);
}
\end{scope}

\begin{scope}[xshift=0cm, yshift=-3cm]{
\draw (0,0) -- (0,3) -- (3,3) -- (3,0) -- (0,0);
\draw[dashed, blue] (4,4) -- (4,1) -- (3,0);
\draw[dashed, blue] (1,4) -- (1,1) -- (0,0);
\draw[dashed, blue] (1,1) -- (4,1);
\draw (0,3) -- (3,0);
\draw[dashed, blue] (4,4) -- (3,0);
\draw[dashed, blue] (0,0) -- (1,4) -- (4,1) -- (0,0);
\draw[dashed, blue] (1,4) -- (3,0);

\begin{scope}[xshift=3cm,yshift=0cm]{
\draw (0,0) -- (0,3) -- (3,3) -- (3,0) -- (0,0);
\draw[dashed, blue] (4,4) -- (4,1) -- (3,0);
\draw[dashed, blue] (1,4) -- (1,1) -- (0,0);
\draw[dashed, blue] (1,1) -- (4,1);
\draw (0,3) -- (3,0);
\draw[dashed, blue] (4,4) -- (3,0);
\draw[dashed, blue] (0,0) -- (1,4) -- (4,1) -- (0,0);
\draw[dashed, blue] (1,4) -- (3,0);
}
\end{scope}

\begin{scope}[xshift=6cm,yshift=0cm]{
\draw (0,0) -- (0,3) -- (3,3) -- (3,0) -- (0,0);
\draw[blue] (4,4) -- (4,1) -- (3,0);
\draw[dashed, blue] (1,4) -- (1,1) -- (0,0);
\draw[dashed, blue] (1,1) -- (4,1);
\draw (0,3) -- (3,0);
\draw[blue] (4,4) -- (3,0);
\draw[dashed, blue] (0,0) -- (1,4) -- (4,1) -- (0,0);
\draw[dashed, blue] (1,4) -- (3,0);
}
\end{scope}
}
\end{scope}
\draw[dashed, line width=0.05cm] (3,3) -- (3,-3);
\draw[dashed, line width=0.05cm] (6,3) -- (6,-3);
\draw[dashed, line width=0.05cm] (3,0) -- (6,0);
\draw[dashed, line width=0.05cm] (3,3) -- (6,3);
\draw[dashed, line width=0.05cm] (3,-3) -- (6,-3);
\node[right] at (6,0) {$\one_c$};
}
\end{scope}
\node at (1.5,-18) {$=$};
\node at (14,-18) {$\oplus$};
\end{tikzpicture}
\caption{The fusion of $\one_c$ and $\one_c$ on the boundary is $\one_c\oplus\one_c$}\label{1_cf1_c}
\end{figure}

\subsection{The rough boundary}

\begin{figure}[htbp!] \label{fig:roughbc}
    \centering
    \begin{tikzpicture}[scale = 0.8]
        \draw[white, fill=blue!20!white, opacity=0.5] (0,3) -- (4,4) -- (3,3) -- (0,3);
        \node at (2.8,3.3) {$p_b$};
        \node at (2,3.5) {$e_1$};
        \node at (3.5,3.5) {$e_2$};
        \draw[line width=0.03cm, dashed] (0,0) -- (0,3) -- (3,3) -- (3,0) -- (0,0);
        \draw[blue] (0,3) -- (1,4) -- (4,4) -- (3,3);
        \draw[dashed, blue] (4,4) -- (4,1) -- (3,0);
        \draw[dashed, blue] (1,4) -- (1,1) -- (0,0);
        \draw[dashed, blue] (1,1) -- (4,1);
        \draw[line width=0.03cm, dashed] (0,3) -- (3,0);
        \draw[dashed, blue] (4,4) -- (3,0);
        \draw[dashed, blue] (0,0) -- (1,4) -- (4,1) -- (0,0);
        \draw[dashed, blue] (1,4) -- (3,0);
        \draw[blue] (4,4) -- (0,3);
    
        \begin{scope}[xshift=3cm,yshift=0cm]
            \draw[line width=0.03cm, dashed] (0,0) -- (0,3) -- (3,3) -- (3,0) -- (0,0);
            \draw[blue] (0,3) -- (1,4) -- (4,4) -- (3,3);
            \draw[dashed, blue] (4,4) -- (4,1) -- (3,0);
            \draw[dashed, blue] (1,4) -- (1,1) -- (0,0);
            \draw[dashed, blue] (1,1) -- (4,1);
            \draw[line width=0.03cm, dashed] (0,3) -- (3,0);
            \draw[dashed, blue] (4,4) -- (3,0);
            \draw[dashed, blue] (0,0) -- (1,4) -- (4,1) -- (0,0);
            \draw[dashed, blue] (1,4) -- (3,0);
            \draw[blue] (4,4) -- (0,3);
        \end{scope}
    
        \begin{scope}[xshift=6cm,yshift=0cm]
            \draw[line width=0.03cm, dashed] (0,0) -- (0,3) -- (3,3) -- (3,0) -- (0,0);
            \draw[blue] (0,3) -- (1,4) -- (4,4) -- (3,3);
            \draw[blue] (4,4) -- (4,1) -- (3,0);
            \draw[dashed, blue] (1,4) -- (1,1) -- (0,0);
            \draw[dashed, blue] (1,1) -- (4,1);
            \draw[line width=0.03cm, dashed] (0,3) -- (3,0);
            \draw[blue] (4,4) -- (3,0);
            \draw[dashed, blue] (0,0) -- (1,4) -- (4,1) -- (0,0);
            \draw[dashed, blue] (1,4) -- (3,0);
            \draw[blue] (4,4) -- (0,3);
        \end{scope}

    \end{tikzpicture}
    \caption{Rough boundary of the 3+1D toric code model: There is no degrees of freedom on the black dashed edges. 
    Hence the $B_p$ operators near the boundary is the product of two $\sigma_z$ operators,
    e.g. $B_{p_b}=\sigma_z^{e_1}\sigma_z^{e_2}$.}
    
\end{figure}
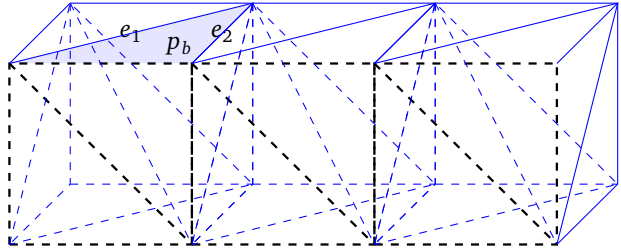

The rough boundary corresponding to the electric charge condensation is realized on the lattice
by projecting the boundary links to the trivial element. 
This is represented as dashed lines in Figure \ref{fig:roughbc}.
The boundary Hamiltonian contains $B_p$ terms that act on ``incomplete'' plaquettes where some links
are the dashed links.  The analysis of excitations is the same as the rough boundary of the 2+1D toric code model and we will not
further belabour about it here. 

\bibliography{anyon_cond}

\end{document}